\documentclass[12pt]{article}
\usepackage{graphicx}
\usepackage{url}
\usepackage{cite}
\usepackage[margin=1.25 in]{geometry}
\usepackage[colorlinks = true, linkcolor = blue, urlcolor = blue,
      citecolor = blue, anchorcolor = blue]{hyperref}

\newcommand\pubnumber{SLAC--PUB--17343}
\newcommand\pubdate{\hfill October, 2018}

\def\SLAC{SLAC,
    Stanford University, Menlo Park, California 94025 USA}
\def\doeack{\footnote{Work supported by the US Department of Energy,
                     contract DE--AC02--76SF00515.}}

\def\Title#1{\begin{center} {\Large #1 } \end{center}}
\def\Author#1{\begin{center}{ \sc #1} \end{center}}
\def\Address#1{\begin{center}{ \it #1} \end{center}}

\def\submit#1{\begin{center}Submitted to {\sl #1} \end{center}}
\newcommand\pubblock{\rightline{\begin{tabular}{l} \pubnumber\\
         \pubdate \end{tabular}}}
\newenvironment{Abstract}{\begin{quotation} \begin{center}
                       ABSTRACT
     \end{center}\bigskip  }{\end{quotation}}

\def\submit#1{\begin{center}Submitted to {\sl #1} \end{center}}
\def\Acknowledgements{\bigskip  \bigskip \begin{center} \begin{large}
             \bf ACKNOWLEDGEMENTS \end{large}\end{center}}

\def\beq{\begin{equation}}
\def\eeq#1{\label{#1}\end{equation}}
\def\eeqn{\end{equation}}

\newenvironment{Eqnarray}%
   {\arraycolsep 0.14em\begin{eqnarray}}{\end{eqnarray}}
\def\beqa{\begin{Eqnarray}}
\def\eeqa#1{\label{#1}\end{Eqnarray}}
\def\eeqan{\end{Eqnarray}}
\def\CR{\nonumber \\ }

\def\leqn#1{(\ref{#1})}

\let\bar=\overbar

\def\VEV#1{\left\langle{ #1} \right\rangle}

\def\ket#1{\left| {#1} \right\rangle}

\def\lsim{\mathrel{\raise.3ex\hbox{$<$\kern-.75em\lower1ex\hbox{$\sim$}}}}
\def\gsim{\mathrel{\raise.3ex\hbox{$>$\kern-.75em\lower1ex\hbox{$\sim$}}}}

\def\D{{\cal D}}

\def\M{{\cal M}}
\def\O{{\cal O}}

\def\tr{{\mbox{\rm tr}}}
\def\half{\frac{1}{2}}

\def\del{\partial}
\def\Dslash{\not{\hbox{\kern-4pt $D$}}}
\def\dslash{\not{\hbox{\kern-2pt $\del$}}}

\def\ee{e^+e^-}

\def\msb{{\bar{\scriptsize M \kern -1pt S}}}

\def\drb{{\bar{\scriptsize D \kern -1pt R}}}

\def\eps{\epsilon}

\makeatletter
\def\section{\@startsection{section}{0}{\z@}{5.5ex plus .5ex minus
 1.5ex}{2.3ex plus .2ex}{\large\bf}}
\def\subsection{\@startsection{subsection}{1}{\z@}{3.5ex plus .5ex minus
 1.5ex}{1.3ex plus .2ex}{\normalsize\bf}}
\def\subsubsection{\@startsection{subsubsection}{2}{\z@}{-3.5ex plus
-1ex minus  -.2ex}{2.3ex plus .2ex}{\normalsize\sl}}

\renewcommand{\@makecaption}[2]{%
   \vskip 10pt
   \setbox\@tempboxa\hbox{\small #1: #2}
   \ifdim \wd\@tempboxa >\hsize     
       \small #1: #2\par          
     \else                        
       \hbox to\hsize{\hfil\box\@tempboxa\hfil}
   \fi}

\makeatother

\def\A{{\cal A}}
\def\J{{\cal J}}

\def\S{{\cal S}} 
\def\Z{{\cal Z}} 

\begin{document}
\begin{titlepage}
\pubblock

\vfill
\Title{Dissection of an $SO(5) \times U(1)$ Gauge-Higgs Unification Model}
\vfill
\Author{Jongmin Yoon and Michael E. Peskin\doeack}
\Address{\SLAC}
\vfill
\begin{Abstract}
We analyze models of electroweak symmetry
breaking in warped 5-dimensional space with gauge bosons and fermions
in the bulk.  The Higgs boson is identified with the 5th
component of a gauge field.  We dynamically generate the Higgs
potential using a competition between the top quark multiplet and
another fermion multiplet to create a little hierarchy characterized
by a small parameter $s =  v/f$.   Using a Green's function method, we
compute  the properties of the model systematically as a power series in $s$.  We
discuss the constraints on this model from the measured value of the
Higgs mass, the masses of top quark partners, and precision
electroweak observables. 
\end{Abstract} 
\vfill
\submit{Physical Review D}
\vfill
\end{titlepage}

\hbox to \hsize{\null}


\tableofcontents

\def\thefootnote{\fnsymbol{footnote}}
\newpage
\setcounter{page}{1}

\setcounter{footnote}{0}

\section{Introduction}

The Standard Model of particle physics  (SM) gives an 
excellent description of elementary particle interactions as observed
today
at particle accelerators.  But, at the same time, this model seems manifestly
incomplete.  The most important qualitative phenomenon in this model,
the spontaneous breaking of its gauge symmetry $SU(2)\times U(1)$, is
put in by hand, by the assumption of a fundamental Higgs scalar field with
negative mass parameter.  This assumption also makes it impossible,
within the model,  to
compute the Yukawa couplings that determine the fermion mass spectrum.

There are two strategies to build a more predictive theory of
$SU(2)\times U(1)$ symmetry breaking.  One is to keep the assumption
that the breaking is due to a fundamental Higgs  field but add a
strong symmetry such as supersymmetry that constrains its
behavior.   The other is to assume that the Higgs field is composite,
formed from some underlying strong dynamics.   The discussions of
these possibilities in the literature contrast greatly.  Since
supersymmetry allows a weak-coupling description, it is possible to
work out the phenomenology in great detail, defining a ``Minimal
Supersymmetric Standard Model'' and canonical non-minimal extensions,
and exploring the properties of these models in every corner of their
parameter spaces~\cite{Godbole,BaerTata,PMSSM}.  

On the other hand, models with a composite Higgs
field are much more difficult to bring under control.  Descriptions of
these models involve strong coupling.  Reviews of the phenomenology of
these models are then necessarily 
qualitative~\cite{Wulzerreview,Csakireview,CGTreview} and
studies of particular models are done by large-scale  parameter
scans~\cite{ACP,Carena,Croon}.

In an attempt to improve this situation, we have been studying the
approach to composite Higgs models given by
 Randall-Sundrum (RS) theory~\cite{RS}.
In this approach, the four-dimensional strong-coupling dynamics is
taken to be dual to a weak-coupling five-dimensional dynamics in a
slice of anti-de Sitter space~\cite{RAP}.  The boundaries of this
slice
define infrared (IR) and ultraviolet (UV) scales for the action of the
new strong forces.
 It is attractive to link this idea to
that of Gauge-Higgs unification, in which the Higgs field is the fifth
component of a five-dimensional gauge field in the bulk
space~\cite{gaugeHiggsone,gaugeHiggstwo}.  Electroweak 
symmetry breaking can be achieved dynamically
from condensation of 5-dimensional fermions~\cite{Hosotani,Toms,CNP}.  
We do not need to introduce any fundamental scalars.
In this way, it is possible to build realistic
theories with a minimal number of free parameters.

Realistic RS models necessarily include hierarchies.   These models
require heavy vectorlike partners of the top quark and heavy
resonances with the quantum numbers of the SM gauge
bosons, and these are not yet observed at the LHC.   In RS models,
these
 heavy particles appear as Kaluza-Klein (KK) recurrences
in the fifth dimension and  have  masses that are several times the IR scale.
The constraints on these particles, especially from precision
electroweak measurements,  are sufficiently strong that their masses
must be well above 1~TeV.  Gauge-Higgs
unification models contain nonlinear sigma model fields whose dynamics
is governed by a decay constant $f$, which is of the order of the RS  IR
scale.  In this paper, we will arrange that there
is a hierarchy between the Higgs field vacuum expectation value $v$
and the nonlinear sigma model scale $f$:  $v/f \ll 1$.   Ideally, this
hierarchy should appear naturally, but here it will be arranged by
fine-tuning.  However, once the hierarchy has been arranged, the KK 
masses are also set to be much larger than $v$.  We can
use $v/f$ as an expansion parameter to organize the corrections from
the new physics present in the RS model.   Using this expansion as a
guide, we will be able to present these effects systematically.

Once this tuning is done, we will be able to focus on the mass ratios
of the heaviest SM particles---the $W$ and $Z$, the top
quark, and the Higgs boson.   In the simplest RS models with gauge-Higgs
unification, the masses of these particles are all approximately
equal.   The ratio of these
masses can be corrected by an idea that fits naturally with the picture
that the RS model is a dual description of a strong-coupling theory in
4 dimensions.  In a complete 4-dimensional model, the electroweak
gauge coupling and the top quark Yukawa coupling will be determined by
dynamics at a very high mass scale, perhaps the grand unification or
string scale.   This boundary condition at a  high mass scale can be
represented  phenomenologically in an RS model by operators on the UV
boundary of the RS interval.

In this paper, we will show how these ideas are realized in the
simplest possible scheme for the bulk gauge group, the $SO(5)\times
U(1)$  gauge symmetry put forward by Agashe, Contino, and Pomarol as
the basis for
the ``minimal composite Higgs model''~\cite{ACP}. 
 We will ignore the dynamics of light
flavors and concentrate on electroweak symmetry breaking driven by top
quark condensation.  With this restriction, the number of parameters
of the  model is small, and the parameter space of the model is
straightforward to describe.

The phenomenology of electroweak symmetry breaking, including 
the computation of precision electroweak corrections, has been studied 
previously in similar models~\cite{Maru, Wagner1, Wagner2, Neubert}.

The outline of this paper is the following: Sections 2--4 will
describe the construction of the model.   In Section 2, we will
recall some basic formalism of RS models and present our notation.
In Section~3, we will discuss the coupling of the top quark to the  $SO(5)$
gauge field in the bulk of the RS space.
 We will introduce the idea of 
competition between 5-dimensional fermion multiplets as a mechanism for
achieving the $v/f$ hierarchy~\cite{YPone}.   This mechanism will
give us a simple tuning parameter and, at the same time, will provide
a robust Higgs quartic term.  We will then discuss the problem of
obtaining the correct ratios 
 of the $W$, Higgs, and top quark masses.
In Section 4, we will introduce UV boundary kinetic terms for the $W$,
$Z$, and  top quark and explain how to adjust these boundary terms to
fit the mass ratios that are observed in nature.

Section 5 presents  the heart of our analysis.  In this section, we will
write the full Higgs potential in our model and minimize it.  We will
show that, after applying constraints from the $W$, $Z$, $t$ and Higgs
masses and other constraints from precision electroweak measurements,
we are left with only 3 parameters to vary.  One of these controls the
$v/f$ hierarchy and the scale of the KK resonance masses.  One controls the degree of
compositeness of the top quark and its competing vectorlike top
multiplet.
The final parameter  turns out to be 
almost irrelevant, affecting the physics of the model
only very weakly.   Thus, the model gives us essentially a 2-parameter
space to explore.

The last sections of this paper will analyze the effects of new
physics on precision electroweak observables in this 2-parameter space
of models.
 In Section 6, we will compute  the precision electroweak 
parameters $S$ and $T$~\cite{PT} and explain the constraints on the
parameter
space that these imply.  To analyze $S$ and $T$, we will introduce 
a systematic expansion of electroweak amplitudes in powers of $v/f$.
This expansion applies to a broad class of RS models beyond the 
specific models constructed in this paper.   In Section 7, we will
study 
the effect of new
physics on the partial width for 
$Z\to b\bar b$.  Although this observable provides a strong constraint
on some classes of composite Higgs models~\cite{Selipsky}, we will find
that the constraint on our RS models is relatively weak.
Section 8 will give some conclusions. 

In this paper, we will ignore the masses of all SM fermions
except  the top quark.   There are more issues to discuss in the quantum
number assignments for generating
masses for the lighter quarks and leptons.  Most immediately, this RS
model predicts deviations from the SM in $\ee\to f\bar f$
reactions at higher energy  that depend on the detailed scheme for generating the 
light fermion masses.   We will present these in the next paper in
this series~\cite{YPthree}.

\section{$SO(5) \times U(1)$ Model}

This section establishes the basic formalism and notation for our
discussion of RS models with gauge-Higgs unification.   The notation
follows that of \cite{YPone}.

\subsection{Overview}

We consider a model of gauge and fermion fields living in the interior
of a slice 
of 5-dimensional anti-de Sitter space 
\beq
      ds^2 =  {1\over (kz)^2} [ dx^m dx_m - dz^2 ] 
\eeq{metric}
 with nontrivial boundary conditions at $z = z_0$ and $z = z_R$, with
 $z_0 < z_R$. 
Then $z_0$ gives the position of the ``UV brane'' and
$z_R$ gives the position of the ``IR brane''.     The perhaps more
physical metric
\beq
     ds^2 = e^{-2k x^5} dx^m dx_m  -   (dx^5)^2
\eeq{RSmetric}
is related by $kz = \exp[ k x^5]$.  We
take the  size of the interval in $x^5$ to be $\pi R$.  Then
\beq
        z_0 = {1/k}  \qquad       z_R  = 1/k_R \equiv    e^{\pi k R}/k \ . 
\eeq{zzeroRdef}
The scales $k$ and $k_R$ set the ultraviolet and infrared boundaries
of the dynamics described by the 5D fields.  

 For concreteness, we  will 
be interested in values of $k_R$ of order 1~TeV and values of $k$ of
order 100~TeV.  Thus, we imagine that $z_0$ is at a flavor dynamics scale
rather than at the Planck scale.

The bulk action of gauge fields and fermions in RS  is 
\beq 
S_{bulk} = \int d^4 x dz\ \sqrt{g}  \biggl[ - {1\over 4} g^{MP} g^{NQ}
  F^a_{MN}
F^a_{PQ} + \bar \Psi  \bigl[ i e^M_A \gamma^A {\cal D}_M 
- m_{\Psi} \bigr] \Psi  \biggr]    \ .
\eeq{bulkaction}
We will notate gauge fields as $A_M^A$, where $M = 0,1,2,3,5$, with
lower case $m = 0,1,2,3$. Fermion fields are 4-component Dirac
fields.  We parametrize the 5D Dirac mass as
\beq
m_{\Psi}=c\,k \ ,
\eeq{cdefin}
defining a dimensionless parameter $c$ for each Dirac multiplet. 
 In our formalism, the Higgs field is a background gauge
field, so we will quantize in the Feynman-Randall-Schwartz  background
field gauge~\cite{RandallSchwartz}.   

Green's functions in RS will be important to our analysis.
Since \cite{YPone} is devoted to the calculation of the
Coleman-Weinberg potential in RS models, formulae for Green's
functions are given there for Euclidean momenta.   In this paper, we will work
with Green's functions with  Minkowski momenta.  Euclidean Green's
functions,
 where they appear, will be denoted $G_E$.

The solutions of field equations in the RS geometry with Minkowski
momenta  are given in terms of 
Bessel functions in the form~\cite{GW,DHR,GP}
\beq
     \Phi =   z^a [ A  J_\nu(pz) + B Y_\nu(pz) ] e^{-ip\cdot x} \ . 
\eeq{Besselsolutions}
It is useful to define  combinations of the Bessel functions so that
the solutions \leqn{Besselsolutions}, as a function of $z = z_1$,  have 
definite
 boundary conditions at a point $z  = z_2$.   Thus we set
\beq
    G_{\alpha\beta}(z_1,z_2) = {\pi \over 2} \left[ J_{\alpha}(p z_1) 
 Y_{\beta} (pz_2)
   -  Y_\alpha(p z_1)  J_\beta (pz_2) \right] \ , 
\eeq{Gdef}
where $\alpha, \beta = \pm 1$. For solutions to the Dirac equation, 
the orders of the Bessel functions
depend on the parameter $c$ according to 
\beq
  \mbox{for} \ \alpha,\beta =+1\  : \ \nu_+ =  c+\half \ ; \qquad 
 \mbox{for} \ \alpha,\beta =-1\  : \  \nu_- = c-\half    \ .
\eeq{alphabetac}
Then $G_{++}(z,z_R)$, $G_{--}(z,z_R)$ give solutions of the Dirac
equation with Dirichlet boundary conditions on the IR brane:
$\Phi(z,z_R) = 0$ at $z = z_R$.
 Similarly, $G_{+-}(z,z_R)$, $G_{-+}(z,z_R)$ will give solutions with 
Neumann boundary conditions on the IR brane. The solutions to
Maxwell's equations are given similarly by these Green's functions for
$c = 1/2$.  Further properties of these Green's functions are given in
Appendix A. The Euclidean Green's function $G_{E\alpha\beta}$ is analogously defined in \leqn{euclideanG}. 

\subsection{Group structure and boundary conditions}

We choose the bulk gauge symmetry to be $G = SO(5) \times U(1)_X$
\cite{ACP}. Boundary conditions
 break the bulk symmetry to the SM gauge symmetry $G_{SM}
 = SU(2)_L \times U(1)_Y$ on the UV brane and to $H = SO(4) \times
 U(1)_X = SU(2)_L \times SU(2)_R \times U(1)_X$ on the IR brane. This
 model can be viewed as a dual description of an approximately
 conformal dynamics between the scales $k_R=1/z_R$ and $k = 1/z_0$ in
 four dimensions. In the dual 4D interpretation, the system
 has a global symmetry $G$, 
of which the subgroup $G_{SM}$ is gauged to a local symmetry. The strongly
interacting theory spontaneously breaks $G$ to the subgroup $H$ at the
scale $k_R$. The extra $SU(2)$ factor in $H$ is a custodial symmetry
that protects the relation $m_W =  c_w m_Z$ from receiving large 
corrections~\cite{custodial}.  The study 
of the 5D model gives a calculable approach to the 4D theory.

We decompose the adjoint representation of the $SO(5)\times U(1)$
gauge group as described in Appendix B.   The 10 generators of $SO(5)$
are labelled  $T^a_L$, $T^a_R$, $T^{a5}$, $T^{45}$, with $a = 1,2,3$.
Consider first a pure $SO(5)$ model, with $SO(5)$ broken to an $SO(4)$ 
containing the 4D local gauge group $SU(2)_L$.
The boundary conditions for the $M=m$ components of the gauge fields 
would be
\beqa 
A^{aL}_{m} &\sim & \pmatrix{+&+} \CR
A^{aR}_m &\sim &  \pmatrix{-&+}   \CR
A^{a5}_m, A^{45}_m &\sim & \pmatrix{-&-}
\eeqa{gaugebc}
with $a=1,2,3$  and  $+$ ($-$) indicates 
Neumann (Dirichlet) boundary conditions on the left (UV) and right
(IR) boundaries.  The zero modes of the 
$A^{aL}_m$ fields would be the 4D $SU(2)_L$ gauge bosons.
 The components $A^A_5$ have the opposite
boundary conditions to those of the $A^A_m$, so $A^{a5}_5$, $A^{45}_5$
have zero modes that can be identified with the Goldstone bosons.
To set up an 
$SO(5)\times U(1)$ model containing the 4D local gauge group 
 $SU(2)\times U(1)$ on the
UV boundary, we introduce a $U(1)$ gauge field $A^X_M$ and
mix it with the field $A^{3R}_M$.   Let $g_5$ and $g_X$ be the
5D gauge couplings of $SO(5)$ and $U(1)$.  
Introduce an angle $\beta$ such that 
\beq
c_\beta \equiv \cos \beta = {g_5 \over \sqrt{g_5^2 + g_X^2}}, \qquad
s_\beta \equiv \sin \beta = {g_X \over \sqrt{g_5^2 + g_X^2}} \ .
\eeqn
We assign the combinations
\beq  \pmatrix{ Z'_m \cr B_m } =
\pmatrix{ c_\beta & -s_\beta \cr s_\beta & c_\beta} \pmatrix{ A^{3R}_m
  \cr A^X_m } 
\eeq{BZ}  
to have the boundary conditions
\beqa
B_m &\sim&  \pmatrix{+ & +} \CR
Z'_m &\sim&  \pmatrix{-& +}   \   .
\eeqa{BZbc}
The zero mode of the field $B_m$ is the 4D $U(1)$ gauge boson.
In terms of the gauge fields with definite boundary conditions, the 5D
 covariant derivative is
\beq
D_M = \partial_M -i\left[ g_5 A_M^{bL} T^{bL} + g_{5Y}  B_M Y
 + g_5 A_M^{aR} T^{aR} + {g_5 \over c_\beta} Z'_M (T^{3R} 
- s^2_\beta Y)  + g_5 A_M^{c5} T^{c5}  \right] \, ,
\eeq{DM}
summed over $a = 1,2,3$, $b = 1,2$, and  $c = 1,2,3,4$. The 5D hypercharge coupling is given by $g_{5Y} = g_5 s_\beta$. The hypercharge and electric charge are given by
\beq
Y = T^3_R + X \qquad \textrm{and} \qquad Q = T^3_L + T^3_R + X
\eeq{charges}
where $X$ is the $U(1)_X$ charge.

\subsection{Identification of the Higgs field}
 
The four zero-modes $A^{a5}_5, A^{45}_5$ transform as a complex doublet under
$SU(2)_L$. In the dual picture, they correspond to massless Goldstone
bosons of the broken global symmetry $G/H$. We identify them as the
Higgs doublet. Since  $A^A_5$ zero modes are proportional to $z$, we
can represent these zero modes as
\beq
    A^{c5,0}_5 (z,x^m) = N_h\, z\, h^c (x^m)\ ,
\eeq{Afiveform}
 with $c=1,2,3,4$ and $N_h$  a normalization constant:  
\beq
N_h = [(z_R^2  - z_0^2)/2k]^{-1/2}  \   . 
\eeq{normofAfive}
 Because
 the Higgs fields appear as components of gauge fields, we can gauge away
their vacuum expectation values in the central region of $z$. However,
 in a 5D system with boundaries, we cannot gauge away these
 background fields completely.  Instead, such a gauge transformation leaves
singular fields at $z_0$ or $z_R$. We can
 parametrize the gauge-invariant information of the background fields
 in terms of a Wilson line element from $z_0$ to $z_R$.  The 
Coleman-Weinberg potential of the Higgs field will depend
 on this variable~\cite{YPone}.

We can  align the expectation value along the $A^{45}$
direction, $\VEV{h^c} = \VEV{h} \delta^{c4} 
\neq 0$ in \leqn{Afiveform}.  Then the Wilson line element becomes
\beq
     U_W = \exp \left( - i g_5 \int_{z_0}^{z_R} dz \, N_h z 
   \VEV{h} T^{45} \right) 
= \exp \left( - \sqrt{2} i { \VEV{h} \over f} T^{45} \right)  \  .
\eeq{UW}
This equation introduces the Goldstone boson
decay constant  $f$, analogous to the pion decay constant in QCD. Explicitly,
\beq
{1 \over f} = {g_5 \over  \sqrt{2} } \int_{z_0}^{z_R} dz \, N_h z 
 = \sqrt{{g_5^2 k \over 4} (z_R^2 - z_0^2)} \simeq {\sqrt{g_5^2
     k}\over 2}
 z_R \ .
\eeq{f}
The magnitude of  $f$ is determined by the IR scale $k_R$ and the 5D gauge
coupling $g_5$.   The standard identification of a dimensionless
 4D gauge coupling
in RS is~\cite{RandallSchwartz}
\beq
         g^2   =    {  g_5^2 \over  \pi R}  = {g_5^2 k \over \log
           z_R/z_0} \ .
\eeq{firstgfiveid}
Then for our models with the parameter choice $k/k_R = z_R/z_0 = 100$, we have
\beq
      f =   {0.93 \over g}\  k_R \ .
\eeq{valoff}
The size of $f$ relative to the IR cutoff depends on the strength of
the 5D coupling $g_5$.

 For each type of field, its representation under the
bulk gauge group $G$ will determine the exact form of the $T^{45}$
matrix. Therefore, the Coleman-Weinberg potential will depend on the
choice of fermion representation. More details of the needed
$SO(5)$ group theory can be found in Appendix B.  

From the form of $U_W$ in \leqn{UW}, it is natural to expect that the 
potential for $h$ will be minimized either for 
$\VEV{h} =0 $ or  for $\VEV{h} \sim f$ for most of the parameter
space. However, our vacuum should satisfy $0 < \VEV{h} \ll f$. This
implies that we must be near a second-order phase transition in the
phase diagram of the system. To implement this, the field content of our system
should provide competing contributions to the Coleman-Weinberg
potential so that the vacuum 
is in the vicinity of the phase transition with a hierarchy $\VEV{h} \ll f$.

\section{$W$/Higgs/top masses in reference models}

Before introducing a complete model, we describe some aspects of our 
model-building approach and present some estimates
of the model parameters.  Complete RS models typically invoke some
boundary interactions in addition to interactions in the bulk of 5D
AdS.  In this section, we will explain why such boundary interactions
are needed in our models by estimating the RS coupling values 
in simplified models in which these terms are absent.

\subsection{Fermion competition and
 electroweak symmetry breaking}

In this paper, we consider models in which the 5D multiplet containing
the top quark is the main
driving force for the electroweak symmetry breaking (EWSB). In composite
Higgs models, it is well known that the Coleman-Weinberg potential
from gauge fields always prefer the symmetric point $\VEV{h} = 0$,
while fermion fields, particularly the top quark, can give negative
contributions to the potential and therefore trigger the EWSB. If the
Higgs field pairs up two massless Weyl fermions with opposite
chirality, it is energetically favorable to give an expectation value
to the Higgs and form a massive Dirac fermion.  We call this an 
`attractive' fermion multiplet.

In models with only gauge fields and the top quark multiplet $\Psi_t$, it is
possible to fine-tune the (mass)$^2$ of the Higgs boson to a small
value compared to the scale $k_R$. But typically this also results in
a small value for the Higgs quartic term, due to cancellations of the
contributions to the quartic from the two sources,  and a minimum of the
potential at $\VEV{h}\sim f$.   To achieve $\VEV{h} \ll f$, we will
introduce a second multiplet of vector-like fermions $\Psi_T$ that gives a
positive contribution to the Higgs potential and opposes the fermion
condensation.  We call this a 
`repulsive' fermion multiplet.
 In \cite{YPone}, we gave examples of the competition
between fermion multiplets in simple $SU(2)$ models and showed how
these can lead to $\VEV{h} \ll f$.  In this more realistic setting,
the multiplet $\Psi_T$ will include a vector-like top partner $T$ that is
naturally light compared to the KK scale $k_R$. 

Our analysis in this paper will explore the interplay of these two
fermion mutiplets and their consequences for precision electroweak       
observables and the properties of the top quark and the Higgs boson.
We will not discuss here the inclusion of light fermions and the issues of
flavor and flavor-changing transitions.  We believe that it is
possible to build an acceptable theory of flavor based on this model
by introducing additional fermion multiplets with $c > 1/2$, peaked near the
ultraviolet boundary~\cite{GrossNeu}.  However, a full
analysis of the flavor dynamics  is beyond the scope of this paper.

\subsection{Top quark embeddings}

Our first task is to embed the top quark into an $SO(5)$ multiplet.
This multiplet must contain the $(t_L,b_L)$ doublet, so that $SU(2)$
gauge bosons can link these states, and the $t_R$, so that the $U_W$
matrix can link this state to the $t_L$.   Before electroweak symmetry
breaking, the spectrum of states in each multiplet depends on the 
boundary conditions.   Our conventions for fermion boundary conditions
are given in Appendix A.2. 

In our models, the $(t_L, b_L)$ will be
left-handed zero modes, requiring $(++)$ boundary conditions.   The $t_R$
will be a right-handed zero mode, with $(--)$ boundary conditions.
If the $t_L$ and $t_R$ are to be linked by a Higgs field, all three
fields should belong to the same 5D multiplet.   In the models we
consider here, 
we will not include the $b_R$ in this multiplet.   This gives the
bottom quark zero mass in the approximation used in this paper.  However, 
it also explicitly breaks the $SO(4)$
custodial symmetry.  We will see later that this produces a
loop-suppressed contribution to the electroweak 
$T$ parameter~\cite{PT}.

There is a strong possibility of confusion between the labels $L,R$ used
for the 4D chirality components of a 5D Dirac field in Appendix A.2 
and the SM  labels such as $t_L$ and $t_R$ for the 4D zero mode fields.
Despite this, we will use the labels $t_L$, $b_L$, $t_R$ to denote the 5D
Dirac fields that contain the 4D $t_L$, $b_L$, $t_R$ as zero modes.
At points of possible confusion, we will be explicit about which label
we are applying.

There are several possibilities for the embedding of the $t_L$, $b_L$,
and $t_R$
states into $SO(5)$ multiplets.  The simplest  is to embed
these three states in the {\bf 4} of $SO(5)$~\cite{4rep},
\beq 
\Psi_t = \left[ \matrix { t_L (++) \cr b_L (++) \cr t_R (--) \cr b'
    (-+)} 
\right], \qquad 
\Psi_T = \left[ \matrix { T (+-) \cr B (+-) \cr T' (-+) \cr B' (-+)} \right] \ .
\eeq{t4}
Another possibility is to embed these states in the {\bf 5} of $SO(5)$,
\beq 
\Psi_t = \left[ \matrix { \pmatrix { \chi_f (-+) & t_L (++) \cr \chi_t
      (-+) & b_L (++)} \cr t_R (--)} \right], \qquad 
\Psi_T = \left[ \matrix { \pmatrix { \chi_F (-+) & T (+-) \cr
 \chi_T (-+) & B (+-)} \cr T' (-+)} \right] \ .
\eeq{t5}
The display of the {\bf 5} here is as in \leqn{fivepresent};  
the matrix in parentheses is a bidoublet, with
$SU(2)_L$ acting vertically and $SU(2)_R$ acting horizontally.  The
fields labelled $f$, $F$ have charge $Q={5\over 3}$. The
embedding  of $t$ and $b$ in the {\bf 5} 
was suggested by Agashe, Contino, Da Rold, and
Pomerol to provide a custodial symmetry constraining the $Zb\bar b$
coupling~\cite{Zbb,5rep}.   For each of 
these choices, we have also put the competing repulsive multiplet $\Psi_T$ into
an $SO(5)$ representation of the same structure.

In this schema, the $t_L$ and $t_R$ fields necessarily belong to the
same $SO(5)$ multiplet and have the same value of the parameter $c$.  
We will set up the model in such a way that the $t_L$ and $b_L$ zero
modes are in the UV, to satisfy precision electroweak constraints on
the $b_L$.   This implies   $ c \gsim 1/2$.   That in turn implies
that the $t_R$ zero mode is in the IR. Some observable
 implications of the composite $t_R$ are  presented in~\cite{YPthree}.

\subsection{Expected mass ratios}

We are now in a position to estimate the mass ratios of $W$, Higgs,
and $t$.    We assume that it is possible to engineer a $v/f$
hierarchy by competition between the $\Psi_t$ and $\Psi_T$ multiplets,
as described above.   In this simplified analysis, we will ignore the
contribution of the gauge bosons to the Coleman-Weinberg potential.
We will see later that this will be a good approximation in our complete
model.

Before we compute the mass ratios, we might ask what values these
ratios have in nature.  In the calculations of this paper, we will not
strive for high precision.  That would require a renormalization
program for loop diagrams in 5D, which is beyond the scope of this
paper.    However, we should take into account
SM renormalizations that have a large influence on the
numerical results.   The most important of these is the QCD
renormalization of the top quark Yukawa coupling from the scale 
$m_{t ,\msb}$ to the 1--3 TeV scale of 5D top quark condensation.
A top quark pole mass of 173~GeV gives an $\bar{MS}$ mass of  163~GeV.
From this value, we can use the 2-loop beta functions to estimate the 
top quark Yukawa coupling at higher mass scales~\cite{Buttazzo}.  We
find 
\beqa  
   y_t &=&  \quad 0.94 \quad (\mbox{at} \ m_{t,\msb})\ , \qquad \quad 0.84 \quad (\mbox{at}\  2\
     \mbox{TeV} )\CR
  m_{t,\msb} &= &  163 \mbox{ GeV} \ (\mbox{at} \ m_{t,\msb})\ ,  \qquad  147\mbox{ GeV} \
  (\mbox{at} \ 2\
     \mbox{TeV}) 
\eeqa{ymtvalues}
The difference between the 1- and 2-loop extrapolations is about
1.5\%.   Other SM corrections are of the order of the
error term.  For example, the rescaling of the Higgs boson mass from 2
TeV to $v = 246$~GeV due to Higgs field strength rescaling is
\beq
   Z(v)^{1/2} = \exp\biggl[ -\half  \int^{2 \, \mbox{\scriptsize TeV}}_v {dQ\over Q}  \
     {3y_t^2(Q)\over (4\pi)^2} \biggr]  =    1 -   1.5\%\ .
\eeqn
Taking $m_t = 147$~GeV, $m_W = 80.4$~GeV, and $m_h = 125$~GeV, we have
in nature
\beq
      m_t/m_W =   1.83      \ , \quad  m_h/m_W =   1.55  \ , \quad
        m_t/m_h =  1.18
\eeq{massratios}
for dynamical electroweak  symmetry breaking at the 2~TeV mass scale.

\subsection{Mass ratios in simple models}

How do these mass ratios compare to those in our models?
The $W$ and $t$ masses can be computed without reference to the form
of the potential by solving for the relevant poles in the 5D Green's
functions.   For definiteness, consider the gauge fields
\leqn{gaugebc} with $a=1$.   The representation  of $T^{45}$ on the triplet 
$(A^{1L}, A^{1R}, A^{15})$ is  given in \leqn{Tinfifteen} and the
corresponding Wilson line element \leqn{UW} in \leqn{Uinfive}. 
Then the matrix ${\bf C}$ in \leqn{mastereq} is 
\beq
 {\bf C} =    
 \pmatrix{  ( (1+c)/2) G_{--}  &  ( (1-c)/2)   G_{--}  &  (-s/\sqrt{2}) G_{-+} \cr
   ( (1-c)/2 ) G_{+-}  &  ( (1+c)/2)  G_{+-}  &  (s/\sqrt{2} ) G_{++} \cr 
(s/\sqrt{2}) G_{+-} & (-s/\sqrt{2}) G_{+-} & c G_{++} \cr } \ ,
\eeq{CforW}
where $s = \sin \theta$, $c = \cos \theta$, and 
 $G_{\alpha\beta} \equiv  G_{\alpha\beta}(z_0,z_R, p)$, evaluated
with $c = \half$. 
The $W$ masses are the zeros of the determinant of ${\bf C}$, given by 
\beqa
  \det {\bf C }  & = &  G_{+-} \biggl[  {1+ c^2\over 2}    G_{++} G_{--}  + 
 { s^2\over 2}  G_{+-} G_{-+}     \biggr]  \CR &=&
G_{+-} \biggl[  p^2 z_0 z_R  G_{++} G_{--}  -
   s^2/2 \biggr] / p^2 z_0 z_R  \ ,
\eeqa{firstWpoles}
 The second step  uses the identity
\leqn{myWronsk}.

We can analyze \leqn{firstWpoles} in the limit $p/k_R \ll 1$.   The
function $G_{+-}$ has its first zero at $p/k_R =  2.41 $; this is a KK
boson.
For $\VEV{h} = 0$, the quantity in brackets has a zero at $p^2 = 0$;
this is the massless $W$ boson of the theory with unbroken $SU(2)_L$. 
Turning on a small value of $\VEV{h}$ moves this zero to 
\beq
   p^2 =   s^2 {1\over \log z_R/z_0 } {1 \over (z_R^2 - z_0^2)}
\eeq{firstWmass}
where we have evaluated the $G$ functions using \leqn{Gpvalshalf}.
The Green's fuctions have a pole at this value that should be
identified with the massive $W$ boson.

Following  \leqn{UW}  we set
\beq 
     s  = \sin \VEV{h}/f   \ ;  \quad \mbox{also}  \quad   s_2  = \sin
   \VEV{h}/ 2f  \ .
\eeq{sval}
We define the parameter $v$ by 
\beq
   v  \equiv  f \sin { \VEV{h} \over f } \quad \mbox{or} \quad   v/f
   \equiv   s \ . 
\eeq{hvindentif} 
With this identification, $v$ will correspond closely to the SM Higgs
vacuum expectation value, equal to 246~GeV.  For example, combining 
\leqn{f} and \leqn{firstWmass}, we find, to leading order
in $v/f$,
\beq 
          m_W^2 =  {1\over 4}  {  g_5^2 k \over \log z_R/z_0 } v^2
\eeqn
Using  the identification of the 4D coupling \leqn{firstgfiveid},
this gives  the  SM formula
\beq
 m_W^2 =  {1\over 4}  g^2 v^2 \ ,
\eeqn
up to corrections of order   $v^2/f^2$.

The top quark mass can be determined in a similar way.  For the 
scenario \leqn{t4},  assuming again  $v/f \ll 1$,   the mixing of
$t_L$ and $t_R$ in \leqn{t4} gives
\beqa
   m_t^2 &=&   s_2^2 {1\over z_0 z_R G_{++} G_{--}}\CR
     & = & \Upsilon(c_t) \,  { g^2\over 8} v^2  \ ,
\eeqa{firstmtinfour}
where
\beq
   \Upsilon(c_t) = \log (z_R/z_0) \left( 1-(z_0/z_R)^2 \over 2 \right)
   \left( 1+2c \over 1-(z_0/z_R)^{1+2c} \right)
   \left( 1-2c \over 1-(z_0/z_R)^{1-2c} \right)   \ .
\eeq{Upsvalue}
For the scenario \leqn{t5}, we have a mixing problem that involves the
three fields $(t_L, \chi_b, t_R)$.   The lowest mass eigenvalue is
\beq
 m_t^2 = \Upsilon(c_t)\,   { g^2\over 4} v^2 \ . 
\eeq{firstmtinfive}
For $z_0/z_R = 0.01$, $\Upsilon(c_t)$  spans the range 
$ 1.75 - 0.42 $ as $c_t$
is varied from $0.3$ to $0.7$. 

Thus, we find
\beq
 \begin{tabular}{lccc}
  & $c_t$  &   {\bf 4}   &   {\bf 5}  \\ \hline
$m_W$  &  -- &    $gv/2 $   &    $ gv/2  $\\ \hline
$m_t$    &  0.3    &    $ 0.94\  gv/2 $ &    $
                                                    1.32\ 
                                                       gv/2 $\\
  &  0.5   &    $ 0.71\  gv/2$ &    $gv/2 $\\
  &  0.7    &    $0.46 \  gv/2$ &    $0.65 \  gv/2 $\\

\end{tabular}
\eeq{Wtrelation}
It is not possible to obtain a $W/t$ mass ratio as large as that seen
in nature, even for values of $c$ down to $c=0$. 
  If we ignore the constraint of the $W$ mass,
we could adjust $g$ to fit the top quark mass in any scenario.
However, this requires large values of $g$,  $g^2/4\pi \sim 1$, for
large $c_t$.

The Higgs boson mass is determined by the curvature of the
Coleman-Weinberg potential at its minimum. As we have described
in Section 2.4, we will obtain a small value of $v/f$ by setting up a
pair of 5D fermions, one with an attractive channel for condensation and
one with a repulsive channel, that compete with one another.  We
choose the values of $c$ for the two fermion representations such that
the quadratic terms in the Coleman-Weinberg potential come close to
cancelling.  If $c_t$ and $c_T$ are the $c$ parameters for $\Psi_t$
and $\Psi_T$, the condition $v/f \ll 1$ is realized in narrow region
near a phase transition in the $(c_t, c_T)$ plane.  Just on the phase
transition line, $v= 0$ and the masses of $W$, $t$, and $h$ all
vanish.  

In this section, just for the purpose of estimation,
we approximate the potential along this line
as having the form
\beq
    V(v) \approx   {1\over 4} \lambda(c_t)  v^4
\eeqn
(In the full expression, there are also $v^4\log 1/v$ terms 
\cite{YPone}).   Then, near the phase transition line, 
 the Higgs mass would be given by 
\beq
         m_h =  \sqrt{2 \lambda(c_t)} \ v \ .
\eeq{Higgsest}
In our method of calculation, the potential is more readily obtained 
in terms
of $s$ or $s_2$ in \leqn{sval}, that is,
in the form
\beq
    V(v) \approx   {1\over 4} \bar \lambda(c_t)  \bigl({v\over
      f}\bigr)^4 \, 
    z_R^{-4} \ .
\eeq{firstVform}
The relation between $\lambda$ and $\bar \lambda$ is
\beq
    \lambda =  \bar \lambda  \cdot {1\over ( f z_R)^4 } = \bar \lambda \cdot \bigl(
  {  g_5^2 k \over 4}\bigr)^2 \ , 
\eeqn
or, for $g$ as in \leqn{firstgfiveid} and $z_R/z_0 = 100$, 
\beq
  \lambda = \bar \lambda \cdot ( g^2)^2 \cdot  (1.3)  \ .
\eeq{lambdabarrel} 
Then 
\beq
   m_h =  1.6  \  g^2 \ \sqrt{\bar \lambda} \  v \  .
\eeq{firstmhform}

For each fermion representation, we can compute the contribution to
the Coleman-Weinberg potential in terms of a finite-dimensional matrix
of RS Green's functions ${\bf C}$ defined in Appendix A.4.  The
result, called Falkowski's 
Theorem~\cite{YPone,Falkowski}, is  
\beq
   V =  - 2\cdot 3 \int {d^4 p_E\over (2\pi)^4} 
 \log \det {\bf C} \   .
\eeqn
The factor $3$ is the number of QCD colors.  (We assume in the
rest of this paragraph
that $\Psi_T$, like $\Psi_t$, is a color ${\bf 3}$.) 
For fermions in the {\bf 4} of $SO(5)$, 
\beqa
    V(\Psi_t) & =&  - 6  \int {d^4 p_E\over (2\pi)^4}  \log \biggl[ 1 + {s_2^2
      \over p_E^2 z_0 z_R G_{E++} G_{E--} } \biggr] \CR
   V(\Psi_T) & =&  - 6  \int {d^4 p_E\over (2\pi)^4}  \log \biggl[ 1 - {s_2^2
      \over p_E^2 z_0 z_R G_{E+-} G_{E-+} } \biggr] 
\eeqa{Vinfour}
For fermions in  the {\bf 5} of $SO(5)$
\beqa
    V(\Psi_t) & =&  - 6  \int {d^4 p_E\over (2\pi)^4}  \log \biggl[ 1
    + { s^2/2 
      \over p_E^2 z_0 z_R G_{E++} G_{E--} } \biggr] \CR
   V(\Psi_T) & =&  - 6  \int {d^4 p_E\over (2\pi)^4}  \log \biggl[ 1 -
{ s_2^2 (2 - s_2^2)
      \over p_E^2 z_0 z_R G_{E+-} G_{E-+} } \biggr]
\eeqa{Vinfive}
The terms of order $s^2$ in these expressions 
are identical between the {\bf
  4} and {\bf 5} up to an overall factor of 2.  Since the vanishing of
the $s^2$ term determines the location of the line of  phase transitions in the
$(c_t,c_T)$ plane, that location  will be the same for the two systems.

Consider first the situation with  $\Psi_t $ in the {\bf 4}. For $c_t
= \half$, the phase transition occurs at $c_T = 0.438$ and, at this
point, the sum of the $\Psi_t$ and $\Psi_T$ potentials is reasonably
approximated by $\bar \lambda(c_t = \half) =   0.0076$.    For $0.3 < c_t
< 0.7$, the value of $\bar\lambda(c_t)$ varies over   the interval
$ 0.019 - 0.0015$. 

For  $\Psi_t $ in the {\bf 5}, the location of the phase transition in
$c_T$ is the same as for the {\bf 4}.  At this point,  
the sum of the $\Psi_t$ and $\Psi_T$ potentials is reasonably
approximated by $\bar \lambda(c_t = \half) =  0.043$.    For $0.3 < c_t
< 0.7$, the value of $\bar\lambda(c_t)$ varies over   the interval
$0.099 - 0.013 $. 

Converting back to $\lambda$ and expressing these results in terms of
a prediction for the Higgs boson mass, we find
\beq
\begin{tabular}{lccc}
     & $c_t$  &   {\bf 4} &   {\bf 5} \\ \hline
$m_h$&   0.3  &   $g^2 \cdot$~55~GeV &  $g^2 \cdot$~130~GeV\\
     &   0.5  &   $g^2 \cdot$~35~GeV &  $g^2 \cdot$~83~GeV\\
     &   0.7  &   $g^2 \cdot$~16~GeV &  $g^2 \cdot$~45~GeV\\
\end{tabular}
\eeq{Higgsmasstable}

It is possible make these values of $m_h$ compatible with the measured
value of 125~GeV, but only by increasing  the coupling constant $g$.  Even in the worst case of $c_t = 0.7$ with $\Psi_t$ in the
{\bf 4}, we need  $g^2/4\pi  = 0.62$, a coupling that is strong but
not prohibitively so.  However, across the table, the value of $g$
required to fit the Higgs boson mass is different from that required
to fit the $t$ mass except at specific (tuned) values of $c_t$.

In  the simple model presented in this section, a single value of $g_5$
was expected to explain the  $W$, $t$, and Higgs masses.  We saw that
this was overly ambitious.   From the point of view of duality with a 
strongly coupled 4D theory, the assumption also seems excessively
strong.  In a 4D theory, the values of the $SU(2)$ gauge coupling and
the top quark Yukawa coupling would be set at some much larger energy
scale, perhaps at the scale of grand unification.   These settings
would appear in the RS model as boundary conditions on the UV brane.
In the next section, we will show that this effect can be modelled by 
introducing boundary kinetic terms for the $SU(2)_L$ bosons and the
top quark multiplets.  This will allow us the freedom that we need to
fit the $W$, $t$, and Higgs masses and, more generally, represent the
known properties of these particles within our RS model.

Though this can be done with either of the choices for the
representation of $\Psi_t$, from here on we will concentrate on the 
choice of $\Psi_t $ in the {\bf 5} of $SO(5)$, which requires smaller
values of $g_5$ to fit the top quark  and Higgs boson masses.

\section{UV boundary kinetic terms}

To model the UV boundary conditions on the 4D gauge and Yukawa
couplings, we introduce boundary kinetic terms for the $SU(2)\times
U(1)$ bosons and the top quark.   In this section, we will describe
the effects of these boundary terms on the Green's fuctions for
these  5D fields.  These effects are straightforward to understand.
The formal derivation of these results is somewhat involved.
We present it in Appendix C.

\subsection{Boundary gauge kinetic term}

For a spin-1 fields with zero modes corresponding to a 4D gauge field, we
introduce the boundary kinetic term of which size is given by a
dimensionless 
parameter $a$, 
\beq
S_{UV} = \int d^4 x dz \ \biggl(\sqrt{g} 
\biggl[ -  {1 \over 4} a z_0 \delta(z-z_0)    g^{mp} g^{nq} F_{mn} F_{pq} \biggr] \biggr) \ .
\eeq{boundarygauge}
For the zero modes, which have wavefunctions constant in $z$, this
term 
adds to the $dz/kz$ or $dx$ integral  of the standard kinetic term over
the fifth dimension.  Through this, it modifies the formula
\leqn{firstgfiveid} for the 4D gauge coupling to 
\beq 
  g^2   =    {  g_5^2 \over  ( \pi R + a/k)}  = {g_5^2 k \over (\log
           z_R/z_0 + a)} \ .
\eeq{newgfive}
To visualize this result, imagine that the vector boson zero mode, which is constant
in $z$ for $a = 0$, acquires a delta function piece  proportional to
$\sqrt{a}$ at $z = z_0$.
By increasing $a$, we can make this gauge coupling as weak as we need
for those modes $A_m^A$ that correspond to
weakly-coupled 4D gauge fields. The addition of the boundary term can
have 
a relatively large effect on the properties of the zero mode
wavefunctions
while giving only small corrections to the masses and wavefunctions of
the corresponding Kaluza-Klein states.  For the components of $A_m^A$
that do not appear in the 
boundary kinetic term, the effective strength of the 5d gauge
interactions is still given by 
\beq
         g_{RS}^2   =   {   g_5^2 k\over \log z_R/z_0 }   \ . 
\eeq{newgRS}

As shown in Appendix C, the boundary kinetic term for $A_m$ 
adds a component
with $-$ boundary conditions to the original component with  $+$ 
boundary conditions.   In terms of the relevant $G_{\alpha\beta}$
functions, the boundary condition at the UV brane is changed from 
\leqn{UVbcs} according to 
\beq
        G_{-, \beta} (z_0,z_R) = 0\  \to \   G_{-, \beta}(z_0,z_R)  
+ a p z_0  G_{+, \beta} (z_0,z_R)  = 0  \ .
\eeq{changeabc}
(Here the subscript $\beta$ specifies the boundary condition on the
IR brane.)   The boundary condition on the $A_5$ component, which
originally had
a $-$ boundary condition in the UV, is also changed by \leqn{changeabc}.   The
boundary kinetic term does not affect   $A_m$ fields with $-$ boundary
conditions or $A_5$ fields with $+$ boundary conditions.  We will see
that taking $a$ large compared to $\log z_R/z_0$, as we will require for a small
$SU(2)$ gauge coupling, suppresses the influence of the zero modes
 on the Coleman-Weinberg potential. 

In the models in this paper, we introduce separate boundary kinetic
terms with coefficients $a_W$ and $a_B$ for the $SU(2)_L$ and $U(1)_Y$
gauge fields, respectively. Other boundary terms would  have no effect, since
the  corresponding gauge fields have $-$ boundary conditions on the UV brane. 
In the following, we abbreviate
\beq
   L_W = \log {z_R\over z_0} + a_W \ , \qquad  L_B = \log {z_R\over z_0}
   + a_B  \   . 
\eeq{LWBdefin}

\subsection{$W^{\pm}$ and charged KK bosons}

The dynamics of the $W^{\pm}$ bosons and their KK excitations is
encoded in the Green's functions of $A^{aL}_m$, $A^{aR}_m$, and
$A^{a5}_m$ for $a = 1,2$.  The calculation of these Green's functions
is described in Appendix D.1.

The mass eigenvalues in this sector are given by the zeros of the
determinant of the ${\bf C}$ matrix for this problem.   This is 
\beq
  \det {\bf C} = G_{+-} \biggl[   G_{++} (G_{--}  + a_W p z_0
  G_{+-}) -  {s^2\over 2 p^2 z_0 z_R} \biggr] \ , 
\eeq{detCW}
a simple generalization of \leqn{firstWpoles}. The factor $G_{+-}$ has
no zeros near $p^2 = 0$.   To leading order in 
$s^2$, the position of the first zero  in the second factor
\beq
    m_W^2 =  { s^2 \over L_W}{1\over   (z_R^2 - z_0^2) } = { g^2
    v^2\over 4 }\ , 
\eeq{mWwa}
with  $g^2 $ given by \leqn{newgfive}.   The low-momentum behavior of
the propagator  $\VEV{A^{aL}_m (z)  A^{aL}_n(z')}$, at leading order
in $s^2$, works out to 
\beq
g_5^2 \VEV{A^{aL}_m (z)  A^{aL}_n(z')} =   { g^2 \, \eta_{mn} \over p^2 - m_W^2 } \ ,
\eeqn
as it should be.  To leading order in $s^2$, the matrix elements of
gauge bosons between  fermion
zero modes such as $(\nu_L, e_L)$ involve only this Green's fuction.  
The expression for the Green's function is independent of $z$ and
$z'$, so the fermion scattering amplitudes are independent of details
of the fermion wavefunctions in $z$ and depend only on the overall
gauge charges~\cite{RandallSchwartz}.   Then
we recover the structure of the SM weak interactions to
this order,
\beq
 i{\M} =   i   {g^2/2\over p^2 - m_W^2}   (T^{+L} T^{-L} + T^{-L}  T^{+L})
 \ . 
\eeq{firstorderWamp}
 We will discuss the order $s^2$ corrections to this
result in Section 6.

Evaluating $\det{\bf C}$ in Euclidean momentum space and using the results of 
\cite{YPone}, we find the contribution to the Coleman-Weinberg
potential of the Higgs boson from the sector of charged gauge bosons,
\beq
    V_W(h) =  +2 \times {3\over 2} \int{d^4 p_E\over (2\pi)^4}  \log \left[ 1 + 
      {s^2 /2  \over p_E^2 z_0 z_R \, G_{E++} (G_{E--} + a_W p_E z_0 G_{E+-})
} \right] \ , 
\eeq{VW}
The effect of the $a_W$ term in this expression is to suppress the
contribution of this sector.

\subsection{$Z/\gamma$ and neutral KK bosons}

In a similar way, the dynamics of the photon and $Z$ boson and their KK excitations is
encoded in the Green's functions of $A^{3L}_m$, $A^{3R}_m$, $A^X_m$ and
$A^{35}_m$.  The calculation of these Green's functions
is described in Appendix D.2.   In this discussion, we will use the
basis  $(A^{3L}_m, B_m, Z^\prime_m, A^{35}_m)$ defined in \leqn{BZ}.

The mass eigenvalues in this sector are given by the zeros of the
determinant of the ${\bf C}$ matrix.  For this sector,
\beqa
  \det {\bf C} &=& G_{+-} \biggl[   G_{++} (G_{--}  + a_B p z_0
  G_{+-}) (G_{--}  + a_W p z_0
  G_{+-}) \CR & & \hskip 0.2in -  {s^2\over 2 p^2 z_0 z_R} \left( (G_{--}  + a_B p z_0
  G_{+-})+ s_\beta^2  (G_{--}  + a_W p z_0
  G_{+-}) \right) \biggr] \ , 
\eeqa{detCZ}
The factor $G_{+-}$ has
no zeros near $p^2 = 0$. The extra factors of the form $(G_{--} + a p
z_0 G_{+-})$   lead to a pole in the Green's functions at $p^2 =0$ in
addition to the pole at a position of order $s^2 / z_R^2$  that we saw
in the charged vector boson Green's functions.   These poles represent
the photon and the $Z$ boson. The $Z$ pole  is located at the first
zero of the second factor in  \leqn{detCZ}, given to leading order
in $s^2$ by 
\beq
    m_Z^2 =   {s^2 (L_B + s_\beta^2 L_W)\over L_B L_W } {1 \over( z_R^2 -
      z_0^2) } \ .
\eeq{mZwa}

The photon pole at $p^2 = 0$ appears only in the Green's functions
$\VEV{A^{3L}_mA^{3L}_n}$,  $\VEV{A^{3L}_mB_n}$, and $\VEV{B_mB_n}$.
    The $Z$ pole appears in all 2-point functions of the four vector
    fields, but the contributions in the $Z'$ and $A^{35}$ Green's 
    functions are subleading in $s^2$.   To leading order in $s^2$, we
    find
\beqa
g_5^2 \VEV{A^{3L}_m(z)A^{3L}_n(z') }&=&  
	{kg_5^2  \eta_{mn} \over p^2 (p^2-m_Z^2) (L_B + s_\beta^2 L_W)} 
	\biggl[ - m_Z^2 s_\beta^2 +  (m_Z^2/s^2) p^2 z_R^2 L_B \biggr] \CR
g_5 g_{5Y} \VEV{A^{3L}_m(z)B_n(z')} &=&
	{kg_5 g_{5Y}  \eta_{mn} \over p^2 (p^2-m_Z^2) (L_B + s_\beta^2 L_W)}
    \biggl[ - m_Z^2 s_\beta \biggr] \CR
g_{5Y}^2 \VEV{B_m(z)B_n(z') } &=&
	{kg_{5Y}^2  \eta_{mn} \over p^2 (p^2-m_Z^2) (L_B + s_\beta^2 L_W) }
    \biggl[ - m_Z^2 +  (m_Z^2/s^2) p^2 z_R^2 L_W \biggr] \ ,
 \eeqa{firstorderZGs}
Again, the expressions are independent of $z$ and $z'$, and so fermion
matrix elements built with these Green's functions depend only on the
global gauge charges.  Putting these expressions together with the
interaction \leqn{DM}, the pole at $p^2 = 0$ has the form
\beq
     {k\over L_B + s_\beta^2 L_W} (g_5 s_\beta  T^{3L}  + g_{5Y} Y )^2
     \cdot {1\over p^2} \ .
\eeqn
For the pole at $p^2 = m_Z^2$, we can evaluate the residue using
\leqn{mZwa}, to find
\beq
   {k\over L_B + s_\beta^2 L_W} (g_5 (L_B/L_W)^{1/2}  T^{3L}  - g_{5Y}
   (s_\beta^2 L_W /L_B)^{1/2} Y )^2
     \cdot {1\over p^2-m_Z^2} \ .
\eeqn
  Identifying
\beq
    s_w^2 =  { s_\beta^2 L_W \over L_B + s_\beta^2 L_W} \ ,  \qquad
    e^2 = g^2 s_w^2 = {k g_5^2 s_\beta^2 \over L_B + s_\beta^2 L_W} \  ,
\eeq{firstsw}
everything falls into place, and we find the SM
interaction
\beq
 i{\M} =   i \biggl[ { e^2 Q^2 \over p^2}  +  {g^2/c_w^2\over p^2 -
   m_Z^2}  
 (T^{3L}  - s_w^2 Q)^2 \biggr]
 \ ,
\eeq{firstorderZamp}
with $Q = T^{3L} +Y$ as in \leqn{charges}.  We will discuss the order 
$s^2$ corrections to this
result in Section 6.

Evaluating $\det{\bf C}$ in Euclidean momentum space and using the results of 
\cite{YPone}, we find the contribution to the Coleman-Weinberg
potential of the Higgs boson from the sector of neutral  gauge bosons,
\beqa
    V_{Z}(h) &=&  + {3\over 2} \int{d^4 p_E\over (2\pi)^4}  \log \biggl[
      1 \CR
    & & \hskip 0.2in + {( s^2/2) \left( (G_{E--} + a_B p_E z_0 G_{E+-}) +  s_\beta^2 
        (G_{E--} + a_W p_E z_0 G_{E+-}) \right)  \over  p_E^2 z_0 z_R \, G_{++}
        (G_{E--} + a_B p_E z_0 G_{E+-})  
 (G_{E--} + a_W p_E z_0 G_{E+-})}
  \biggr] \ . \CR
\eeqa{VZ} 
Again, the  $a_W$ term serves to suppress the
contribution of this sector.

\subsection{Boundary top quark kinetic term}

We pointed out at the end of Section~3 that, once we arrange for the
little hierarchy between $v$ and the KK scale,  some extra  tuning is
required to  obrain the observed ratio  of masses $m_t/m_h$.   To
allow this freedom in our model, we add a boundary kinetic term for
the top quark.

The formalism for a fermion boundary kinetic term is presented in
Appendix C.3.   For each $SU(2)\times U(1)$ multiplet of fermions, we
can add  a boundary kinetic term either for the left-handed or for the
right-handed components of the 5-d Dirac fermion.   However, this term
has a substantial effect on the dynamics only if we  add a left-handed
boundary term to a fermion with a  UV-dominated
left-handed zero mode  $(c \gsim 1/2)$, or, alternatively, if we add a
right-handed boundary term to a fermion with a UV-dominated
right-handed zero mode ($c \lsim - 1/2)$.  As we have discussed at
the end of Section~2, we choose the $\Psi_t$ multiplet to
have 
$c \gsim 1/2$.   Then the $t_R$ zero mode, which is also contained in
this multiplet, will be IR-dominated.   With this choice, only a
left-handed boundary kinetic term for $\Psi_t$ gives a robust
parameter for the model.   Similarly, the multiplet $\Psi_T$, which
has no zero mode, is not strongly affected by any choice of a boundary
kinetic term.   Thus, we will add only one parameter here,  the
coefficient $a_t$ of the left-handed boundary kinetic term for the
components   $(t_L,
b_L)$ in \leqn{t5}.

Adding the parameter $a_t$, the determinant of the ${\bf C}$ matrix
for the  $(t_L, \chi_b, t_R)$ elements of \leqn{t5} is 
\beq
  \det {\bf C} = G_{+-} \biggl[   G_{++} (G_{--}  + a_t p z_0
  G_{+-}) -  {s^2\over 2 p^2 z_0 z_R} \biggr] \ . 
\eeqn
We must take some further care in expanding this expression for small
$p$, since now the $G$ functions are evaluated at a general value of
$c_t$.  Define
\beq
L_t =   (G_{--}  + a_t p z_0 G_{+-})|_{p=0} = {1\over 2c_t-1} \bigl[ ({z_R\over
  z_0})^{c_t-1/2} -  ({z_0\over
  z_R})^{c_t-1/2} \bigr] +  a_t  ({z_R\over
  z_0})^{c_t-1/2} \CR
\eeq{Ltdefin}
and note that 
\beq
G_{++}(z_0,z_R; p = 0 ) =  {1\over 2c_t+1} \bigl[ ({z_R\over
  z_0})^{c_t+1/2} -  ({z_0\over
  z_R})^{c_t+1/2} \bigr] 
\eeqn
is well approximated by 
\beq
G_{++}(z_0,z_R; p = 0 )  =  {1\over 2c_t+1} {z_R\over z_0} ({z_R\over
  z_0})^{c_t-1/2} 
\eeqn
for $c_t > 0$, $z_0/z_R \sim 0.01$.    Then, 
to leading order in $s^2$, $m_t$ takes the form
\beq
m_t^2 = {2c_t+1\over 2} { s^2 z_R^{-2} \over L_t }
\biggl({z_0 \over z_R }\biggr)^{c_t-1/2} \ .
\eeq{mtfirstval}

With the effect of $a_t$, the contribution to the  
Coleman-Weinberg potential from the $\Psi_t$ multiplet is altered
from \leqn{Vinfive} to
\beq
    V_{t}(h) =  - 6 \int{d^4 p_E\over (2\pi)^4}  \log \left[ 1 +
      {s^2/2 \over p_E^2 z_0 z_R \, G_{E++} ( G_{E--} + a_t p_E z_0 G_{+-}) }
  \right] \ . 
\eeq{Vt} 
The contribution of the multiplet $\Psi_T$ remains 
\beq
  V_T(h)   =  - 6  \int {d^4 p_E\over (2\pi)^4}  \log \biggl[ 1 -
{ s_2^2 (2 - s_2^2)
      \over  p_E^2 z_0 z_R \, G_{E+-} G_{E-+} }\biggr]  \ .
\eeq{VT}

\subsection{UV and IR gauges}

Up to this point in our discussion, we have quoted all Green's
functions in the gauge in which the Wilson line \leqn{UW} is
represented as a boundary condition at the UV boundary.   However, it
is equally well possible to change the gauge and move  the Wilson line 
onto the IR boundary.   We will refer to these two gauges as the ``UV
gauge''
and the ``IR gauge'', respectively.

For the purpose of calculation, it is typically easier to use the UV
gauge.  In the UV gauge, the boundary conditions in the IR are 
simple.  For a gauge field,  for
example,  
the Green's functions are naturally expressed as linear combinations
of  the 
elements 
$G_{+-}(z,z_R)$ and $G_{++}(z,z_R)$ with definite boundary conditions
in the IR.   Physical
quantities computed from the Green's functions will have an explicit
dependence on $z_R$, but this is a good thing, since $z_R$ sets the
scale of the RS dynamics, as we have seen already in this section.  In the IR
gauge, the Green's functions are more naturally written in terms of
elements with definite boundary conditions in the UV, such as
$G_{+-}(z,z_0)$.
Then they will contain explicit dependence on $z_0$ which typically
cancels out to a great extent. 

However, there are some advantages to  working in the IR gauge.  As we
explained at the end of  Section~3 of \cite{YPone}, mixing of fields 
on the boundary has no effect if these fields have the same boundary
conditions.   In our discussion of precision electroweak corrections,
we will find some mixing effects that seem to magically cancel in the
UV gauge.  These cancellations are easier to see in the IR gauge.  The
fermions that mix in the UV gauge have 
identical boundary conditions in the IR, so that the mixing terms have
no effect~\cite{YPone}.   The fields $A_m^{a5}$ have vanishing
boundary values on the IR brane, so the mixing of these fields with
the other gauge fields is also substantially reduced.

Often, the simplest analysis combines these two approaches, by
representing the IR gauge Green's functions in terms of the elements
used in the UV gauge.   This is achieved by writing the relation
between the 
Green's functions in the two gauges as
\beq
     \VEV{A^A_M(z) A^B_M(z')}_{IR} =  (U_W)^{AC}  \VEV{A^C_M(z)
       A^D_M(z')}_{UV}   (U_W^\dagger)^{DB} \ . 
\eeq{UVIR}
   For those who do not consider this equation obvious, we provide an 
explicit proof in  Appendix E.  

\section{Complete model and its parameter space}

We are now in a position to find the ground state of the $SO(5)$ model
and understand the dependence of the spectrum of the model on its parameters.

\subsection{The complete Coleman-Weinberg potential and its implication}

\begin{figure}
\centering
\includegraphics[width=4in]{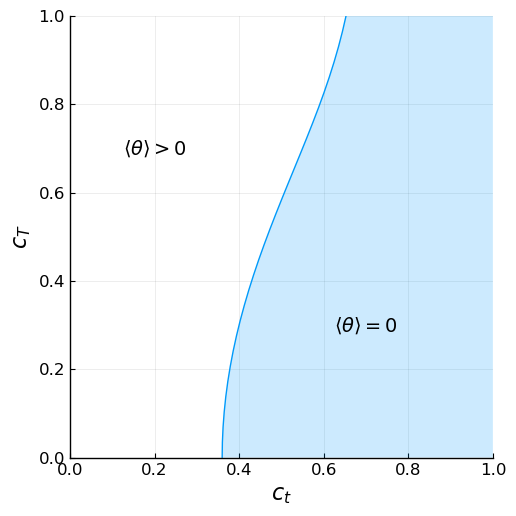}
\caption{Phase diagram of the potential \leqn{approxV} in the $(c_t,c_T)$ plane, with $z_R/z_0=100$, $a_W = a_B = 40$, and $a_t = 10$. The solid blue line corresponds to $A=0$. It should be noted that in realistic models, $a_W, a_B$, and  $a_t$ will be determined by the mass relations \leqn{mWwa}, \leqn{mZwa}, and \leqn{mtfirstval}. }
\label{fig:phase}
\end{figure}

The full Higgs potential can be obtained by summing up the Coleman-Weinberg potentials \leqn{VW}, \leqn{VZ}, \leqn{Vt}, and \leqn{VT}. Our final results will be obtained from a full numerical evaluation of these integrals.  However, we can obtain insight into these result by first examining the expansion of the potential in powers of $s$. Up to $\O (s^4)$, the Higgs potential can be written as
\beq
V(h) = {k_R^4 \over 8\pi ^2} \left[ - A s^2 + {1 \over 2} B s^4 
+ {1 \over 2} C s^4 \log {1 \over s^2 } + \O (s^6) \right] \ .
\eeq{approxV}
where the full expression for the coefficients $(A,B,C)$ is given in
Appendix F. Their values depend on $c$ parameters of fermions as well
as the boundary kinetic terms $a$. The coefficients $B$ and $C$ are
always positive. The line of phase transition is determined by
the condition $A=0$.   Fig.~\ref{fig:phase} shows the phase diagram in the
$(c_t,c_T)$ plane for $z_R/z_0=100$, $a_W = a_B = 40$ and $a_t = 10$. In realistic
models, $c_t$ and $c_T$ should be tuned to be near the line $A=0$ in
order to make
 $v/f \ll 1$. 

Now we compute the mass of the Higgs boson. By differentiating 
$V(h)$ twice, we find
\beq
\left( m_h \over v \right)^2 = {g_5^4 k^2 \over 32\pi^2} \left[ B + \left( -\log {v^2 \over f^2} -{3 \over 2} \right) C \right] \ .
\eeq{mh}
The quartic term in the Higgs potential originates from the box
diagrams of the top and top partner, so we naturally have a factor of
$g_5^4$ in this expression. Composite Higgs models typically predict
a quartic term smaller than what is required for the observed Higgs
mass; however, from \leqn{mh}, we can see how our model can overcome
the challenge. First, the quadratic terms of the Higgs
 potential from $\Psi_t$ and $\Psi_T$ cancels each other, but their
 quartic terms add.
 Therefore, we can tune $A$ near zero without sacrificing
 the quartic term $B$ and $C$. Second, we can push $g_5$ to a larger value. 
As shown in \leqn{newgfive}, the gauge boundary
kinetic term gives us freedom to fit the correct $SU(2)$ 
coupling even with a large $g_5$. Third, some  choices of the $SO(5)$
representation for 
$\Psi_T$ can make a relatively large 
contribution to $B$. This is indeed the 
case for the $\Psi_T$ in the {\bf 5} of $SO(5)$. See Appendix F for details. 
 
We can obtain a futher insight of the parameter space of our model by
studying the relationship between the Higgs mass and the top quark
mass. In Appendix F, we argue that near the phase transition line
$A=0$, the coefficients 
$B$ and $C$ can be estimated as
\beq
B \sim  {3 \over 4} A_t (c_t, a_t) ,  \qquad C \sim 0 \ ,
\eeqn
where $A_t$ is the quadratic term of the top quark contribution
 to the Higgs potential, defined in \leqn{Vtcoeff}. Then, from 
\leqn{mtfirstval} and \leqn{mh}, we have
\beq
\left( m_h \over m_t \right)^2 \sim {g_5^2 k \over 4 \pi^2} \cdot \left[ {
    L_t(c_t, a_t) \over 1+2c_t} \left( z_0 \over z_R \right)^{1/2-c_t}
\right] \cdot  {3 \over 4}  A_t (c_t, a_t) \, .
\eeq{mhmt}
The term in bracket and ${3\over4} A_t(c_t,a_t)$ depend strongly on $c_t$ and $a_t$, but
their product turns out almost constant across a wide range of $c_t$
and $a_t$. Numerically, for $ 0.3 < c_t < 0.7$ and $ 0 < a_t < 20$,
the product stays within the interval $1.2 - 1.5$. This is actually to
be expected, 
since there is a positive correlation between the top quark Yukawa 
coupling and its contribution to the Higgs potential. 

Then the mass ratio \leqn{mhmt} gives a rough estimate of the required
value of RS coupling $g_5$ in our model. With this determined, we choose the
size of the gauge boundary kinetic term $L_W$ which fits the $W$ boson
mass. Using 1.3 for the value
 of the product in \leqn{mhmt}, we have
\beq
g_5^2 k \sim 22 \quad \textrm{and} \quad L_W \sim 51 \ .
\eeqn
This shows that $g_5$ and $L_W$ are pushed to large values in our
model. The full numerical study agrees well with this result. It
gives $L_W$ between 
$35 - 55$ for $ 0.4 < c_t < 0.7$  and 1.5 TeV $ < k_R < $ 3 TeV. 

The large value of $g_5$ results from the relatively small quartic term $B$ of
the Higgs potential. It is possible to increase $B$ by decreasing $c_t$, but this also decreases the term in bracket in \leqn{mhmt}, so that the value of $g_5$ stays large across the entire parameter space. This tension can be relaxed
if there is an additional, large
source of the Higgs quartic term. This will also relieve the degree of
 fine-tuning in our model.
In \cite{YPone}, we showed that there are fermion gauge multiplets
that can provide a positive contribution to the quartic term in the
potential without affecting the quadratic term.  Perhaps adding such a
multiplet here will provide a more attractive set of model
parameters.

\subsection{Allowed region of parameter space}
\label{fitting}

Now we study our parameter space with a full numerical treatment. There are nine parameters in our theory,
\beq
z_0, \, z_R,\, c_t,\, c_T,\, g_5,\, g_X,\, a_W,\, a_B, \, a_t \ ,
\eeqn
or, keeping $k_R = 1/z_R$ as the only dimensionful parameter, we have
\beq
k_R \quad \textrm{and} \quad z_R/z_0,\, c_t,\, c_T,\, g_5^2k,\, g_X^2k,\, a_W,\, a_B, \, a_t \ .
\eeq{params}
These parameters should produce correct values of the five independent observables,
\beqa
G_F &=& 1.166 \times 10^{-5} \, \textrm{GeV}^{-2}, \quad m_t = 147\ \textrm {GeV,} \CR 
&& \hskip -0.7in m_W = 80.4 \ \textrm {GeV,} \quad m_Z = 91.2\ \textrm {GeV,} \quad m_h = 125\ \textrm {GeV.}
\eeqa{observables}
We can also consider these quantities as one dimensionful observable $v$ and four dimensionless number, $e,g,y_t$, and $m_h/v$.

It is easiest to think of this parameter space as parametrized by the KK scale (a few times $k_R$) and the ratio $z_R / z_0$. This latter ratio is constrained by flavor physics, since light flavors will couple to the Higgs sector at the UV boundary. Flavor structure is beyond the scope of this paper, so for the moment we propose $z_R / z_0 = 100$.

Furthermore, we can expect from the small hypercharge coupling that $a_B$ should have little effect on the Higgs potential. This is indeed numerically observed. Therefore, we will assume $a_B = a_W$ throughout the rest of our analysis. This leaves us effectively a 2-dimensional parameter space. 

Our strategy to find the available parameter space is as follows. We first choose values of $(c_t, a_t)$. Then, $m_W/m_t$ determines $a_W$ by \leqn{mWwa} and \leqn{mtfirstval}, and $(g,e)$ determine $(g_5^2k, g_X^2k)$ by \leqn{newgfive} and \leqn{firstsw}. With those parameters fixed, the potetial minimum is now determined by $c_T$. We search for the value of $c_T$ which gives the observed value of $m_h$. Although in the analysis above we have used the small $s$ expansion of the Higgs potential, our numerical analysis is conducted with the full potential before the expansion. The minimum of the full potential differs by about 10\% compared to that obtained from the approximate formulae \leqn{approxV}. 

\begin{figure}
\centering
\includegraphics[width=4in]{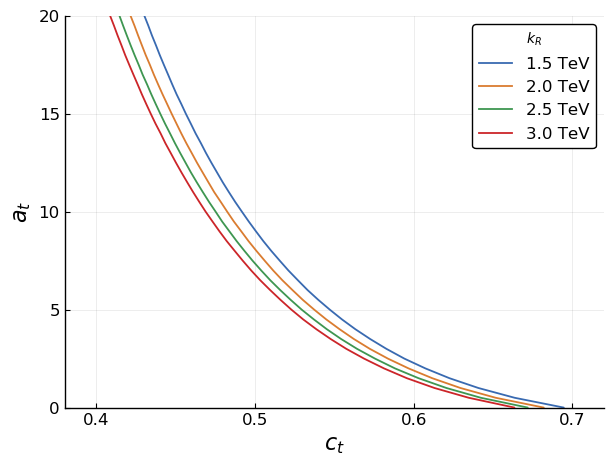}
\caption{Allowed region of parameter space in the $(c_t, a_t)$ plane. Here $\Psi_T$ is not charged under $SU(3)_C$.}
\label{fig:ctat}
\end{figure}

Figure~\ref{fig:ctat} shows the allowed region of parameter space in the $(c_t, a_t)$ plane for different values of $k_R$. Note that parameters do not depend strongly on $k_R$. This implies that $k_R$ can be seen as one of the orthogonal directions of our two-dimensional parameter space and we can consider its effect on observables separately from other parameters. In the following analysis, we choose $c_t$, which represents the degree of compositeness of the top quark, as the other main variable of our parameter space. We show how physical quantities change as we vary $c_t$ at values of $k_R= 1.5 - 3$ TeV. 

\subsection{Mass spectrum of the top partner}

The masses of new particles beyond the SM are determined by $k_R$. Before looking at the masses of new states in our theory, it is instructive to study masses of generic KK states in RS models. For $z_R/z_0=100$, the first KK masses of a gauge field (or a fermion with $c=1/2$) with different boundary conditions are  
\beq
\begin{tabular}{c|cccc}
b.c.       & $(++)$ & $(+-)$ & $(-+)$ & $(--)$ \\ \hline
$m / k_R $ & 2.8    & 0.72   & 2.4    & 3.8  \\ 
\end{tabular}
\eeq{KKmasses}
The UV boundary kinetic term supresses the masses of $(++)$ and $(+-)$
states, but only slightly. It has no effect on $(-+)$ and $(--)$
states. Therefore, except for the $(+-)$ state, the masses of new states will
be a few times $k_R$. In our model, those heavy states correpsond
to $Z'$ and KK states of $W$, $B$ and the top quark. For $k_R >$ 1.5~TeV, 
these particles have masses above  4~TeV.   At the lower end of this
range, we must still consider the observability of these states in LHC
Drell-Yan measurements.  However, the KK vector bosons are IR-dominated
and have suppressed couplings to light fermions associated with UV zero modes.
 Compared to a sequential $W'$ and $Z'$, 
the suppression is a factor of 4 in the couplings, 
or more when the KK boson has a UV boundary kinetic term, and this 
suppression factor is squared in the cross section formula.   Therefore, 
these KK resonances 
 are not yet constrained by LHC searches~\cite{ATLASdilepton,CMSdilepton}.

\begin{figure}
\centering
\includegraphics[width=5in]{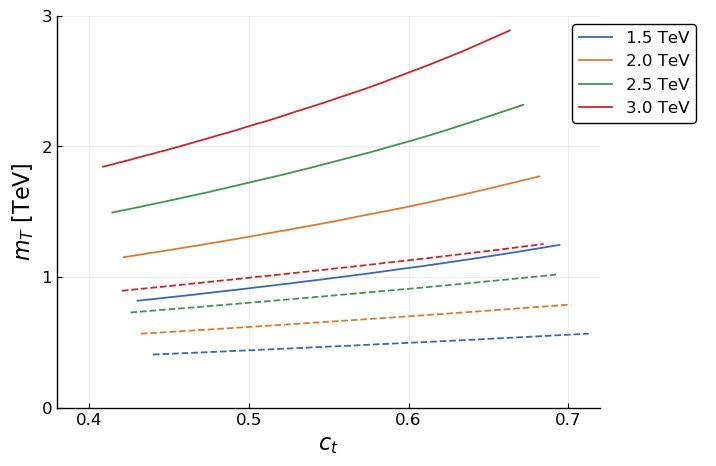}
\caption{Masses of lightest  top partner from the multiplet
  $\Psi_T$. Dashed lines correpond to color triplet $\Psi_T$,  and
  solid lines correspond to color singlet $\Psi_T$.}
\label{fig:toppartner}
\end{figure}

On the other hand, the $(+-)$ states in the top partner multiplet
$\Psi_T$ can have a mass lower than $k_R$.  The dashed lines in
Fig.~\ref{fig:toppartner} show the masses of the top partner for 
different values of $c_t$ and $k_R$.  Searches for  a vectorlike  top
partner  at the LHC currently put the mass of this particle 
above 1.37 TeV~\cite{mT} and
 thus constrains our model for $k_R < 3$ TeV.

The LHC search assumes that the top partner is charged under $SU(3)_C$
and can decay into the top or bottom quark. However, whether the
$\Psi_T$ in our model is colored or not is a model-building choice and
we can proceed in either way, as long as $\Psi_T$ can compete with the top
quark and generate the correct Higgs potential.  In terms of the
experimental constraints, it is much more attractive to assume that $\Psi_T$ is a
singlet under $SU(3)_C$ and its states are heavy  leptons: The strongest
current 
experimental bound on a new heavy lepton is 560~GeV, in a particularly
optimistic scenario \cite{typethree}.   

The hypothesis  that $\Psi_T$ is a color-singlet has much in common
with the idea of ``neutral naturalness'' put forward in 
\cite{NeutralNatone,NeutralNattwo}. In both cases, the Higgs
potential obtains competing contributions from the top quark multiplet and
from color-singlet mirror states at the TeV scale.   However,
conventionally in this framework, a discrete symmetry between these
multiplets
 is used to make
the one-loop contributions to the Higgs potential finite, and then
further fine-tuning is needed to achieve a small value of $v/f$.
Here, the finiteness of the Higgs potential is insured by the RS
structure, so there is no need for mirror symmetry; however, we still
need to tune $v/f$ to a small value. 

The solid lines in 
Fig.~\ref{fig:toppartner} show the mass of the lightest KK state
from $\Psi_T$ in the case where $\Psi_T$ is a color singlet.  
Without the multiplicity from color in $V_T$, we need to lower
the value of $c_T$ for the correct tuning of $\Psi_T$ against $\Psi_t$
to come close to $A \sim 0$ in the Higgs
potential. This leads to larger  values of $m_T$.   We find  $m_T > 820$ GeV
for $k_R \ge 1.5$ TeV, so that in this case $k_R$ is unconstrained by
LHC searches.  In the
 rest of our  analysis, we will use the parameter space of the uncolored $\Psi_T$. 

\subsection{Measure of fine tuning}

\begin{figure}
\centering
\includegraphics[width=5in]{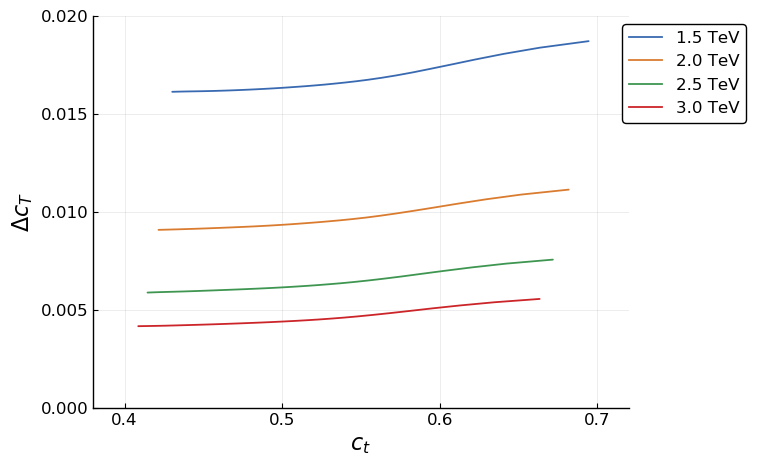}
\caption{Values of the fine-tuning measure 
$\Delta c_T =  c_T - c_{T,critical}$ for varying $c_t$ and $k_R$.}
\label{fig:finetuning}
\end{figure}

\begin{figure}
\centering
\includegraphics[width=4in]{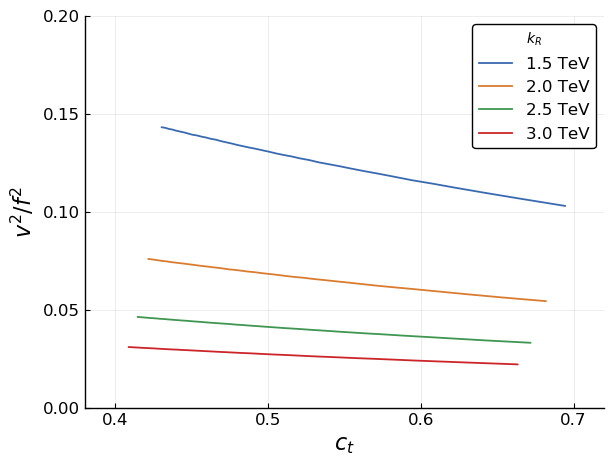}
\caption{Values of  $v^2/f^2$ for varying $c_t$ and $k_R$.}
\label{fig:vf}
\end{figure}

In the composite Higgs literature, it is customary to use $\epsilon =
v^2/f^2$ to quantify the degree of fine-tuning. However, at least in
the class of theories where the Higgs potential is generated
dynamically, $v/f$ is only a derived quantity which is determined by
more fundamental parameters in the theory. In our model, those
parameters are $c_t$ and $c_T$. The little hierarchy $v/f \ll 1$
requires $c_t$ and $c_T$ to be fine-tuned near the line of phase
transition, as illustrated in Fig. \ref{fig:phase}. Therefore, we
propose to use $\Delta c_T = c_T - c_{T,critical}$ as the measure of
fine-tuning, where $c_{T,critical}$ is the value of $c_T$ on the phase
transition line $A=0$. Fig.~\ref{fig:finetuning} shows the value of
$\Delta c_T$ for varying $c_t$ and $k_R$. It should be noted that this
choice does not soften the fine-tuning. For completeness, 
we also include a plot of $v^2/f^2$ in Fig.~\ref{fig:vf}.

\section{Precision electroweak observables}

One of the constraints on the parameter space of RS models
is that from precision electroweak measurements.  The strongest of
these are represented by constraints on the values of the oblique
parameters $S$ and $T$~\cite{PT}.  We have already invoked the small
size of the $T$ parameter to require a symmetry-breaking pattern with
a custodial $SU(2)$ symmetry.  Beyond this,  the $S$ parameter, which
is a measure of the total size of the new physics correction to $W$
and $Z$ vacuum polarization functions, places a lower bound on the KK
scale $k_R = z_R^{-1}$.

\subsection{Simplified $S$ and $T$}

Our discussion of the oblique parameters will be simplified in several
respects.  We will concentrate on
observables involving either no external fermions or only external
light leptons.  In this analysis, we will ignore all masses of light
leptons and assign these particles to appropriate zero modes in the
${\bf 5}$ of $SO(5)$. We will assume that all of these zero modes are
UV-dominated, that is, $c > 1/2$ for left-handed leptons and $c < -1/2$
for right-handed leptons. Realistic models might have different
assignments, especially for the right-handed components of the
quarks and leptons.   We will discuss other possibilities in \cite{YPthree}.  

RS models contain additional vector bosons beyond the SM
gauge bosons $\gamma$, $W^\pm$, and $Z$.   Thus, strictly, an
analysis in terms of the two parameters $S$ and $T$ does not capture
the full complexity of the new physics corrections to precision
electroweak formulae, even to leading order.  Here, we use simplified
formulae for $S$ and $T$ that capture the constraints from the five
best measured observables:  $\alpha(m_Z^2)$, $G_F$, $m_Z$, $m_W$, and
$s_*^2$, the effective value of $s_w^2$ at the $Z$ pole.   It is shown
in  \cite{PandW} that such an approach can put meaningful constraints
on new physics even in models with additional heavy vector bosons.

In this discussion, we define $\Delta A$ to be the new physics
contribution to an observable $A$.  We define  $\delta A$ to be the
fractional deviation from the SM prediction:  $\delta A = \Delta A/
A$. 

The $S$, $T$ formalism defines a reference weak mixing angle
$\theta_0$ by 
\beq
      \sin^2 2 \theta_0 = 4 s_0^2 c_0^2 =   { 4\pi \alpha(m_Z^2) \over
      \sqrt{2} G_F m_Z^2} 
\eeqn 
and then expresses the values of additional electroweak observables in
terms of $s_0^2$ and the oblique parameters.   In this approach,
\beqa
      m_W^2/m_Z^2 - c_0^2 &=& {\alpha c_0^2 \over c_0^2 - s_0^2}
      (-\half\,  S +
      c_0^2\, T) \CR  
 s_*^2 -  s_0^2 &=&    {\alpha \over c_0^2 - s_0^2} ({1\over 4}  \, S -
  s_0^2 c_0^2  \,  T)
\eeqa{STform}
In the current situation, values of $S$ and $T$ are mainly determined by the five
observables~\cite{pdg}.  Then we can find convenient formulae representing  the
measured values of the oblique parameters by solving \leqn{STform} for 
$S$ and $T$.   Choose a reference set of parameters which, in
zeroth order, satisfies the SM relations and let $\Delta
A$ (and $\delta A = \Delta A/A$)  represent the deviation of
observables from the predictions at this parameter set.  Then
\beqa
  \alpha \, S  &=&   4 \biggl[ \Delta s_*^2 +  s_0^2(\delta m_W^2 +
  \delta G_F - \delta \alpha ) \biggr] \CR
 \alpha \, T  &=&   \biggl[ (\delta m_W^2 + s_0^2 \delta G_F)/c_0^2 -
 \delta m_Z^2 + (2 \Delta s_*^2 
- s_0^2 \delta \alpha)/c_0^2 \biggr] \  .
\eeqa{mySTforms}
As a check, note that, if the only corrections to precision
electroweak come in a $q^2$-independent correction to the $W$ mass
(which also affects $G_F$), then these formulae predict  $\alpha S =
0$ and $\alpha T = \delta m_W^2 - \delta m_Z^2$, as desired.

In the next few subsections, we compute the tree-level  ${\cal O}(s^2)$
corrections to the five observables within our model.   We will
discuss the most important loop-level corrections in Section~6.5.

\subsection{$\alpha(m_Z)$, $m_W$, $m_Z$}

We take \leqn{firstsw} to provide the reference values of coupling
constants and express dimensionful parameters in terms of
the mass scale $s^2/z_R^2$.  We then expand the expressions for
observables in powers of $s^2$ around this reference point.  We have
seen in Section 4 that, in  zeroth
order,  the observables satisfy the SM relations.   Then
$S$, $T$ computed from \leqn{mySTforms} will be of order $s^2$.   In
the discussion of this section, we will keep terms only to order $s^2$
and we will  also ignore terms of order $z_0^2/z_R^2$. 

For the electromagnetic coupling, the reference formula in \leqn{firstsw}
gives the exact value at the tree level; there are no ${\cal O}(s^2)$
corrections.  So 
\beq 
    \delta \alpha = 0   \ . 
\eeq{deltaalpha}

Solving for zeros of the expressions \leqn{detCW}, \leqn{detCZ} to one
higher order in $s^2$, we find the corrections to \leqn{mWwa}, \leqn{mZwa}
\beqa
    \delta m_W^2 &= & +  {s^2\over 8 L_W^2}(3 L_W - 2) = { m_W^2 z_R^2
    \over 4} \, \biggl[ {3\over 2} - {1\over L_W} \biggr] \CR
   \delta m_Z^2 &= & +  {s^2\over 8 L_B^2 L_W^2}\,\bigl(3 L_B L_W (L_B +
   s_\beta^2 L_W)- 2  ( L_B^2 + s_\beta^2 L_W^2) \bigr) \CR
    & = &  {m_Z^2 z_R^2\over 4} \biggl[ {3\over 2} - \bigl( {c_w^2
      \over L_W } + {s_w^2\over L_B} \bigr) \biggr] \ .
\eeqa{WZmshifts}

These shifts in $m_W^2$ and $m_Z^2$ imply that their contributions  to
the $T$ parameter largely cancel.  The residue is 
\beq
\alpha\, T\biggr|_{m_W^2, m_Z^2}  = -  { s_w^2 m_Z^2 z_R^2 \over4} ({1\over L_W} - {1\over L_B} )
\eeq{WZtoT}
and this entirely vanishes if $L_W = L_B$ or $a_W = a_B$. 

\subsection{$G_F$}

To compute $G_F$, we consider the matrix element for muon decay $\mu
\to \nu_\mu e \bar \nu_e$.   This is computed from matrix elements of
the $A_m^A$ propagators between zero mode wavefunctions.  At first
sight, it seems that the only contribution comes from the matrix
element of $\VEV{A^{1L}_m(z) A^{1L}_n(z')}$ taken between simple
left-handed zero modes for $\mu_L$, $\nu_{\mu L}$, $e_L$, $\nu_{eL}$.
The Green's function in $(z,z')$ is integrated over 
the two sets of fermion zero mode wavefunctions in $z$ and $z'$.
This Green's function is given by the result \leqn{ALLWval} derived in Appendix D.1, 
\beq
\VEV{A_m^{1L}(z) A_n^{1L}(z')} = -\eta_{mn} { k z_R^2\over s^2} 
\left( 1 -  {s^2\over 2} \left(1 - {z_<^2\over z_R^2} \right) \right) \  ,
\eeqn
where $z_<$ is the smaller of $z$, $z'$ under the
integrals. We would find $G_F$ as the matrix element of this
expression multiplied by the coupling constant $g_5^2/2$. 
Note that the ${\cal O}(s^2)$ corrections depend on the
form of the zero mode wavefunctions and not simply on the total
normalizations times global charges.

However, there is a subtlety here.  We might assign a left-handed
lepton multiplet  to a ${\bf 5}$ according to 
\beq 
\Psi_e = \left[ \matrix { \pmatrix { E(-+) & \nu_e (++) \cr
     N (-+) & e_L (++)} \cr
 N'(-+)} \right], 
\eeq{Nt5}
as in \leqn{t5}.  Here we have chosen the $N$ and  $N'$ to have $(-+)$ boundary
conditions so that neither has a right-handed zero mode that can
combine with  
the $\nu_e$ zero mode to give a massive fermion.   Nevertheless, the $U_W$
matrix generated by top quark condensation will have the form
\leqn{Uinfive} in Appendix B, and this will mix the $\nu_e$ and $N'$
fields on the UV boundary.  As a result, the zero mode will be a
mixture
\beq
 {  (1 + c)\over 2} \ket{\nu_e} +  {  (1 - c)\over 2}\ket{N}
- { s\over \sqrt{2}} \ket{N'} \ .
 \eeq{nNzeromixing}

 The matrices $T^{a5}$ in \leqn{Tinfive}, for
$a = 1,2$ have
matrix elements between $\ket{\nu_e}$ and $\ket{N'}$.  The gauge field
Green's functions that can take advantage of these matrix elements
are, at $p=0$ and 
to the leading order in $s$,
\beqa
\VEV{A_m^{1L}(z) A_n^{15}(z')} &=&   \eta_{mn} { k z_R^2\over s^2} {s\over \sqrt{2}}\Bigl( 1 - 
{z^{\prime 2}\over z_R^2} \Bigr) \  ,\CR
\VEV{A_m^{15}(z) A_n^{1L}(z')} &=&   \eta_{mn}{ k z_R^2\over s^2} {s\over \sqrt{2}}\Bigl( 1 - 
{z^2\over z_R^2} \Bigr) \ .
\eeqa{Lfivematelems}
The piece of the matrix element $\VEV{A^{15}(z) A^{15}(z')} $
containing the $W$ boson pole is  proportional to
$p^2/(p^2-m_W^2)$.  It vanishes at $p^2 = 0$ and so does not
contribute to $G_F$. 
Assembling the pieces, including for each the square of the
coefficient in \leqn{nNzeromixing}, we find that there is a cancellation, so
that  $G_F$ is finally given by 
\beq
{4 G_F\over \sqrt{2}} =  {  g_5^2\over 2}{k z_R^2\over s^2} \biggl[ 1 - {s^2\over 2} {
 \VEV{ z_>^2 \over z_R^2}} \biggr] \ .
\eeq{fulldeltaGF}
Then
\beq
\delta G_F =     - {s^2\over 2} { \VEV{ z_>^2 \over z_R^2}}  \ . 
\eeq{deltaGF}
The evaluation of $\VEV{ z_>^2 / z_R^2}$ is discussed in Appendix C.3.  It is less than 0.2 for zero modes with $c = 1/2$ and exponentially suppressed for $c > 1/2$. Therefore, for UV-dominated light leptons, $\delta G_F$ is negligible.

There is an easier way to obtain this result.  Since all of the fields
in $\Psi_e$ have the same boundary conditions in the IR, the Wilson
line $U_W$ has no effect on the state when it is applied at the IR
boundary.
Then, in the IR gauge, the neutrino zero mode is purely
$\ket{\nu_e}$.   So,  in this gauge, only the  $\VEV{A^{1L}_m(z)
  A^{1L}_n(z')}$ matrix element contributes.   Using \leqn{UVIR}, we 
find 
\beq
        \VEV{A^{1L}_m(z) A^{1L}_n(z')}_{IR} =  -\eta_{mn} {k z_R^2 \over s^2}
        \biggl(1 -{s^2\over 2}  {z_>^2\over z_R^2} \biggr) \ , 
\eeqn
and the result \leqn{deltaGF} follows immediately.

\subsection{$s_*^2$}

The parameter $s_*^2$ appears in the ratio of the amplitudes for
$\ee\to \mu^+\mu^-$ in the different helicity states. 
We now calculate $s_*^2$, defined by the formula
\beq 
  {    g_Z(e^-_R)\over g_Z(e^-_L)} =    {-2 s_*^2\over 1- 2s_*^2} \ ,
\eeq{sstardef}
which corresponds to the tree-level SM relation.

The $Z$ couplings to the $e^-_L$ and $e^-_R$ zero modes are computed by
taking the matrix element of the $Z$ propagator---or, rather, the $Z$
boson 
pole terms in the $(A^{3L}, B, Z', A^{35})$
propagators---between fermion zero modes.   As in our discussion of
$G_F$, it avoids some difficulty to work in the IR gauge where the
zero modes are unmixed.  Then the zero modes have matrix elements
only with $(A^{3L}, B, Z')$, proportional to the $T^{3L}$, $Y$, and $T^{3R}$ charges as
they appear in the covariant derivative~\leqn{DM}. Furthermore, 
it should be noted that UV-dominated fermions have suppressed coupling 
to the $Z'$, since  this field has a $-$  UV boundary condition. This implies that
the leading  corrections to $s_*^2$ should have 
no explicit dependence on $T^{3R}$.  We will see this explicitly below. 
Since  $s_*^2$ depends only on the $T^{3L}$ and $Y$ charges, 
our result for $s_*^2$ actually 
holds for any assignments of $e^-_L$ and $e^-_R$ to $SO(5)$
representations. 

We construct the propagators in the UV gauge and then apply
\leqn{UVIR}. 
The three  fields $(A^{3L}, B, Z')$, 
have $+$ boundary conditions in the IR brane.  Then, following the
general formula \leqn{Greensconstruct},  all of their Green's
functions take the form 
\beq
\VEV{ A^A_m(z)  A^B_n(z')} = \eta_{mn}\  k p z_R z z' \biggl[  {\bf A}^{AB}
G_{+-}(z,z_R) G_{+-}(z',z_R)  + \cdots \bigr]\ .
 \eeq{ZGreens}
The $Z$ boson pole $(p^2-m_Z^2)$ is contained in the matrix ${\bf
  A}^{AB}$, and so the terms omitted in \leqn{ZGreens} contain $Z$
pole terms that include the factor
$G_{++}(z,z_R)$.   This factor   will appear in the matrix
elements of $(A^{3L}, B, Z')$ when we convert to the IR gauge using 
\leqn{UVIR}, however, always with a coefficient of order $s^2$.   Then
we will need $G_{++}(z,z_R)$ only to leading order
\beq
            G_{++}(z,z_R) = {z_R\over 2z} \biggl( 1 - {z^2\over
              z_R^2}\biggr) \ , 
\eeq{Gppexpand}
while we will need $G_{+-}(z,z_R)$ to the next order, 
\beq
G_{+-}(z,z_R) =  {1\over p z} \cdot \biggl[ 1 + {(p z_R)^2 \over 4}
\biggl(-1 + {z^2 \over z_R^2} + 2 {z^2 \over z_R^2} \log {z_R \over z}
\biggr) \biggr] \ . 
\eeq{Gpmexpand}

The calculation of the matrix ${\bf A}^{AB}$ is
described in Appendix D.2.  The expression for this matrix contains an
overall factor
\beq 
[\det {\bf C}]^{-1} = {2 p^3 z_0^2 z_R m_Z^2 /s^2 \over (L_B + L_W s_\beta^2)  }
{1\over p^2 - m_Z^2}  \cdot  {\cal Z}_Z(p^2)\ . 
\eeq{detCdecomp}
up to corrections of higher order in $s^2$. 
The factors of  $m_Z^2$ in \leqn{detCdecomp}  include the order $s^2$
corrections shown in \leqn{WZmshifts}.

The terms of order $p^2$, evaluated at the $Z$ pole, 
 contribute 
corrections of order $s^2$  to the residue.  However, ${\cal Z}_Z$
gives a correction to normalization factor  that is 
common to all of the Green's
functions we will discuss, and one that cancels out of the ratio of
couplings.  The $(-1)$ term in \leqn{Gpmexpand} also contributes to the
common overall factor. 
The $z$-dependent terms are very small for
 fermion zero modes that are
peaked in the UV and therefore we omit this correction here.
 Similarly, we ignore $z^2/z_R^2$ in $G_{++}$. We will 
return to consider those terms in Section~7. Aside from
 these factors, we keep below all corrections of ${\cal O}(s^2)$.

With this understanding, we can write the poles at $p^2 = m_Z^2$ in 
the vector field Green's functions. Up to terms of order $s^2$, we find
\beqa
\VEV{A^{3L}_m A^{3L}_n} &=& {k\eta_{mn} \over (p^2 - m_Z^2)(L_B +
   s_\beta^2 L_W)} \biggl[  {L_B\over L_W} -{ s^2\over 4}{(L_B+
 s_\beta^2 L_W)\over L_B L_W^3} (L_B - L_W + 2 L_B L_W^2) \biggr] \CR
\VEV{A^{3L}_m B_n} &=& {k \eta_{mn}  s_\beta\over (p^2 - m_Z^2)(L_B +
 s_\beta^2 L_W)} \biggl[ -1 +  { s^2\over 4}{(L_B+
s_\beta^2  L_W)\over L_B L_W} (L_B + L_W) \biggr]  \CR
\VEV{B_m B_n} &=& {k \eta_{mn} s_\beta^2 \over (p^2 - m_Z^2)(L_B +
   s_\beta^2 L_W)} \biggl[  {L_W\over L_B}  - { s^2\over 4} {(L_B+
 s_\beta^2 L_W)\over L_B^3  L_W} (L_W - L_B + 2 L_B^2  L_W) \biggr]  \CR
\VEV{A^{3L}_m Z^\prime_n} &=& {k \eta_{mn}  c_\beta \over (p^2 - m_Z^2)(L_B +
 s_\beta^2 L_W)} \biggl[  + { s^2\over 4}{(L_B+
  s_\beta^2  L_W)\over  L_W} \biggr] \CR
\VEV{B_m Z^\prime_n} &=& {k \eta_{mn} s_\beta c_\beta \over (p^2 - m_Z^2)(L_B +
 s_\beta^2 L_W)} \biggl[  - { s^2\over 4} {(L_B+
 s_\beta^2  L_W)\over  L_B} \biggr] \CR
\VEV{Z^\prime_m Z^\prime_n} &=& 0 \CR
\VEV{A^{3L}_m A^{35}_n} &=& {k \eta_{mn}  \over (p^2 - m_Z^2)(L_B +
  s_\beta^2 L_W)} \biggl[  -{ s\over \sqrt{2}} {(L_B+
  s_\beta^2  L_W)\over L_W} \biggr] \CR
\VEV{B_m A^{35}_n} &=& {k \eta_{mn} s_\beta \over (p^2 - m_Z^2)(L_B +
 s_\beta^2 L_W )} \biggl[  + { s\over \sqrt{2}} {(L_B+
 s_\beta^2 L_W)\over  L_B} \biggr] \CR
\VEV{Z^\prime_m A^{35}_n} &=& 0 \CR
\VEV{A^{35}_m A^{35}_n} &=& {k \eta_{mn}  \over (p^2 - m_Z^2)(L_B +
 s_\beta^2 L_W)} \biggl[  + { s^2\over 2} {(L_B+
 s_\beta^2 L_W)^2 \over L_WL_B} \biggr] \ .
\eeqa{relevantGreens}
The expressions factorize onto the pole of a single vector meson, as
required, giving the coupling between lepton zero modes  1 and 2
\beq
       {g_{Z}(1) \   g_{Z} (2) \over p^2 - m_Z^2}  \ .
\eeq{finalZcoupling}

From these expressions, and using \leqn{firstsw} to make some
simplifications, we can write the $Z$ wavefunction in the UV
gauge (for $z \ll z_R$) as
\beqa
   \ket{Z}  = \left({k\over L_W c^2_w }\right)^{1/2} \times 
 && \Biggl\{  c_w^2 \left( 1 - {s^2\over 8} 
 {(L_B + s_\beta^2 L_W) \over L_B^2 L_W^2} (L_B - L_W + 2 L_B L_W^2) \right) \ket{A^{3L}} \CR
& & \hskip 0.3in - {s_w^2 \over s_\beta} \left( 1 - {s^2\over 8} 
{(L_B+ s_\beta^2 L_W)\over L_B^2 L_W^2} (L_W - L_B + 2 L_B^2 L_W) \right) \ket{B}\CR
& &  \hskip 0.3in  + c_\beta {s^2\over 4} \ket{Z'} - { s\over \sqrt{2}} \ket{A^{35}}\Biggr\}
\ . 
\eeqa{Zwfone}
To obtain the $Z$ wavefunction in the IR gauge, apply $U_W$ to this
wavefunction as indicated in \leqn{UVIR}.   There is a nice
cancellation, and we find
\beqa
   \ket{Z}_{IR}  = \left({k\over L_W c^2_w }\right)^{1/2} \times 
 && \Biggl\{  c_w^2 \left( 1 - {s^2\over 8} 
 {(L_B + s_\beta^2 L_W) \over L_B^2 L_W^2} (L_B - L_W) \right) \ket{A^{3L}} \CR
& & \hskip 0.3in - {s_w^2 \over s_\beta} \left( 1 + {s^2\over 8} 
{(L_B+ s_\beta^2 L_W)\over L_B^2 L_W^2} (L_B - L_W) \right) \ket{B}\Biggr\}
\ . 
 \eeqa{ZwfoneIR}
with no $Z'$ or $A^{35}$ components. Then we can read off the $Z$
coupling to a massless fermion as
\beqa
g_Z &= &  g c_w \biggl[   T^{3L}  \ \bigl\{ 1 -  { s^2\over 8} { (L_B + L_W
s_\beta^2)\over  L_B^2 L_W^2} (L_B - L_W) \bigr\} \CR
& & \hskip 0.8in  -  {s_w^2\over c_w^2}  Y\ \bigl\{ 1 +  { s^2\over 8}{ (L_B + L_W
s_\beta^2)\over  L_B^2 L_W^2} (L_B - L_W)\bigr\}  \biggr]\ .
\eeqa{finalgZform}
This formula applies to any zero-mode fermion that is unmixed in the
IR gauge and strongly localized in the UV. 
 Note that it contains no separate dependence on $T^{3R}$.
It is an interesting exercise to collect the extra terms that appear in the UV
gauge for $(\nu,e)_L$ in the ${\bf 5}$ and also in the ${\bf 4}$ and
see how the $T^{3R}$ terms cancel in all of these cases.

Finally, as in \leqn{WZtoT},
the precision electroweak correction is proportional to $(L_W -
L_B)$.   Computing \leqn{sstardef}, we find
\beq
 \Delta s_*^2 = (s_*^2 - s_w^2) =  {s^2 s_w^2 c_w^2 \over 4L_B^2 L_W^2} (L_B + 
 s_\beta^2 L_W)  =  { m_Z^2 z_R^2 s_w^2c_w^2\over 4} ({1\over L_W} - {1\over
   L_B} )  \  . 
\eeq{finaldetalsstar} 

\subsection{Loop corrections to $T$}

The formulae that we have derived so far represent the formally
leading new physics corrections to $S$ and $T$.  However, it has been
shown in other investigations of precision electroweak corrections to
composite Higgs models, that loop effects on $T$ from the top quark and top
partners can also make significant
contributions~\cite{PT,LittleHiggsone,LittleHiggstwo}.  In this
section, we will make an estimate of the contribution to $T$ from
fermion loop effects, dealing as best we can with the
non-renormalizability of this 5D  theory.

We will work from the original formula for $T$~\cite{PT}, 
\beq
       \alpha T = { e^2\over s_w^2 c_w^2 m_Z^2}  \bigl(  \Pi_{1L,1L}(0) -
  \Pi_{3L,3L}(0) \bigr)  \ , 
\eeq{starterforT}
where $\Pi_{aL,aL}$ is the vacuum polarization amplitude for the
currents of the 
$a$ component of weak isospin.   The expression for $T$ in
\leqn{mySTforms}
 involves
contributions at $q^2 =0$ and at $q^2 = m_W^2, m_Z^2$.  In this section, we will
simplify the calculation of the loop integral by working at $q^2 = 0$
only.  We will calculate in the IR gauge, in which the contribution of
$A^{a5}$ to the $W$ and $Z$ wavefunctions is, if not completely zero,
at least highly suppressed. 

The vacuum polarization amplitudes in \leqn{starterforT} involve loops
with the $t_L$ and $b_L$ field in $\Psi_t$ and the corresponding
fields in $\Psi_T$.  The currents involve only the 4D left-handed
components of these fields.   Then the propagators, in Euclidean
space, 
can be written as
\beqa
     \VEV{ (t_L)_L(z,p) ( t_L)_L^\dagger(z',p) } &=&   \sigma\cdot p \
    \S_t (z,z',p) \CR
    \VEV{ (b_L)_L(z,p) ( b_L)_L^\dagger(z',p) } &=&   \sigma\cdot p \
    \S_b (z,z',p) \ . 
\eeqa{tbEforms}
Here the $L$ inside the parentheses labels the species in \leqn{t5}
while the $L$ outside the parentheses indicates a projection onto the
2-component
fermion with left-handed chirality.   Using \leqn{tbEforms} to
evaluate the $\Psi_t$ contribution to $T$, we find 
\beq
 \alpha T =  { 3 e^2\over s_w^2 c_w^2 m_Z^2} \int
 {dz \over (kz)^4} { dz'\over (kz')^4} \int
  {d^4 p \over (2\pi)^4} \   {1\over 4}p^2 \  (\S_t - \S_b)^2 \ .
\eeq{Texpress}
The integral $d^4p$ is over Euclidean momentum space.
Note that $(\S_t - \S_b)$ is of order $s^2$, so this
contribution to $T$  is of order $s^4$. 

We proceed, then, to evaluate $T$ from the formula \leqn{Texpress}.
A complete evaluation of this expression \leqn{Texpress} is beyond
 the scope of this
paper. Instead, we will estimate the integral from its 
low-momentum behavior of the integrand.
The $t_L$ and $b_L$ propagators are given by their SM formulae, plus
corrections of order $m_t z_R$ and $p z_R$.  We can write these as
\beqa
\S_t  &=&  f_L(c)^2 (z z')^{2-c} \cdot 
 {1\over p^2 + m_t^2}\biggl[ 1 +  A\, (m_t z_R)^2+
B\, (p z_R)^2 + \cdots \biggr]\CR
\S_b  &=&   f_L(c)^2 (z z')^{2-c}\cdot  {1\over p^2}\biggl[ 1 +  
C\, (p z_R)^2 + \cdots \biggr] \ .
\eeqa{Stbdecomp}
The first factor is the form of the zero mode wavefunctions as
functions of $z$ and $z'$; see~\leqn{Greenpsiepssmall}.   The correction terms are summarized in 
coefficients $A$, $B$, $C$.  These coefficients may contain additional dependence on
$z$, $z'$. To linear order in the coefficients, this is treated by
taking  the expectation
values of the $z$, $z'$-dependent terms as indicated by the $dz$ integrals.

Using \leqn{Stbdecomp},  the 
$d^4p$ integral in \leqn{Texpress} becomes
\beqa
\int  {d^4 p \over (2\pi)^4} \  p^2 \ & &  \biggl[ {m_t^4\over p^4 (p^2 +
   m_t^2)^2   } - 2  A { m_t^4 z_R^2\over p^2 (p^2+ m_t^2)^2 } \CR & &
 \hskip 0.7in -2 B
 {m_t^2z_R^2 \over
( p^2 + m_t^2)^2} +2 C{  m_t^2 z_R^2 \over p^2 (p^2 + m_t^2)} + \cdots  \biggr]
\eeqan
The integral  of the first term is convergent.  
This is  proportional to   $m_t^4/m_t^2$ and so actually of order $s^2$ due to
the infrared behavior of the integral.    The integrals of the
correction terms give cutoff-dependent contributions of order $m_t^4
z_R^2$.  Higher-order terms in $p^2$ in \leqn{Stbdecomp}  also
contribute at this order, and we expect that the sum leads to an
expression that is at worst log divergent in the ultraviolet.  But
these terms in the integral also contain 
infrared-enhanced terms   of order $m_t^4 z_R^2 \log (
1/m_t^2)$.  Using
either dimensional regularization or an explicit cutoff on the
integral, we find
\beq 
\alpha T =  \alpha \cdot {3 m_t^2 \over 16 \pi s_w^2 c_w^2 m_Z^2}
\biggl[   1 + 2\,  ( 2B - A - C)\, m_t^2 z_R^2\, \log
(\Lambda^2/m_t^2)\  +\ 
{\cal O} (m_t^2 z_R^2) \biggr] \  ,
\eeq{nearfinalT}
where $\Lambda$ is an ultraviolet scale. There is also a 
contribution to $T$ from the vacuum
polarization of $\Psi_T$, but this contains no light fermions and so
contributes only the hard, non-logarithmic, term in
\leqn{nearfinalT}.   The leading term is the
usual SM contribution to $T$ from the $(t,b)$ doublet.   The usual
convention is that $T$ parametrizes a deviation from the $SM$, so we
will now drop this term.
We claim that the RS
contribution to $T$ can be estimated from the expression
\beq 
\alpha T =  \alpha \cdot {3 m_t^2 \over 16 \pi s_w^2 c_w^2 m_Z^2}
\cdot 2\, ( 2B - A - C)\,  m_t^2 z_R^2\, \log (\Lambda^2/m_t^2) 
\eeq{finalT}
by ignoring the hard corrections and 
 varying $\Lambda$ over  the interval  $1/z_R$ to  $1/z_0$. 

The complete expressions for the coefficients $A$, $B$, $C$ in the IR
gauge are  given in  Appendix~G.  In the parameter
discussion in Section 5, we found that  the top quark boundary
kinetic term $a_t$  and the related value $L_t =  G_{t--}(z_0,z_R)$
must be large. Then we can simplify the full expression for our
estimate by keeping only the terms leading in $a_t$.  This gives
the relatively simple estimate
\beq
T \approx {3 m_t^4 z_R^2  \over 16 \pi s_w^2 c_w^2 m_Z^2}\  
s^2 \  \VEV{ ( {z\over
     z_R}\bigr)^{2c_t+1}  +  ( {z'\over
     z_R}\bigr)^{2c_t+1}} 
\log (\Lambda^2/m_t^2)\    ,
\eeq{finalTresult}
where the indicated expectation value is taken in the zero mode
wavefunction using the measure \leqn{zeromodemeasure}. 
However, because the indicated expectation values of $z$ and $z'$ are
small, this parametrically dominant term is not actually larger than
the other pieces, so we quote it here mainly for illustration.   The
full result for our estimate of  $T$ is given in Appendix G in
\leqn{fullTvalue}.

\subsection{Phenomenological implications}

We must now sum all of these contributions as indicated in
\leqn{mySTforms}.
We may omit the small correction $\delta G_F$.  Then, for $S$
\beq
\alpha S = m_Z^2 z_R^2 s_w^2 c_w^2 \biggl({3\over 2}  -
{1\over L_B}  \biggr) \ . 
\eeq{finalSval}
For large $L_B$ as is found in the parameter space of Section~5, a
limit of $S < 0.135$ gives the constraint
\beq
           k_R > 1.5~\mbox{TeV} \  .
\eeq{kRzRlimit}
For $T$, we find the tree-level  RS correction
\beq
\alpha T = {m_Z^2 z_R^2s_w^2 \over 4}
\biggl[ {1\over L_W} - {1\over L_B} \biggr]
  \eeq{finalTval}
plus the loop correction estimated by \leqn{finalTresult}.

\begin{figure}
\centering
\includegraphics[width=5in]{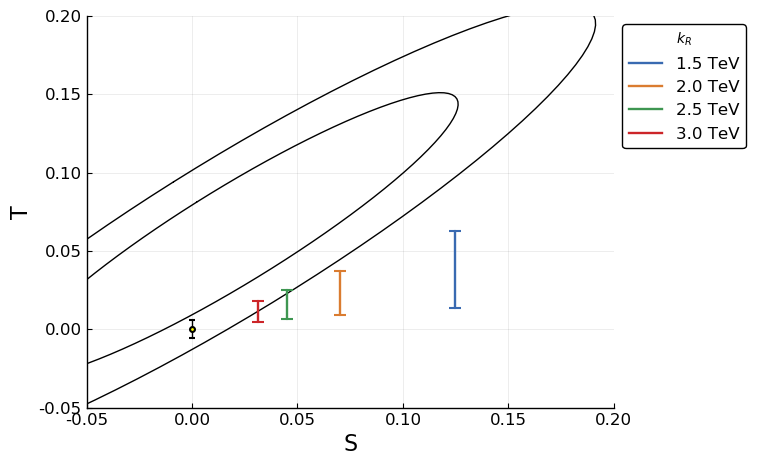}
\caption{Corrections to $S,T$ parameters of our model. The inner and
  outer contours are for 68\% and 95\% confidence level, respectively,
  from \cite{pdg}.  
The black error bar at the origin corresponds to current top quark
mass uncertainty. Each colored line represents the estimated range of 
$T$ parameter by varying the cutoff $\Lambda$ from $k_R$ to $k$.   }
\label{fig:ST}
\end{figure}

Fig.~\ref{fig:ST} shows the mapping of our parameter space onto the region of
$S$ and $T$ allowed by experiment~\cite{pdg}. In view of the
uncertainties in our estimate of the $T$ parameter, we regard the
parameter region of our model with $k_R > 1.5$ TeV to be in reasonable
agreement with the current values of the precision electroweak
observables. 

\section{$Z\to b\bar b$}

In the analysis of the coupling of the $Z$ to fermions, we assumed that
all of the relevant quarks and leptons are associated with fermion
zero modes that are highly peaked in the UV.    However, this is not
the case for the $b$ quark.    The $b_L$ is the $SU(2)$ partner of the
$t_L$, and so it must share  the same value of $c$.   For
$b_R$, the story is somewhat more involved.  The $b_R$ zero
mode is not included in either of the multiplets $\Psi_t$, $\Psi_T$
that we have considered so far in our analysis.  However, models for
generating the $b$ quark mass typically require $b_R$ to have a {\it
  positive}
value of $c$, pushing the zero mode  wavefunction to the IR and
potentially giving large effects~\cite{YPthree}.  In this
section, we provide general formulae for the special influence of the 
$b$ quark zero modes on the relevant precision electroweak observables. 
As in our discussion of  $\Delta
s_*^2$, we will not need to assume the
particular model studied in Section~5, because we will work from the
simple, general formulae for $Z$ boson couplings derived in Section~6.4.

The $b$ quark couplings to the $Z$ boson are tested with precision by
the ratio of yields
\beq    R_b =   {\Gamma(Z\to b\bar b) \over \Gamma(Z \to
  \mbox{hadrons}) }
\eeq{Rbdefin}
and the polarization asymmetry
\beq
       A_b =     {\Gamma(Z \to b_L\bar b_R) - \Gamma(Z\to b_R \bar
         b_L) \over \Gamma(Z \to b_L\bar b_R) +\Gamma(Z\to b_R \bar
         b_L) }
\eeq{Abdefin}
Looking back at the discussion following \leqn{detCdecomp}, we see
that the factor $  {\cal Z}_Z$   cancels out of both ratios,
while $z$-depdendent terms in $ G_{+-}(z,z_R)$ and $z^2/z_R^2$ in $G_{++}(z,z_R)$ will make a contribution if the zero mode
wavefunction extends into the IR. Including those factors, the $Z$ wavefunction in the IR gauge can be written as
\beqa
   \ket{Z}_{IR}  &=& \left({k\over L_W c^2_w }\right)^{1/2} \CR 
 && \times \Biggl\{  c_w^2 \left[ 1 + {m_Z^2 z_R^2 \over 4} 
  \left(-{1 \over 2 L_W} + {1 \over 2 L_B} + (1-2L_W){z^2 \over z_R^2} + 2 {z^2 \over z_R^2} \log {z_R \over z}\right) \right] \ket{A^{3L}} \CR
& & \hskip 0.3in - {s_w^2 \over s_\beta} \left[ 1 + {m_Z^2 z_R^2 \over 4} 
  \left({1 \over 2 L_W} - {1 \over 2 L_B} + (1-2L_B){z^2 \over z_R^2} + 2 {z^2 \over z_R^2} \log {z_R \over z}\right) \right] \ket{B} \CR
& & \hskip 0.3in  + c_w^2 c_\beta {m_Z^2 z_R^2 \over 4} \cdot 2 L_W {z^2 \over z_R^2} \ket{Z'} + c_w {m_Z z_R \over 2} \cdot  \sqrt{2L_W} {z^2 \over z_R^2} \ket{A^{35}} \Biggr\} \ . 
 \eeqa{ZwfoneIRfull}
The $A^{35}$ contribution will be utimately suppressed by $m_b^2 z_R^2$. Then the correction to the coupling of $Z$ boson to the bottom quark is given by
\beqa
   \Delta g_{Z b} &=& g c_w \cdot {m_Z^2 z_R^2\over 4} \times 
   \Biggl\{(T^{3L} - {s_w^2\over c_w^2} Y) \VEV{ 2 {z^2\over z_R^2} \log{z_R\over z} + {z^2\over z_R^2} } \CR
 && \hskip 0.5in   + (T^{3L} + {s_w^2\over c_w^2} Y ) \left( {1 \over 2 L_B} - {1 \over 2 L_W} \right) 
 + 2 L_W \VEV{ {z^2\over z_R^2} } (-T^{3L} + T^{3R} ) \Biggr\} \CR
 &=& g_{Zb} \cdot {m_Z^2 z_R^2\over 4} \left( \VEV{ 2 {z^2\over z_R^2} \log{z_R\over z} + {z^2\over z_R^2} } + 2 L_W \VEV{ {z^2\over z_R^2} } {-T^{3L} + T^{3R} \over T^{3L} - (s_w^2 / c_w^2 ) Y } \right)  \, .
\eeqa{gzbcorrect}
where $g_{Zb}$ is the SM $Z$ coupling to $b_L$ or $b_R$  and the expectation value 
of $z$ must be computed in the appropriate zero mode wavefunction.  
Note that in the final line we omitted the terms suppressed by $1/L_{W,B}$. 

The second term in \leqn{gzbcorrect} is enhanced by a large $L_W$ and
can cause a large deviation in $g_{Zb}$. However, specifically for
$\Psi_t$ in the {\bf 5} representation of $SO(5)$, we have $T^{3L} =
T^{3R} = -1/2$ for $b_L$, and the $L_W$-enhanced term in
\leqn{gzbcorrect} vanishes identically. This shows that  the custodial
symmetry proposed in \cite{5rep} to protect the $Zbb$ vertex is
working correctly. Although the formula~\leqn{gzbcorrect} is a
general result which applies to any assignment of $b$ quark in
$SO(5)$, we focus on the {\bf 5} representation in the remaining of
this section and  study whether the remaining correction in $g_{Zb}$
gives constraints on the parameter space. 

For the case of a $(t_L
, b_L)$ doublet in {\bf 4} as in \leqn{t4}, the $L_W$-enhanced term
will give a dominant correction 
 to $g_{Zb}$.  Such models can still be viable for higher values of
 $k_R$ or for assignments of both $b_L$ and $b_R$ to UV-dominated zero
 modes~\cite{YPthree}. 

For the evaluation of \leqn{gzbcorrect}, 
the computation of the $z$
expectation values is discussed in Appendix~C.3.   For a left-handed
zero mode with positive $c$, taking $a= 0$ for reference,
\beq
\VEV{ 2 {z^2\over z_R^2}
        \log{z_R\over z} + {z^2\over z_R^2} } =   (   0.36, 0.11, 0.047    )
\eeqn
for $c =( 0.3, 0.5, 0.7)$ and $z_0/z_R = 0.01$. 
   The value is exponentially decreasing with
increasing $c$.    For a right-handed zero mode with positive $c$,  we find 
\beq 
  \VEV{ 2 {z^2\over z_R^2}
        \log{z_R\over z} + {z^2\over z_R^2} } =  { (1+2c)( 5+ 2c)\over (3
        + 2c)^2} =
 (0.69, 0.75, 0.79)
\eeqn
for $c =( 0.3, 0.5, 0.7)$.

The $b_L$ and $b_R$ $Z$ couplings have different effects on $R_b$
and $A_b$ due to the very different sizes of these couplings,
\beq
        {   g_{ZbR}^2 \over g_{ZbL}^2} =   0.0331  \ .
\eeqn
The correction to $R_b$ is dominated by the shift of
$g_{ZbL}$,
\beq
        \Delta R_b \approx 2  R_b (1- R_b) \bigl[ \delta g_{ZbL}  +
        {g_{ZbR}^2 \over g_{ZbL}^2}  \delta g_{ZbR} \bigr] \ .
\eeqn
Then
\beqa
    \Delta R_b  &\approx   & \half  R_b (1- R_b)   \cdot m_Z^2 z_R^2 \VEV{ 2 {z^2\over z_R^2}
        \log{z_R\over z} + {z^2\over z_R^2} }_L  \CR
   & = &    (3.7\times 10^{-4})  \VEV{ 2 {z^2\over z_R^2}
        \log{z_R\over z} + {z^2\over z_R^2} }_L  \ .
\eeqa{finalRberr}
The last line here is evaluated using the limit in  \leqn{kRzRlimit}
for $z_R$.   The expectation value is to be taken in the $b_L$ zero
mode wavefunction. 

On the other hand, the correction to $A_b$ comes equally from
$g_{ZbL}$ and 
$g_{ZbR}$, 
\beq
\Delta A_b \approx   4  {g_{ZbR}^2 \over g_{ZbL}^2}\,  (\delta g_{ZbL}
- \delta g_{ZbR}) \ .
\eeqn
Then
\beqa
\Delta A_b  &\approx &   {g_{ZbR}^2 \over g_{ZbL}^2}  m_Z^2 z_R^2  \biggl( \VEV{ 2 {z^2\over z_R^2}
        \log{z_R\over z} + {z^2\over z_R^2} }_L  - \VEV{ 2 {z^2\over z_R^2}
        \log{z_R\over z} + {z^2\over z_R^2} }_R \biggr) \CR
&\approx & - (1.4\times 10^{-4})  \VEV{ 2 {z^2\over z_R^2}
        \log{z_R\over z} + {z^2\over z_R^2} }_R  \ . 
\eeqan
where again we used \leqn{kRzRlimit} for $z_R$.  The expectation value
in the last line is to be taken in the $b_R$ zero
mode wavefunction, which, to obtain as large as possible a value of
$\Delta A_b$, would be larger than the corresponding expectation value
for $b_L$. 

The experimental measurements of these quantities are \cite{LEPEWWG}
\beq
    R_b  = 0.216\pm 0.00066 \, , \quad   A_b = 0.923 \pm 0.020
\eeqn
so the predicted deviations from this class of RS models are well
within the errors. Although it is typical in composite Higgs models that
the experimental measurement of $R_b$ places a very strong
constraint, that is not true with the custodial symmetry protecting 
the $Zbb$ vertex.

\section{Conclusions}

In this paper, we developed and examined a realistic model of a composite Higgs boson based on the gauge-Higgs unification framework and $SO(5) \times U(1)$ gauge symmetry. The top quark multiplet $\Psi_t$ triggers electroweak symmetry breaking.  A new Dirac fermion $\Psi_T$ competes with the top quark and allows us to tune the value of the Higgs boson (mass)$^2$ term. We can achieve the hierarchy $v/f \ll 1$ by arranging the 5D mass parameters of the top quark and top partner to be close to the second-order phase transition in the plane of these parameters. We also introduced UV boundary kinetic terms for the gauge fields and top quark, which give us the freedom to fit the $SU(2)$ gauge coupling and the top quark Yukawa coupling. 

After applying constraints from the $W$, $Z$, $t$, and Higgs masses, our model has an effectively two-dimensional parameter space. We computed the full Higgs potential and studied the allowed region in this space. It turns out that our minimal model requires large values for the UV boundary terms. An additional source of the quartic term in the Higgs potential could relax the tension that leads to these large terms.

Our model is not strongly constrained by current experimental results.
Although the mass of the top partner $\Psi_T$ is significantly smaller than the scale of the new composite sector, we can avoid constraints from LHC searches if $\Psi_T$ is color-neutral. This solution is similar to the idea of “neutral naturalness” but is distinct in important respects.   The main constraint on our parameter space comes from precision electroweak measurements. To analyze this constraint, we use the small 
value of $v/f$ required in our model as an expansion parameter.  This strategy allows us to write 
general formulae for the precision electroweak corrections due to the new composite sector. A lower limit on the RS scale of 1.5~TeV allows our model to be consistent with current electroweak data. 

In this paper, we left open the question of how the lighter quarks and
leptons receive their masses. A particularly interesting question is that of how we can  generate the bottom quark mass in this framework.  In a forthcoming paper, we will study possible scenarios of light flavor mass generation and their implication for observable effects in  $\ee \to f {\bar f}$ processes~\cite{YPthree}.

\appendix

\section{Properties of Minkowski-space Green's functions}

The computations done in this paper make use of Green's functions for
spin~1/2 and spin~1 fields in the RS background with Dirichlet or
Neumann boundary conditions on the IR brane.   The formalism for
computing these Green's functions was reviewed in \cite{YPone}.
  However, since \cite{YPone} was mainly devoted to the
computation of the Coleman-Weinberg potential, the equations for 
Green's functions were given for Euclidean time, and the full
expressions for the Green's functions were not needed.   In this
Appendix, we present the formulae for Minkowski-space Green's
functions in a  notation consistent with the conventions of
\cite{YPone}.   In this section, and in the rest of the Appendix, we
will  work in the UV gauge defined in Section~4.5 unless it is
explicitly noted otherwise. 

\subsection{Building blocks}

Green's functions for fields in the RS background are built up from
Bessel functions with definite boundary conditions at the UV and IR
branes.  In Minkowski space, we will choose as out basic building
blocks the combinations $G^{(c)}_{\alpha\beta}(z_1,z_2,p)$,  defined in
\leqn{Gdef}.     These
functions depend on two 5th-dimension coordinates $z_1$, $z_2$, the
parameter $c$, and $\alpha,
\beta = \pm$. For a massive spin 1/2 field in RS, $c = m/k$.   The 4-vector
$p$ is the 4D momentum. 
  When combined with a prefactor $z^a$, where
$a = 1$ for spin 1 field and $a= 5/2$ for spin 1/2 fields, 
$G_{\alpha\beta}(z,z_R)$ satisfies Dirichlet or Neumann boundary conditions 
on the IR brane at $z = z_R$.  Typically, we will keep the
dependence on $c$ and $p$ implicit.  In this paper, we will work with the $G$
functions for Minkowski $p^\mu$, that is $p^2 > 0$, $p = (p^2)^{1/2}$. 
   Analogous formulae for the $G$
functions at Euclidean $p$,  $p^2 < 0$, which we will denote
$G_{E\alpha\beta}(z_1,z_2)$,
 are  given in \cite{YPone}.

The  $G$ functions  (at fixed $c$ and $p$) manifestly satisfy
\beq
       G_{ab}(z_1,z_2) = - G_{ba}(z_2,z_1) \ . 
\eeq{Greverse}
Less trivially, they  satisfy the
Wronskian  identity 
\beq
    G_{\alpha + }(z_1,z_3) G_{\beta -}(z_2,z_3) -  G_{\alpha -} 
(z_1,z_3) G_{\beta +}(z_2,z_3) =   {1\over p z_3}
    G_{\alpha\beta}(z_1,z_2) \ .
\eeq{genWronsk}
An important special case is
\beq
    G_{+ + }(z_1,z_2) G_{--}(z_1,z_2) -  G_{+ -
    }(z_1,z_2) G_{-+}(z_1,z_2) =   {1\over p^2 z_1 z_2}\ .
\eeq{myWronsk}

To explore the properties of particles with masses much less than the
KK scale $k_R$, we will need the expansions of
$G_{\alpha\beta}(z_1,z_2)$ for small $p$.  For general $c$ in the range
$-1 < c < 1$, 
\beqa
 G_{++}(p) &\approx &  (z_1z_2)^{-c-1/2} (z_2^{2c+1} - z_1^{2c+1})/(2c+1)\CR
G_{+-}(p) &\approx & z_2^{c-1/2 }  z_1^{-c-1/2} / p \CR
G_{-+}(p) &\approx & - z_1^{c-1/2 }  z_2^{-c-1/2} / p \CR
G_{--}(p) &\approx &  (z_1z_2)^{-c + 1/2} (z_2^{2c-1} -
z_1^{2c-1})/(2c-1)\   .
\eeqa{Gpvals}
For the special case of $c = \half$, 
\beqa
 G_{++}(p) &\approx &  (z_1z_2)^{-1} (z_2^{2} - z_1^{2})/2 \CR
G_{+-}(p) &\approx &  z_1^{-1} / p \CR
G_{-+}(p) &\approx & - z_2^{-1} / p \CR
G_{--}(p) &\approx & \log (z_2/z_1)  \   .
\eeqa{Gpvalshalf}

\subsection{Spin 1 fields}

For spin 1 fields, $c = 1/2$.  The solutions of the gauge-fixed
Maxwell equations
in $z$ 
are $z \, G_{\alpha \beta}(z,z')$,  with 
$\alpha = +$ for $A^A_m$, $m = 0,1,2,3$, and $\alpha = -$ for
$A^A_5$.  The solutions for the ghost fields also have $\alpha = +$.

We will construct solutions with definite Neumann $(+)$ or Dirichlet $(-)$
boundary conditions on the IR brane at $z = z_R$.   The solutions
to the Maxwell equation satisfying these boundary conditions contains
\beq
\begin{tabular}{l|cc}
   &    $+$ b.c.  at $z_R$      &   $ -$ b.c. at $z_R$  \\ \hline
$A^A_m$   &    $ G_{+-}(z,z_R)   $  & $   G_{++}(z,z_R) $\\ 
 $A^A_5$   &   $ G_{-+}(z,z_R)  $   &  $  G_{--}(z,z_R) $\\
\end{tabular}
\eeq{IRbcs}
For a consistent definition of $F^A_{m5}$ on the boundary, $A^A_5$ must
satisfy $-$ boundary conditions if $A^A_m$ satisfies $+$ boundary
conditions, and vice versa. 
We will also need to impose the condition that our solution 
satisfies Neumann $(+)$ or Dirichlet $(-)$
boundary conditions on the  UV  brane at $z = z_0$.  These conditions are
\beq
\begin{tabular}{l|cc}
   &   $+$ b.c.  at $z_0$      &   $ -$ b.c. at $z_0$  \\ \hline
$A^A_m$   &   $  G_{-,\beta}(z_0,z_R)  = 0  $ &  $  G_{+,\beta}(z_0,z_R)  =
  0 $\\ 
 $A^A_5$   &  $  G_{+,\beta}(z_0,z_R) = 0    $  &
                                           $     G_{-,\beta}(z_0,z_R)
                                                = 0 $ \\
\end{tabular}
\eeq{UVbcs}
That is, the first index of $G$ should be appropriately raised or
lowered to apply the Neumann condition.

Then the Green's functions of spin $\half$ fields are given by the following
formula:  For the Green's functions of fields $A^A_m, A^B_n$ obeying
$\beta,\gamma$ boundary conditions on the IR brane
\beqa
\VEV{ A^A_m(z)  A^B_n(z')}& = &\eta_{mn}\  k p z_R z z' \biggl[  {\bf A}^{AB}
G_{+,-\beta}(z,z_R) G_{+,-\gamma}(z',z_R) \CR
  & & \hskip 0.4in  -  \delta^{AB} \cases{ \widetilde G_{+,-\beta}(z,z_R)
    G_{+,-\gamma}(z',z_R) & $ z < z'$ \cr G_{+,-\beta}(z,z_R)
    \widetilde G_{+,-\gamma}(z',z_R) & $ z> z'$ \cr } \biggr] \ ,
\eeqa{Greensconstruct}
where the $\widetilde G$ are defined by
\beq 
   \matrix{    \widetilde G_{++}(z,z_R) = +  G_{+-}(z,z_R)  & 
   \widetilde G_{+-}(z,z_R) = -  G_{++}(z,z_R)  \cr
 \widetilde G_{-+}(z,z_R) = +G_{--}(z,z_R)  & 
   \widetilde G_{--}(z,z_R) = -G_{-+}(z,z_R)  \cr  }
\eeq{tildeGs}
The term in the second line of \leqn{Greensconstruct} 
satisfies the discontinuity of the Green's
function at $z = z'$.  It is present only in the diagonal correlation
function.
The Green's functions of $A^A_5$ fields are constructed similarly,
with $ G_{+,-\beta} \to  G_{-,-\beta}$. 

The choice of starting from definite $+$ or $-$ boundary conditions on
the IR brane comes from our convention of choosing the UV gauge,  in which
the Wilson line $U_W$ is implemented as a boundary condition on the UV
brane.   There is an equivalent formalism for Green's functions in the
IR gauge,  in which the Wilson line is
moved to the IR brane and implemented there as an IR boundary
condition.  In that case, we would choose definite $+$ or $-$ boundary
conditions on the UV brane.  The solution for the Green's function in
this case is completely analogous, starting from the formula
\beqa
\VEV{ A^A_m(z)  A^B_n(z')}_{IR}& = &- \eta_{mn}\  k p z_0 z z' \biggl[  {\bf A}^{AB}
G_{+,-\beta}(z,z_0) G_{+,-\gamma}(z',z_0) \CR
  & & \hskip 0.2in  -  \delta^{AB} \cases{ G_{+,-\beta}(z,z_0)
  \widetilde  G_{+,-\gamma}(z',z_0) & $ z < z'$ \cr \widetilde G_{+,-\beta}(z,z_0)
  G_{+,-\gamma}(z',z_0) & $ z> z'$ \cr } \biggr] \ ,
\eeqa{GreensconstructIR}
with $z_R \leftrightarrow z_0$ in \leqn{IRbcs}, \leqn{UVbcs}, and
\leqn{tildeGs}.

\subsection{Spin 1/2 fields}

Wavefunctions of spin 1/2  fields depend on the parameter 
 $c = m/k$, where $m$ is the 5D Dirac
mass.   We will decompose 4-component Dirac fields into 2-component
4D chirality eigenstates,
\beq
            \Psi =   \pmatrix{  \psi_L \cr \psi_R \cr } \ .
\eeqn
The Dirac equation couples these components.   The solution of the
Dirac equation contains $G^{(c)}_{\alpha,\beta}(z,z')$ with 
$\alpha = +$ for $\psi_L, \psi^\dagger_L$ and $\alpha = -$ for
$\psi_R, \psi^\dagger_R$. 

Canonical boundary conditions for the spin 1/2 fields have  $\psi_R =
0$ on the boundary ($+$ b.c.) or  $\psi_L =
0$ on the boundary ($-$ b.c.).
We will construct solutions with definite $+$ or $-$
boundary conditions on the IR brane at $z = z_R$.  These
solutions are 
\beq
\begin{tabular}{l|cc}
   &     $+$ b.c.  at $z_R$      &   $ -$ b.c. at $z_R$  \\ \hline
$\psi_L $   &    $ G_{+-}(z,z_R)  $   &    $G_{++}(z,z_R) $\\ 
 $\psi_R $   &   $ G_{--}(z,z_R) $    &  $  G_{-+}(z,z_R) $\\
\end{tabular}
\eeq{IRbcspsi}
We will also need to impose the condition that our solution 
satisfies $+$ or $-$
boundary conditions on the  UV  brane at $z = z_0$.  These conditions are
\beq
\begin{tabular}{l|cc}
   &     $+$ b.c.  at $z_0$      &   $ -$ b.c. at $z_0$  \\ \hline
 $\psi_{L,R} $  &   $  G_{-,\beta}(z_0,z_R)  = 0 $  &   $ G_{+,\beta}(z_0,z_R)  =
  0 $\\ 
\end{tabular}
\eeq{UVbcspsi}

Then the Green's functions of spin $\half$  fields are given by the following
formula:  For the Green's functions of fields
$\psi_L^A,\psi_L^{\dagger B}$ obeying
$\alpha, \beta$ boundary conditions on the IR brane
\beqa
\VEV{ \psi^A_L(z) \psi^{\dagger B}_L(z')}&
 = &(\sigma \cdot p) k^4  p z_R( z z' )^{5/2}\biggl[  {\bf A}^{AB}
G_{+,-\alpha}(z,z_R) G_{+,-\beta}(z',z_R) \CR
  & & \hskip 0.2in  -  \delta^{AB} \cases{ \widetilde G_{+,-\alpha}(z,z_R)
    G_{+,-\beta}(z',z_R) & $ z < z'$ \cr G_{+,-\alpha}(z,z_R)
    \widetilde G_{+,-\beta}(z',z_R) & $ z> z'$ \cr } \biggr]\ ,
\eeqa{Greensconstructpsi}
where the $\widetilde G$ are defined in \leqn{tildeGs}.
The term in the second line satisfies the discontinuity of the Green's
function at $z = z'$.  It is present only in the diagonal correlation
function.
The Green's functions $\VEV{ \psi^A_L(z) 
\psi^{\dagger B}_R(z')}$, 
$\VEV{ \psi^A_R(z) \psi^{\dagger B}_L(z')}$, and $\VEV{ \psi^A_R(z)
  \psi^{\dagger B}_R(z')}$  are constructed similarly,
with $ G_{+,-\alpha} \to  G_{-,-\alpha}$ for each $\psi_R$. 

\subsection{Solution for ${\bf A}^{AB}$}

To complete the solution for Green's functions, we need to solve
for the matrix ${\bf A}^{AB}$.   With the boundary conditions at $z =
z_R$ and $z = z'$ already imposed, we determine ${\bf A}^{AB}$ by
imposing the boundary condition at $z = z_0$.

 If a field $A^A_5$ obtains
an expectation value, the corresponding Wilson line element, a unitary
matrix $U$ defined by 
\leqn{UW},  is applied to the multiplet of Green's functions before
imposing this boundary condition.  We then find a linear
equation for the elements of ${\bf A}^{AB}$ that has the form
\beq
   U_{AC} \bigg[ A^{CB} G_{-\alpha,-\gamma}(z_0,z_R) -  \delta^{CB}
 \widetilde G_{-\alpha,-\gamma}(z_0,z_R) \bigg] = 0 
\eeqn
where $\alpha, \gamma  = \pm$ are the  boundary conditions of the 
$A$ field at $z= z_0$ and the $C$ field at $z = z_R$, respectively.
If fields of different $c$ are involved, the Green's functions are
evaluated at the value corresponding to the field $C$.  Let 
\beqa
      {\bf C}_{AC}  &=&   U_{AC}  G_{-\alpha,-\gamma}(z_0,z_R) \CR
      {\bf D}_{AC}  &=&   U_{AC}  \widetilde
      G_{-\alpha,-\gamma}(z_0,z_R) \ .
\eeqa{CDvals}
Then ${\bf A}^{CB}$ is the solution of the equation
\beq
    {\bf C}_{AC} {\bf A}^{CB} =  {\bf D}_{AB} \ .
\eeq{mastereq}
The matrix ${\bf C}_{AC}(p)$ defined here is the analytic continuation
  of the similar matrix defined in \cite{YPone} to Minkowski momenta
  $p$.
The zeros of $\det{\bf C}(p)$ give the mass spectrum associated with
the fields.

From its use in representing the Green's function, we see that the
matrix ${\bf A}$ must be Hermitian.  This is certainly not obvious
from \leqn{mastereq}, and actually it is a nice check that ${\bf A}$
has been computed correctly from this formula.   We sketch a proof of
the Hermitian nature of ${\bf A}$ in Appendix E.

\section{$SO(5)$ Generators}

In this Appendix, we provide our choice of basis for  the generators
of $SO(5)$.  We will choose representations in which the decomposition
\beq
         SU(2)_L \times SU(2)_R = SO(4) \subset  SO(5)  
\eeqn
is explicit.   We will identify $SU(2)_L$ with the weak interaction
$SU(2)$ gauge group and $SO(4)$ with the custodial symmetry group.
For this purpose, we write 
\beq    
    T^{aL} =  \half (\eps^{abc} T^{bc} + T^{a4}) \qquad   T^{aR} = \half 
  (\eps^{abc} T^{bc} -  T^{a4}) \ . 
\eeqn
with $a,b, c = 1,2,3$.  Then the $SO(5)$ generators are labelled
$T^{aL}$, $T^{aR}$, $T^{a5}$, and $T^{45}$.  It will be convenient to
rescale $T^{a5}$ and $T^{45}$ such that all generators
have a uniform normalization, so that   $\tr[(F_{MN}^A T^A)^2]= c\ (F_{MN}^A)^2 $.  

The {\bf 4} spinor representation decomposes under $SU(2)_L\times
SU(2)_R$
\beq
          {\bf 4 } \to   (2,1) \oplus (2,1)  \ .
\eeqn
The corresponding representation matrices are 
\beqa
         T^{aL} = \pmatrix{  \tau^a & 0 \cr 0 & 0\cr} &\qquad 
      T^{aR} = \pmatrix{  0 & 0 \cr 0 & \tau^a\cr} \CR
        T^{a5} = {1\over \sqrt{2}}\pmatrix{  0& \tau^a \cr \tau^a & 0\cr} &\qquad
      T^{45}= {1\over 2 \sqrt{2} }\pmatrix{  0 & -i \cr i& 0 \cr} 
\eeqa{Tinfour}
where $\tau^a = \sigma^a/2$.  In the {\bf 4} representation, we have $\tr  (T^A)^2 = \half$. 

The {\bf 5} fundamental representation decomposes under $SU(2)_L\times
SU(2)_R$
\beq
          {\bf 5} \to   (2,2) \oplus (1,1) 
\eeqn
The corresponding representation matrices are 
\beq
    T^{aL} = \pmatrix{  \tau^a \otimes {\bf 1} & 0 \cr  0 & 0\cr }
   \qquad\qquad\qquad
      T^{aR} = \pmatrix{  {\bf 1 }\otimes \tau^a & 0 \cr  0 & 0\cr} 
\eeqn
and 
\beq
\matrix{
        T^{15} = \half \pmatrix{  0 & \pmatrix{-1\cr 0\cr 0\cr
       1 \cr} \cr \pmatrix{-1 & 0 & 0&
            1\cr}  & 0 \cr } & 
        T^{25} =\half  \pmatrix{ 0&\pmatrix{i\cr 0\cr 0\cr
      i\cr} \cr \pmatrix{-i
       &0&0&-i \cr}
         & 0 \cr} \cr
      T^{35} = \half \pmatrix{  0 & \pmatrix{0\cr 1\cr
        1\cr 0\cr}
         \cr \pmatrix{ 0 &1& 
           1&0\cr} &0\cr }  &
   T^{45} =\half \pmatrix{  0 & \pmatrix{0\cr i\cr -i
        \cr 0\cr}
         \cr \pmatrix{0 &-i&
        i  &0\cr} & 0\cr } \cr }
  \eeq{Tinfive}
with the normalization $\tr  (T^A)^2 = 1$.   
In this basis, the elements of the ${\bf 5} $ multiplet are
\beq
\pmatrix{\xi_{++} \cr \xi_{-+} \cr \xi_{+-} \cr \xi_{--} \cr \xi_{00}
\cr} \ ,
\eeq{firstfivepresent}
with the subscripts indicating the $T^3_L$ and $T^3_R$ quantum numbers
$+\half$, $-\half$, or 0.  We will also write this multiplet as 
\beq 
 \pmatrix{ \pmatrix{\xi_{++} \cr \xi_{-+} \cr } \pmatrix{\xi_{+-} \cr
     \xi_{--} \cr }
 \cr \xi_{00}\cr } \ . 
\eeq{fivepresent}

We will find it useful to have explicit representations of 
\beq
     U =  \exp \left( -\sqrt{2} i\theta T^{45} \right) \ ,
\eeq{myfavoriteU}
in the ${\bf 4}$ and ${\bf 5}$   representations.  In the ${\bf 4}$, 
\beq
     U_{(4)}  =   \pmatrix{  c_2   & -s_2 \cr s_2 & c_2  } \ , 
\eeq{Uinfour}
where $s_2 = \sin\theta/2$, $c_2 = \cos\theta/2$.
   In the ${\bf 5}$,  $U$
mixes three rows of the 5-vector.   The
$3\times 3$ mixing matrix acting on $(\xi_{+-}, \xi_{-+}, \xi_{00})$
(the third, second, and fifth entries, respectively, of
\leqn{firstfivepresent}) is
\beq
    U_{(5)} =  \pmatrix{   (1+c)/2  &   (1-c)/2     &  -s/\sqrt{2} \cr
    (1-c)/2  &   (1+c)/2     &   s/\sqrt{2} \cr  s/\sqrt{2} &
   -s/\sqrt{2} & c \cr } \ ,
\eeq{Uinfive} 
where $s = \sin \theta$, $c = \cos\theta$.  

Finally, we consider the adjoint (${\bf 15}$) representation.   The
elements of $T^{A}$ in the adjoint representation are computed as the 
commutators of the $T^A$ matrices above.  In particular, it is
straightforward to show that 
\beq
  T^{45}_{\bf Adj}  \pmatrix{ T^{aL} \cr T^{aR} \cr T^{a5} \cr } = \half
\pmatrix{ 0 & & -i \cr  & 0 & i \cr
                     i   & -i & 0 \cr }
 \pmatrix{ T^{aL}\cr T^{aR} \cr T^{a5} \cr }
\eeq{Tinfifteen}
The corresponding mixing matrix is again the $3\times 3$ matrix 
\leqn{Uinfive}.

\section{Formalism for boundary kinetic terms}

In this appendix, we describe how the boundary
 kinetic terms for gauge fields and fermion fields modify the 
Green's functions for these fields. Our discussion generalizes the
presentation of Green's functions in Appendix A.  

\subsection{Boundary kinetic term for gauge fields}

For the description of gauge fields, we begin with the gauge-invariant 
bulk action in RS,
\beq 
S_{bulk} = \int d^4 x dz\ \biggl(\sqrt{g}  \biggl[ - {1\over 4} g^{MP} g^{NQ} F^a_{MN}
F^a_{PQ} \biggr] -   {\cal J}^M A_M  \biggr)   \ .
\eeq{gaugeaction}
The quantization of this action is described in Appendix B of \cite{YPone}.
Now  add a UV localized boundary kinetic term,
\beq 
S_{UV} = \int d^4 x dz \ \biggl(\sqrt{g} \biggl[ -  {1 \over 4} a z_0
\delta(z-z_0)    g^{mp} g^{nq} F^a_{mn} F^a_{pq} \biggr] \biggr) \  .
\eeq{bgaugeaction}
Note that we parametrize the coefficient of the boundary term in units 
of $z_0=1/k$.

In our formalism, the Higgs field is a background gauge field, so we
will quantize in the Feynman-Randall-Schwartz 
 background field gauge~\cite{RandallSchwartz}.
Expand
\beq    
   A_M^a \to  A_M^a(z)  + \A_M^a  \ , 
\eeqn
where, on the right,  $A_M^a$ is a fixed background field, 
\beq
    A_M^a(z) = (0,0,0,0, A_5^a(z) )  
\eeqn
and $\A_M^a$ is a fluctuating field.  Let $A_M = A_M^a t^a$ and 
$F_{MN} =
F^a_{MN} t^a$, where $t^a$ are the generators of 
the gauge group. Let  $D_M$ be the covariant
derivative containing the background field only.   Then the linearized
form for the field strength is
   ${\cal F}_{MN}  =   D_M \A_N -   D_N \A_M $.  In the backgrounds we
   consider in this paper, $F_{MN} = 
D_M A_N - D_N A_M = 0$ and $[D_M,D_N] = 0$. 
Inserting the metric \leqn{metric}, the action becomes
\beqa
   S_{bulk}+S_{UV} &=&  \int d^4 x dz \Biggl\{
  {1\over kz} \ \bigg[ - {1\over 4}
 \Big( (1+ a z_0 \delta(z-z_0))(\D_m\A_n - D_n\A_m)^2 \CR
& & \hskip 0.4in  -\half  (D_m\A_5 -  D_5 \A_m)^2\bigg]   -
\J^m\A_m + \J_5 \A_5 \Biggr\} \ . 
\eeqa{spinonebaction}

In the 5D bulk, following \cite{RandallSchwartz}, we introduce the 
gauge-fixing term  
\beq
    S_{GF} =   \int d^4 x dz  {1\over kz} \biggl[ -\half
\biggl( D^m \A_m - kz D_5 {1\over kz}
     \A_5\biggr)^2 \biggr] \ ,
\eeq{RSGF}
where we set the gauge parameter $\xi = 1$ for simplicity.
On the UV boundary, the gauge fixing term must be changed in accord
with the addition of the surface term.   The presence of the delta
function in 
\leqn{spinonebaction} requires some regularization.
  One possible way to do this, which we will follow
here, is to expand the boundary to an interval $[z_0, z_0+\eps]$ in
which the coefficient of the first term in \leqn{spinonebaction} is 
$(1 + a z_0/\eps)$.   A compatible gauge-fixing term on this interval is 
\beq
    S_{GF}^{UV} =   \int d^4 x\int^{z_0+\eps}_{z_0} 
 dz  {1\over kz} \biggl[ - {(1 + az_0/\eps)\over 2}
\biggl( D^m \A_m -   {1\over (1 + az_0/\eps)} \,kz D_5 {1\over kz}
     \A_5\biggr)^2 \biggr] \ .
\eeq{RSGFb}
After some integrations by parts, the action in the boundary region
comes into the form
\beqa
   S_{bulk}+S_{UV} + S_{GF}^{UV} &=&  \int d^4 x \int^{z_0+\eps}_{z_0} dz \ 
 \Bigg\{ {1\over kz} \bigg[ \half \A_m \eta^{mn}
 \Bigl( (1+ a z_0 /\eps) D^2 - kz D_5 {1\over kz} D_5 \Bigr) \A_n \CR
& & \hskip-0.3in  -\half  \A_5 \bigl( D^2  
- D_5{ kz \over (1+az_0/\eps)} D_5 {1\over  kz} \bigr)  \A_5 \bigg]   
- \J^m\A_m + \J_5 \A_5 \Biggr\} \ . 
\eeqa{gbfaction}
and the action in the bulk has the same form with  the $az_0/\eps$
terms removed.  Here and in the following, raised and lowered indices 
are contracted
with the Lorentz metric $\eta^{mn}$ and $D^2 = D^m D_m$.  It is
convenient to define ${\bf D}_5 = kz D_5 (1/kz)$.

The surface terms from integration by parts should not be ignored.
They are
\beqa
  S_{surface} &=& \int d^4x \half \Biggl\{  {1\over kz} \biggl[   \A^m D_5 \A_m
 +  2 D^n \A_n \A_5  -  {1\over  (1 + az_0/\eps)} \A_5  {\bf D}_5 \A_5
\biggr] \biggr|^{\eps -} _0 \CR
&&  \hskip 0.7in   + \biggl[   \A^m D_5 \A_m
 +  2 D^n \A_n \A_5  -  \A_5 {\bf D}_5 \A_5
\biggr] \biggr|^R_{\eps+} \Biggr\}\ ,
\eeqa{boundaries} 
with $0$, $\eps - $,  $\eps+$, $R$ denoting the boundaries at $z_0$,
$(z_0+ \eps )$ in the boundary region, $(z_0 + \eps)$ in the bulk
region, and $z_R$, respectively.
Requiring these expressions to vanish, we learn that $\A_m$, $\A_5$
obey the boundary conditions:
\beq
\begin{tabular}{lccc} 
at $z_0$:  &   $ D_5 \A_m |_0 =0\ , \ \A_5|_0  = 0 $  & or &  $ \A_m|_0
                                                            = 0 \ , \  
                                                {\bf  D}_5  \A_5|_0 =
                                                  0 $  \\ 
at $z_R$:  &   $ D_5 \A_m |_R =0\ , \ \A_5  |_R= 0 $  & or & $   \A_m
                                                            |_R= 0\ , \
                  {\bf  D}_5  \A_5|_0 =  0 $  \\ 
at $(z_0+\eps)$: & $  \A_{m}  |_{\eps-} = \A_m |_{\eps+} $  & and &  
$  D_5\A_m  |_{\eps-}  =D_5 \A_m  |_{\eps+} $ \\
 & $  \A_5  |_{\eps-}  = \A_5  |_{\eps+} $ & and &  
$  (1+az_0/\eps)^{-1} {\bf D}_5 \A_5 |_{\eps-} 
 =    {\bf D}_5 \A_5  |_{\eps+} $ \  . \\
\end{tabular}
\eeq{bcforAa}
The first two lines are the now-familiar $+$ and $-$ boundary
conditions for the spin 1 field. 

In the boundary region, $\A_m$ and $A_5$ obey the equations
\beqa
  \Bigl[ (1 + az_0/\eps) p^2 + kz D_5 {1\over kz} D_5 \Bigr]  \A_m(z,p)& =& 0 \CR
 \Bigl[ p^2 + D_5{ kz\over (1+az_0/\eps)}  D_5 {1\over kz} \Bigr] \A_5(z,p)&
 =& 0 
\eeqa{twospinoneeq} 
Since the region is very narrow, both equations can be approximated by 
\beq
\Bigl[  {a z_0\over \eps} p^2 +   \del_5^2 \Bigr]  \A(z,p)  = 0  \ . 
\eeq{onespinoneeq} 
Then a solution satisfying $ \A/D_5 \A = 0 $ at $z_0$ has 
\beq
         \Bigl( \A \ / \ D_5 \A \Bigr) |_{\eps-} =  \eps 
\eeq{minusbcweps} 
and a solution satisfying   $D_5\A / \A$ = 0 at $z_0$ has 
\beq
         \Bigl(  D_5\A \ / \A \Bigr) |_{\eps-} =  - a z_0 p^2  \ . 
\eeq{plusbcweps} 
Then the boundary conditions at $\eps+$ for the solutions in bulk are 
(with $\epsilon \rightarrow 0$)
\beq
\begin{tabular}{l|cc}
 &     $+$ b.c. at $z_0$     &   $-$ b.c. at $z_0$  \\ \hline
 $\A_m$ &    $ D_5 \A_m  / \A_m = - az_0 p^2 $&   $ \A_m /D_5 \A_m = 0 $\\
 $\A_5$ &    $ {\bf D}_5\A_5  / \A_5 = 0 $&   $ \A_5 /{\bf D}_5\A_5= a
  z_0 $\\
\end{tabular}
\eeq{finalbcsforA}
Using the property of the $G$ functions
\beq
    \del_z( z G_{+,\beta})=  pz\, G_{-,\beta}\ , \quad   
\del_z G_{-,\beta} =  -p\,  G_{+,\beta}\ , 
\eeq{derivGs}
these boundary conditions are implemented by imposing 
\beq
\begin{tabular}{l|cc}
   &    $+$ b.c.        &   $-$ b.c.   \\ \hline
$\A^a_m$   &   $  G_{-,\beta}(z_0,z_R) + az_0 p \,
             G_{+,\beta}(z_0,z_R) = 0$ & 
 $  G_{+,\beta}(z_0,z_R)  =
  0 $\\ 
 $\A^a_5$   &  $  G_{+,\beta}(z_0,z_R) = 0    $  &
                                           $    G_{-,\beta}(z_0,z_R) + az_0 p \,
             G_{+,\beta}(z_0,z_R) = 0$ \\
\end{tabular}
\eeq{UVbcswaeps}
instead of \leqn{UVbcs}.  $+$ boundary conditions for $\A_m$
require $-$ boundary conditions for $\A_5$, and vice versa.   Since the
boundary conditions on the Green's functions are the same for these
cases, the Laplacians for compatible $\A_m$ and $\A_5$ will have the
same spectrum, just as in the case of $a= 0$. The ghosts $c$
have the same spectrum as $\A_m$.   It is necessary for the $\A_5$
fields to have the same spectrum as the ghosts so that the
determinant of the $\A_5$ Laplacian 
 can cancel the determinant of the
ghost Laplacian.   This  allows the complete functional integral over $\A$
to be gauge-independent.

It is illuminating to compute the Green's function for $\A_m$ in the
case of $++$ boundary conditions.  Before imposing the UV boundary
condition, the Green's function takes the form in
\leqn{Greensconstruct}.  For $z < z'$, 
\beqa
\VEV{ \A_m(z)  \A_n(z')}& = &\eta_{mn}\,  k p z_R z z' \biggl[  {\bf A}
G_{+-}(z,z_R) G_{+-}(z',z_R) \CR
  & & \hskip 0.4in  + G_{++}(z,z_R)
    G_{+-}(z',z_R) \biggr] \ ,
\eeqa{Greensconstructwaeps}
Imposing the $+$ boundary condition on the UV brane with the
modification due to the boundary kinetic term, we find
\beq
   {\bf A} (G_{--} + a z_0 p \, G_{+-}) + (G_{-+} + a z_0 p \, G_{++})
 = 0 \  . 
\eeqn
This is easy to solve for ${\bf A}$.   Using \leqn{genWronsk}, the
Green's function for $z < z'$ can be rewritten as 
\beq 
\VEV{ \A_m(z)  \A_n(z')} = \eta_{mn}\, k z z' \biggl[  
  { G_{+-}(z,z_0) + a z_0 p \, G_{++}(z,z_0)\over G_{--} + a z_0 p \,
  G_{+-}} \biggr]  G_{+-}(z', z_R) \ . 
\eeq{Greensconstructwatwo}
Taking the $p\to 0 $ limit using \leqn{Gpvalshalf} 
\beq 
\VEV{ \A_m(z)  \A_n(z')} \to  \eta_{mn}\  {k\over p^2} {1\over (\log
  z_R/z_0 + a)} \ . 
\eeq{Greensaepssmall} 
This equation shows exactly that the 4D coupling of $\A_m$ is modified 
according to \leqn{newgRS}. 

To compute the Coleman-Weinberg potential, we need to redo this
analysis for Euclidean momenta.  For $p_E^2 = -p^2$, there are minus
sign changes in the formulae \leqn{twospinoneeq} and in
\leqn{derivGs}.   At the end of the analysis, we find, $+$ and $-$
boundary conditions for the Euclidean Green's functions are
implemented by 
\beq
\begin{tabular}{lcc}
   &    $+$ b.c.     at $z_0$       &   $-$ b.c.   at $z_0$     \\ \hline
$\A^a_m$   &   $  G_{E-,\beta}(z_0,z_R) + az_0 p_E \,
             G_{E+,\beta}(z_0,z_R) = 0$ & 
 $  G_{E+,\beta}(z_0,z_R)  =
  0 $\\ 
 $\A^a_5$   &  $  G_{E+,\beta}(z_0,z_R) = 0    $  &
                                           $    G_{E-,\beta}(z_0,z_R) + az_0 p_E \,
             G_{E+,\beta}(z_0,z_R) = 0$ \ . \\
\end{tabular}
\eeq{UVbcswaepsEucl}
This result makes it straightforward to derive the expressions  for the
$W$ and $Z$ boson Coleman-Weinberg potentials in 
\leqn{VW} and
\leqn{VZ}. 

\subsection{Boundary kinetic term for fermion fields}

For the description of fermion fields, we begin with the gauge-invariant 
bulk action in RS,
\beq 
S_{bulk} = \int d^4 x dz\  \biggl(\sqrt{g} \ \bar \Psi [ i
e^M_A \gamma^A D_M - m] \Psi -
\bar{\cal K} \Psi - \bar\Psi {\cal K} \biggr)   \ .
\eeq{fermionaction}
The quantization of this action is described in Appendix A of
\cite{YPone}.  After specializing to the metric \leqn{metric} and
dividing $\Psi$ into its 4D chiral components, this action becomes 
\beqa
S_{bulk} &=& \int d^4 x dz\ \biggl(  {1\over (kz)^4}  \biggl[
  \psi_L^\dagger
  i  \bar\sigma^m D_m \psi_L +   \psi_R^\dagger
  i \sigma^m D_m \psi_R +  \CR
 & &\hskip 0.4in  + \psi_L^\dagger {\bf D} \psi_R - \psi_R^\dagger
 \bar{\bf D} \psi_L \biggr] - 
\bar{\cal K} \Psi - \bar\Psi {\cal K} \biggr)   \ ,
\eeqa{fermionactiontwo}
where 
\beq
  {\bf D} =  D_5 - {2+c\over z} \qquad   \bar{\bf D} =  D_5 -
  {2-c\over z} \ . 
\eeq{Ddefforfermi}

The fermion fields in \leqn{fermionactiontwo} obey equivalent Laplace
equations 
\beqa
     (   p^2 + {\bf D} \bar{\bf D} ) \psi_L(z,p)& = & 0  \CR
    (   p^2 + \bar {\bf D} {\bf D} ) \psi_R(z,p)& = & 0  
\eeqa{fermionLap}
and are linked by the equations of motion
\beqa
     \sigma\cdot p \psi_R &=& \bar {\bf D} \psi_L\CR
     \bar \sigma\cdot p \psi_L &=&-  {\bf D} \psi_R  \ .
\eeqa{LRrelation}
There are two ways to add a boundary kinetic term to
\leqn{fermionactiontwo}.   We can add either a kinetic term for
$\psi_L$ or a kinetic term for $\psi_R$.   (Adding both terms leads to
unnecessary complexity.)  We will describe the first alternative in
detail and then quote the results for the second.

Then, add to \leqn{fermionactiontwo} the UV boundary term
\beq 
S_{UV} = \int d^4 x dz \ {1\over (kz)^4}  a z_0 \delta(z - z_0)
\psi_L^\dagger
 i \bar\sigma^m D_m \psi_L  \ . 
\eeq{bfermionaction}
The delta function requires regularization, and again we will
regularize it by spreading its influence over a small interval of size
$\eps$ at the UV brane.   The equations of motion in the boundary
region become
\beqa
  \sigma\cdot p \psi_R &=& \bar {\bf D} \psi_L\CR
    (1 +   a z_0/\eps)   \bar \sigma\cdot p \psi_L &=&-  {\bf D} \psi_R  \ .
\eeqa{bLRrelation}
 In the narrow boundary region, the Laplace equations for $\psi_L$ and
 $\psi_R$ are both well approximated by           
\beq
\Bigl[  {a z_0\over \eps} p^2 +   \del_5^2 \Bigr]  \psi_{L,R}  = 0  \ . 
\eeq{onespinhalfeq} 
Deriving the equations of motion for $\psi_L^\dagger$,
$\psi_R^\dagger$ requires an integration by parts.   The boundary term
in $z$ is 
\beq 
  \int  d^4x {1\over (kz)^4} [ \psi_L^\dagger \psi_R - \psi_R^\dagger
\psi_L ] 
\eeqn
and is not altered by the addition of \leqn{bfermionaction}.   So the
boundary conditions on $\psi_L$, $\psi_R$ are the standard ones,
\beq
\begin{tabular}{lccc} 
at $z_0$:  &   $ \psi_R =0 $  & or &  $ \psi_L = 0 $ \\ 
 at $z_R$: &   $ \psi_R =0 $  & or &  $ \psi_L = 0 $ \\ 
 at $(z_0+\eps)$: & $  \psi_R |_{\eps-} = \psi_R |_{\eps+}$  & and &  
$  \psi_L |_{\eps-} = \psi_L |_{\eps+}$ \ .
\end{tabular}
\eeq{bcforpsis}

Consider first the $+$ boundary condition $\psi_R= 0$ at $z = z_0$.
Then, in the boundary region,
\beqa
    \psi_R &=&  C \   \sin \biggl[\bigl( {a z_0\over \eps} \bigr)^{1/2} p(z-z_0)\biggr]
    \CR
    \psi_L &=&  - {\eps\over a z_0} {\sigma\cdot p\over p}  \cdot C \ \Bigl(
 {a z_0\over \eps} \Bigr)^{1/2} \cos \biggl[\bigl( {a z_0\over \eps} 
\bigr)^{1/2} p(z-z_0)\biggr]
 \eeqan
At $z = (z_0+ \eps)_-$, 
\beq
     \psi_R/\psi_L  =    -  a p z_0   {\bar\sigma\cdot p\over p} \ . 
\eeqn
This condition is very similar to that in the $+$ case for $A_m$
above.   The boundary condition is imposed on the Green's functions by requiring
\beq
    G_{-,\beta}(z_0,z_R) + az_0 p \,   G_{+,\beta}(z_0,z_R) = 0 \ .
\eeqn
In the case of $-$ boundary conditions,  $\psi_L = 0$ at $z= z_0$, 
\beq
     \psi_L/\psi_R  =    {\cal O} (\eps)
\eeqn
at $z = (z_0+ \eps)_-$, and so the boundary condition is unchanged.
In all, the boundary conditions for fermion fields with the boundary
kinetic term \leqn{bfermionaction} are 
\beq
\begin{tabular}{l|cc}
   &    $+$ b.c.    at $z_0$      &   $-$ b.c.    at $z_0$ \\ \hline
 $\psi_{L,R} $  &   $  G_{-,\beta}(z_0,z_R) + az_0 p \,
                  G_{+,\beta}(z_0,z_R) = 0 $  & 
  $ G_{+,\beta}(z_0,z_R)  =
  0 $\\ 
\end{tabular}
\eeq{UVbcspsimodone}
instead of \leqn{UVbcspsi}.

Similarly, we can modify \leqn{fermionactiontwo} by adding the UV boundary term
\beq 
S_{UV} = \int d^4 x dz \ {1\over (kz)^4}  a z_0 \delta(z - z_0)
\psi_R^\dagger
 i \sigma^m D_m \psi_R  \ . 
\eeq{bfermionactiontwo}
In this case, the UV boundary conditions become 
\beq
\begin{tabular}{l|cc}
   &    $+$ b.c.   at $z_0$      &   $-$ b.c.    at $z_0$ \\ \hline
 $\psi_{L,R} $  &   $  G_{-,\beta}(z_0,z_R) = 0 $  & 
  $ G_{+,\beta}(z_0,z_R)  -  az_0 p \,
                  G_{-,\beta}(z_0,z_R) = 0 $ \ . 
\end{tabular}
\eeq{UVbcspsimodtwo}

To illustrate the effect of the UV boundary kinetic term, we can work
out the Green's function $\VEV{\psi_L(z,p) \psi_L^\dagger(z',p) } $  for
the case of $++$  boundary conditions and the modification
\leqn{UVbcspsimodone}.
   This is the Green's function
that contains the  zero mode for a 4D left-handed chiral fermion.
 Before imposing the UV boundary
condition, the Green's function takes the form in
\leqn{Greensconstructpsi}.  For $z < z'$,  
\beqa
\VEV{ \psi_L(z) \psi^{\dagger}_L(z')}&
 = &(\sigma \cdot p) k^4  p z_R( z z' )^{5/2}\biggl[  {\bf A}
G_{+,-}(z,z_R) G_{+,-}(z',z_R) \CR
  & & \hskip 0.2in  + G_{+,+}(z,z_R)
    G_{+,-}(z',z_R) \biggr]\ .
\eeqan
Imposing the $+$ boundary condition on the UV brane, including 
the effect of the boundary term, we find
\beq
   {\bf A} (G_{--} + a z_0 p G_{+-}) + (G_{-+} + a z_0 p G_{++}) = 0 \
   . 
\eeq{findAforpsippwa}
This is easy to solve for ${\bf A}$.   Using \leqn{genWronsk}, the
Green's function for $z < z'$ can be rewritten as 
\beq 
\VEV{ \psi_L(z)  \psi_L^\dagger(z')} = \ (\sigma \cdot p) k^4
 (z z')^{5/2} \biggl[  
  { G_{+-}(z,z_0) + a z_0 p G_{++}(z,z_0)\over ( G_{--} + a z_0 p
  G_{+-})} \biggr]  G_{+-}(z', z_R) \ . 
\eeq{Greensconstructwathree}
Taking the $p\to 0 $ limit using \leqn{Gpvals} 
\beq 
\VEV{ \psi_L(z)  \psi_L^\dagger(z')}  \to   {\sigma\cdot p\over p^2}
f_L^2(a) \   k^4 ( z z')^{2-c}\ .
\eeq{Greenpsiepssmall} 
Here $f_L^2(a)$ is the normalization factor for the zero mode, 
which is 
altered from its standard form by the inclusion of a term involving the
boundary factor $a$.   The new expression for the zero mode is 
\beq
f_L(a) \ z^{2-c} =\biggl[{z_R^{1-2c} - z_0^{1-2c} \over 1-2c }  +   a \,
  z_0^{1-2c}\biggr]^{-1/2}
 \  z^{2-c} \ . 
\eeq{newftwo}
The $a$ term is always a suppression for $a>0$.   This suppression is
small if the zero mode is dominantly in the IR  ($c < \half$), but it becomes
significant when the zero mode is dominantly in the UV ($c > \half$). 

The Green's function that contains the right-handed 4D
chiral fermion  is  $\VEV{\psi_R(z)\psi_R^\dagger(z')}$, for a fermion
  field with $--$ boundary conditions.  In a similar way, we can
  compute this Green's function and take the $p\to 0$ limit.  The
  result is 
\beq 
\VEV{ \psi_R(z)  \psi_R^\dagger(z')}  \to   {\bar \sigma\cdot p\over p^2}
f_R^2(a) \   k^4 ( z z')^{2+c}\ .
\eeq{GreenpsiepssmallR} 
Here $f_R^2(a)$ is the normalization factor for the right-handed 
zero mode, which is also 
altered from its standard form.    The new expression for
 the zero mode is 
\beq
f_R(a) \ z^{2+c} =\biggl[{z_R^{1+2c} - z_0^{1+2c} \over 1+2c }  +   a \,
  z_0^{1+2c}\biggr]^{-1/2}
 \  z^{2+c} \ . 
\eeq{newftwoR}
Again, the $a$ term suppresses the normalization of the zero mode.
Again, this suppression is large only when the zero mode is dominantly
in the UV, which occurs for $c < -\half$ in this case.

To compute the Coleman-Weinberg potential, we need to redo this
analysis for Euclidean momenta.  For $p_E^2 = -p^2$, there are minus
sign changes in the formulae \leqn{onespinhalfeq} and in the formulae
for derivatives of the $G$ functions.
 At the end of the analysis, we find, the $+$ and $-$
boundary conditions for the Euclidean Green's functions, with boundary
kinetic terms for $\psi_L$,  are
implemented by 
\beq
\begin{tabular}{lcc}
          &    $+$ b.c.   at $z_0$     &   $-$ b.c.   at $z_0$ \\ \hline
$\psi_{L,R}$:
&   $  G_{E-,\beta}(z_0,z_R) + az_0 p_E \,
                               G_{E+,\beta}(z_0,z_R) = 0$ & 
                       $  G_{E+,\beta}(z_0,z_R)  =  0 $\\ 
\end{tabular}
\eeq{UVbcswaepsEpsi}
This result makes it straightforward to derive the expression for the
top quark Coleman-Weinberg potential in \leqn{Vt}.

\subsection{Moments of fermion zero modes}

To compute some corrections we consider in this paper, it is necessary
to evaluate moments of  $z^2/z_R^2$ in fermion zero modes.  For a
single left-handed fermion zero mode, and for $a = 0$, this is straightforward to
evaluate using the $z$ wavefunction of the zero mode
\beq
  \VEV{A(z)} =   \int{ dz\over (kz)^4} \ | \psi(z)|^2\  A(z) =   f^2 _L(0)  \int_{z_0}^{z_R}  dz \
    z^{-2c} \ A(z) \ . 
\eeq{zeromodemeasure}
with $f^2_L(a)$ given by \leqn{newftwo}.   Then
\beq
   \VEV{\bigl( {z\over z_R}\bigr)^\beta} = {(1-2c) \over (1 + \beta - 2c) } {(z_R^{1 +
       \beta - 2c} - z_0^{1+\beta -2c})\over z_R^\beta (z_R^{1-2c} - z_0^{1-2c})}
   \ .
\eeq{zeromodemom}
For $a > 0$, part of the zero mode is concentrated at $z = z_0$.
Then moments would be evaluated with the measure
\beq
    \int { dz\over (kz)^4} | \psi(z)|^2  =   f^2 _L(a)  \int_{z_0}^{z_R}  dz \
   \bigl[  z^{-2c}  +  a z_0^{1-2c} \delta(z - z_0) \bigr] \ ,
\eeq{azeromodemeasure}
adding an extra term to \leqn{zeromodemom},
\beq
   \VEV{\bigl( {z\over z_R}\bigr)^\beta} =
 {(z_R^{1 +  \beta - 2c} - z_0^{1+\beta -2c})/(1+ \beta - 2c) 
+ a \, z_0^{1 + \beta - 2c} \over z_R^\beta [ ( z_R^{1-2c} - z_0^{1-2c})/(1-2c)  +   a  \,
     z_0^{1-2c} ] }\ .
\eeq{zeromodemomwa}
Note that these moments go to zero exponentially when the zero modes
are UV-localized, that is, when $c > 1/2$. 

\begin{figure}
\centering
\includegraphics[width=5in]{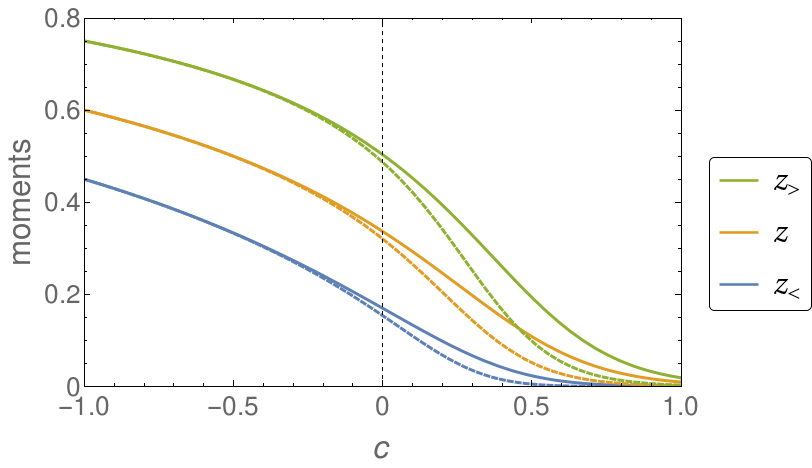}
\caption{Values of $\VEV{z^2/z_R^2}$ for left-handed fermion zero mode
wavefunctions, plotted as functions of $c$, for (bottom to top)
$z_<$, $z$, and $z_>$.    Solid lines: $a$ = 0; Dashed lines: $a$ = 5. }
\label{fig:zvals}
\end{figure}

In the evaluation of matrix elements that involve Green's functions,
we encounter these moments for pairs of coordinates $(z, z')$. For example, 
\beq
\VEV{z_<^2\over z_R^2} = \int {dz \over (kz)^4} \ | \psi_1(z)|^2 \int {dz' \over (kz')^4} \ | \psi_2(z')|^2 \left( z_<^2 \over z_R^2 \right)
\eeqn
where $z_<$/$z_>$ the smaller/larger of $z$ and $z'$.
To get a feel for this, we quote the values of the expectation values
of these constrained at $c = 1/2$, $a = 0$, 
\beq
   \biggl( \VEV{z_<^2\over z_R^2}, \VEV{z^2\over z_R^2},
   \VEV{z_>^2\over z_R^2}
 \biggr) =   (  0.024, 0.109,  0.194)  \ ,
\eeq{somezvals}
for $z_0/z_R = 0.01$. These values decrease with $a$ and decrease
steeply with $c$.  So typically, for left-handed zero modes, 
terms  with $z_<$ will be
negligible while terms with $z_>$ might make a noticeable correction.
In Fig.~\ref{fig:zvals}, we plot values of $\VEV{z^2/z_R^2}$ as a
function of $c$ for $a = 0$ and $a = 5$. Note that the boundary term has effects only when $c \gsim 0$.

For right-handed zero modes, the situation can be different. The
formulae for the evaluation of $\VEV{z^\beta}$ are changed by the
substitution $c \to -c$.  Thus, if  $c> 0$, the zero modes are
strongly shifted to the IR, and so $\VEV{z^\beta}$ can take large
values. For right-handed zero modes with $c > 0$ and $z_0/z_R \ll 0.1$, it is a good approximation to ignore the factors with $z_0$.   Then
\beqa
  \VEV{ z^2/ z_R^2  }&=& {1 + 2c \over 3 + 2c}\CR
 \VEV{ z^2 \log (z_R/z) / z_R^2} &=& {1+2c \over (3 +
   2c)^2}
\eeqa{zVEVs}
We will need these formulae in Section~7.

\section{Construction of the $W$, $Z$, and $t$ propagators}

Using the formalism of Appendices A and C, it is almost automatic to
construct the gauge boson and top quark propagators.  We present the essential
formulae here.

\subsection{$W$ propagator}

The 5D $W$ boson is a mixture of the three fields $A_m^{aL}, A_m^{aR},
A_m^{a5}$,
$a = 1,2$, with boundary conditions shown in \leqn{gaugebc}.   The 5D
propagator for these fields is given by \leqn{Greensconstruct}.  In
this equation, 
${\bf A}^{AB}$ is a $3\times 3$  matrix  given by 
\beq
     {\bf A} = {\bf C}^{-1} {\bf D} \ , 
\eeq{compAW}
with 
\beq 
{\bf C} = \pmatrix{  {(1+c)\over 2} G_{W--}
   &  {(1-c)\over 2} G_{W--} &
 -{s\over \sqrt{2}} G_{W-+}\cr
 {(1-c)\over 2} G_{+-} &   {(1+c)\over 2} G_{+-}  & {s\over \sqrt{2}}  G_{++} \cr
{s\over \sqrt{2}}  G_{+-} & -{s\over \sqrt{2}}  G_{+-} &   c \ G_{++} \cr } 
\eeq{CaWforW}
and
\beq 
{\bf D} = \pmatrix{ - {(1+c)\over 2}G_{W-+}
   & -  {(1-c)\over 2} G_{W-+} &
 -{s\over \sqrt{2}} G_{W--}\cr
 -{(1-c)\over 2} G_{++} & -  {(1+c)\over 2} G_{++}  & {s\over \sqrt{2}}  G_{+-} \cr
-{s\over \sqrt{2}}  G_{++} & {s\over \sqrt{2}}  G_{++} &   c \ G_{+-}
\cr } \ , 
\eeq{DaWforW}
where
\beqa
   G_{W--} & = & G_{--} + a_W p z_0 G_{+-} \CR
  G_{W-+} & = & G_{-+} + a_W p z_0 G_{++}  \ . 
\eeqa{GWdefs}
Below, we will also need a similar modification for the $U(1)$ gauge field,
\beqa
   G_{B--} & = & G_{--} + a_B p z_0 G_{+-} \CR
  G_{B-+} & = & G_{-+} + a_B p z_0 G_{++}  \ .
\eeqa{GBdefs}

The mass eigenvalues in this sector  and the contribution to the
Coleman-Weinberg potential are controlled by the determinant of ${\bf
  C}$, which has the form,
\beq
  \det {\bf C} = G_{+-} \biggl[   G_{++} G_{W--} -  {s^2\over 2 p^2 z_0 z_R} \biggr] \ . 
\eeq{Wdeterminant}

 For our discussion of precision electroweak constraints, we will need
 the expansion of ${\bf A}$ including terms of order  $s^0$ for the leading
 term in $p$ in each matrix element as $p\to 0$.  It will suffice
 to ignore terms of order $z_0^2/z_R^2$.   Then
\beq
{\bf A} = -{ 2p z_R\over s^2}\pmatrix{ 
1/2                   &   {s^2/ 8}  & {-s/ \sqrt{2} p z_R } \cr
{s^2/ 8}              &    s^2/4    &  0 \cr
{-s/ \sqrt{2} p z_R } &      0      &  (\log z_R/z_0 + a_W )\cr }\ .
\eeq{reduxofAW}
From the definition \leqn{Greensconstruct}, {\bf A} must be
symmetric.  This is not obvious from \leqn{compAW}, but it is true, 
and this is reflected in \leqn{reduxofAW}. The general proof of the Hermitian nature of {\bf A} is given in Appendix E.

Now we find 
\beqa
\VEV{A_m^{1L}(z) A_n^{1L}(z')}  &\to & \eta_{mn} kz_R^2 \, p^2 zz' \biggl[ -
{1\over s^2} G_{+-}(z,z_R) G_{+-}(z',z_R) \CR
& & \hskip 1.2in + {1\over p z_R} G_{++}(z_<, z_R)
G_{+-}(z_>,z_R) \biggr] \CR
 &\to &   -\eta_{mn} {kz_R^2\over s^2}\biggl[ 1 - {s^2\over 2}\Bigl(1 -
 {z_<^2\over z_R^2} 
\Bigr)\biggr] 
\eeqa{ALLWval}
in the limit $p\to 0$, where $z_<$, $z_>$ are the smaller and larger
of  $z$, $z'$.  Similarly,
\beqa
 \VEV{A_m^{1L}(z) A_n^{15}(z')}  &\to &\eta_{mn} kz_R^2 \, p^2 zz' \biggl[ 
{2s\over \sqrt{2} s^2 p z_R} G_{+-}(z,z_R) G_{++}(z',z_R)  \biggr]\CR
 &\to &   \eta_{mn} {kz_R^2\over s^2}\biggl[  {s\over \sqrt{2}}\Bigl(1-
 {z^{\prime 2}\over z_R^2} 
\Bigr)\biggr] \ .
\eeqa{ALRWval}

\subsection{$Z$ propagator}

The $Z$ propagator is derived in a similar way.  In the basis
$(A^{3L}, B, Z', A^{35})$ defined in \leqn{BZ} and \leqn{DM}, the
matrix $U_W$ has the form
\beq
    U_W =  \pmatrix{  
(1+c)/2  & s_\beta (1-c)/2 & c_\beta (1-c)/2  & -s/\sqrt{2} \cr
s_\beta (1-c)/2 & c_\beta^2 + s_\beta^2(1+c)/2 & 
-c_\beta s_\beta (1-c)/2 & s_\beta s/\sqrt{2} \cr 
c_\beta (1-c)/2 & -c_\beta s_\beta (1-c)/2     &
s_\beta^2 + c_\beta^2 (1+c)/2 & c_\beta s/\sqrt{2} \cr
s/\sqrt{2} & -s_\beta s/\sqrt{2} & -c_\beta s/\sqrt{2} & c }  \   .  
\eeq{UinZ}
Then the {\bf C} and {\bf D} matrices are 
\beq 
{\bf C} = \pmatrix{  {(1+c)\over 2} G_{W--}
   & s_\beta { (1-c)\over 2} G_{W--} & c_\beta {(1-c)\over 2}  G_{W--} & 
 -{s\over \sqrt{2}} G_{W-+}\cr
s_\beta  {(1-c)\over 2} G_{B--} &
 (c_\beta^2 + s_\beta^2  {(1+c)\over 2}) G_{B--}  & 
- c_\beta s_\beta {(1-c)\over 2} G_{B--}  &   s_\beta {s\over \sqrt{2}}
G_{B-+} \cr 
  c_\beta  { (1-c)\over 2} G_{+-}  &  -c_\beta s_\beta {(1-c )\over 2}
  G_{+-}
 & ( s_\beta^2 + c_\beta^2
{ ( 1+c )\over 2}) G_{+-}      &   c_\beta
{ s\over \sqrt{2}} G_{++}  \cr 
{s\over \sqrt{2}}  G_{+-} & -s_\beta {s\over \sqrt{2}}  G_{+-} & 
-c_\beta {s\over \sqrt{2}}  G_{+-} &  c \ G_{++} \cr } 
\eeq{CaZforZ}
and 
\beq 
{\bf D} = \pmatrix{ - {(1+c)\over 2} G_{W-+}
   & -s_\beta { (1-c)\over 2} G_{W-+} & -c_\beta {(1-c)\over 2}  G_{W-+} & 
 -{s\over \sqrt{2}} G_{W--}\cr
-s_\beta  {(1-c)\over 2} G_{B-+} &
 -(c_\beta^2 + s_\beta^2  {(1+c)\over 2} )G_{B-+}  & 
c_\beta s_\beta {(1-c)\over 2} G_{B-+}  &  s_\beta {s\over \sqrt{2}}
G_{B--} \cr 
- c_\beta  { (1-c)\over 2} G_{++}  &  c_\beta s_\beta {(1-c) \over 2}
  G_{++}
 &- ( s_\beta^2 + c_\beta^2
{ ( 1+c )\over 2}) G_{++}      &   c_\beta
{  s\over \sqrt{2}} G_{+-}  \cr 
-{s\over \sqrt{2}}  G_{++} &s_\beta {s\over \sqrt{2}}  G_{++} & 
c_\beta {s\over \sqrt{2}}  G_{++} &  c \ G_{+-} \cr } 
\eeq{DaZforZ}

The mass eigenvalues in this sector  and the contribution to the
Coleman-Weinberg potential are controlled by the determinant of ${\bf
  C}$.  This is given by 
\beq
  \det {\bf C} = G_{+-} \biggl[   G_{++}G_{W--} G_{B--}  -  {s^2\over
    2 p^2 z_0 z_R} 
(G_{B--} + s_\beta^2 G_{W--})\biggr] \ . 
\eeq{Zdeterminant}

\subsection{$t$ propagator}

The 5D $t$ quark is a mixture of the three fields $t_L$, $\chi_b$, $t_R$
in  \leqn{t5}.
 The 5D
propagator for these fields is given by \leqn{Greensconstructpsi}.  In
the basis $(t_L, \chi_b, t_R)$ used in \leqn{Uinfive}, the ${\bf C}$
and ${\bf D}$ matrices take the form 
\beq 
{\bf C} = \pmatrix{  {(1+c)\over 2} G_{t--}
   &  {(1-c)\over 2} G_{t--} &
 -{s\over \sqrt{2}} G_{t-+}\cr
 {(1-c)\over 2} G_{+-} &   {(1+c)\over 2} G_{+-}  & {s\over \sqrt{2}}  G_{++} \cr
{s\over \sqrt{2}}  G_{+-} & -{s\over \sqrt{2}}  G_{+-} &   c \ G_{++}
\cr } \ ,
\eeq{Catfort}
and 
\beq 
{\bf D} = \pmatrix{ - {(1+c)\over 2} G_{t-+}
   & - {(1-c)\over 2} G_{t-+} &
 -{s\over \sqrt{2}} G_{t--}\cr
 -{(1-c)\over 2} G_{++} &  - {(1+c)\over 2} G_{++}  & {s\over \sqrt{2}}  G_{+-} \cr
-{s\over \sqrt{2}}  G_{++} & {s\over \sqrt{2}}  G_{++} &   c \ G_{+-}
\cr } \ ,
\eeq{Datfort}
where
\beq
   G_{t-\pm}  =  G_{-\pm} + a_t p z_0 G_{+\pm} \  .
\eeq{Gtdef}
The mass eigenvalues in this sector  and the contribution to the
Coleman-Weinberg potential are controlled by the determinant of ${\bf
  C}$, which has the form,
\beq
  \det {\bf C} = G_{+-} \biggl[   G_{++}G_{t--}  -  {s^2\over 2 p^2 z_0 z_R} \biggr] \ . 
\eeq{tdeterminant}

\section{Relation of the UV and IR gauges}

In Section 4.5, we claimed that Green's functions in the UV and IR
gauges are related by the formula
\beq
  (U_W^\dagger)^{AC}    \VEV{A^C_m(z) A^B_n(z')}_{IR}  = \VEV{A^A_m(z)
       A^C_n(z')}_{UV}  (U_W^\dagger)^{CB}  \ . 
\eeq{UVIRagain}
In this section, we prove this relation from the representations for
the UV and IR gauge Green's functions given in Appendix A. The idea of
the proof is to use the identity \leqn{genWronsk} to relate
$G_{ab}(z,z_0)$ and $G_{ab}(z,z_R)$.   For
definiteness,  we
consider the representations of the Green's functions of the 4d
components of a spin~1 field given in 
\leqn{Greensconstruct} and 
\leqn{GreensconstructIR}.     It will be clear from the derivation
that the result for all other RS Green's functions can be carried out
with the same logic. 

We need to  be very explicit about the
boundary conditions on the various fields.  We  assign
 the field with gauge index $A$ the
boundary conditions  $a_{UV} = \pm$ and $a_{IR} = \pm$ in the UV and
IR, respectively.  

It suffices to consider the case $z > z'$.  In this case, the  Green's
function on the 
right-hand side of \leqn{UVIRagain} takes the form 
\beqa
   \VEV{A^A_m(z) A^B_n(z')}_{UV} &=&  \eta_{mn} kp z_R z z' 
                  \biggl[ G_{+,-a_{IR} }(z,z_R)  {\bf A}_{UV}^{AB}
   G_{+,-b_{IR}} (z',z_R) \CR & & \hskip 0.4in  
- G_{+,-a_{IR} }(z,z_R)  \delta^{AB}
    \tilde  G_{+,-b_{IR}}(z',z_R)\biggr] \ .
\eeqa{firstUV}
From \leqn{tildeGs}, the $\tilde G$ functions are given by 
\beq
         \tilde G_{c,-b_{IR}} = (-b_{IR}) G_{c,+b_{IR}} \ . 
\eeq{tildeGdefinite}
In the second line of \leqn{firstUV}, we can put $a_{IR} = b_{IR}$. 

In the UV gauge, the matrix ${\bf A}$ is computed as
\beq
    {\bf A}_{UV} =  {\bf C}^{-1}_{UV}  {\bf D}_{UV}  \ ,
\eeqn
The matrix elements of ${\bf C}$ and ${\bf D}$ are 
\beqa
 {\bf  C}^{AB}_{UV} &=&   U^{AB}  G_{-a_{UV}, -b_{IR}} (z_0, z_R)\CR
 {\bf  D} ^{AB}_{UV} &=&   U^{AB}\, (-b_{IR} )\, G_{-a_{UV}, +b_{IR}} (z_0, z_R) \ .
\eeqan
The formula \leqn{firstUV}  then factorizes as
\beqa
   \VEV{A^A_m(z) A^B_n(z')}_{UV} &=&  \eta_{mn} kp z_R z z' 
    \biggl[ G_{+,-a_{IR} }(z,z_R) {\bf C}^{-1}_{UV} {}^{AC} \CR
   & & \hskip - 0.4in\cdot  \biggl\{ {\bf D}_{UV}^{CB}   G_{+,-b_{IR}}
   (z',z_R) -   {\bf C}_{UV}^{CB} (-b_{IR})  G_{+,+b_{IR}} (z',z_R)
   \biggr\} \biggr]\CR
\eeqa{secondUV}
The term in braces is 
\beqa
 & & \biggl\{ G_{+, -b_{IR}}(z',z_R) (-b_{IR})
 G_{-c_{UV},+b_{IR}}(z_0,z_R)\CR 
& & \hskip 0.5in -\  G_{+, +b_{IR} } (z',z_R) (-b_{IR})
 G_{-c_{UV},-b_{IR}} (z_0,z_R) \biggr\}
U^{CB} \CR
& & =\biggl\{  G_{+, +}(z',z_R) G_{-c_{UV},-}(z_0,z_R)
- G_{+, - } (z',z_R)  G_{-c_{UV},+} (z_0,z_R) \biggr\}U^{CB}  \CR
& & =  {1\over p z_R} G_{+, -c_{UV}}(z,z_0) U^{CB} \ ,
\eeqa{thirdUV}
where, in the last line, we have used \leqn{genWronsk}. 

The UV gauge Green's function then reassembles into 
\beq
   \VEV{A^A_m(z) A^B_n(z')}_{UV} = \eta_{mn} k z z' 
        \big[   G_{+,-a_{IR} }(z,z_R)  {\bf C}^{-1}_{UV}{}^{AC}
           G_{+,-c_{UV}}(z',z_0) \big]  U^{CB} \ .
\eeq{finalUV}

The IR gauge Green's function can be rearranged in a similar way.
\beqa
   \VEV{A^A_m(z) A^B_n(z')}_{IR} &=&  -\eta_{mn} kp z_0 z z' 
 \biggl[ G_{+,-a_{UV} }(z,z_0)  {\bf A}_{IR}^{AB}
 G_{+,-b_{UV}} (z',z_0) \CR & & \hskip 0.4in  
- \tilde G_{+,-a_{UV} }(z,z_0)  \delta^{AB}
  G_{+,-b_{UV}}(z',z_0)\biggr] \ .
\eeqa{firstIR}
The $\tilde G$ functions are given by 
\beq
         \tilde G_{c,-a_{UV}} = (-a_{UV}) G_{c,+a_{UV}} \ . 
\eeq{tildeGdefiniteIR}

In the IR gauge, the matrix ${\bf A}$ is computed as
\beq
    {\bf A}_{IR} =   {\bf D}_{IR}  {\bf C}^{-1}_{IR}  \ ,
\eeqn
The matrix elements of ${\bf C}$ and ${\bf D}$ are 
\beqa
 {\bf  C}^{AB}_{IR} &=&   U^{AB}  G_{-b_{IR}, -a_{UV}}(z_R,z_0) \CR
 {\bf  D} ^{AB}_{IR} &=&   U^{AB}\,  (-a_{UV})\, G_{-b_{IR}, +a_{UV}} (z_R,z_0) \ .
\eeqan
The formula \leqn{firstIR}  then factorizes as
\beqa
   \VEV{A^A_m(z) A^B_n(z')}_{IR} &=&  -\eta_{mn} kp z_0 z z' \CR
  & & \cdot   \biggl[\biggl\{  G_{+,-a_{UV} }(z,z_0) {\bf D}^{AC}_{IR} -
  G_{+, +a_{UV}}(z,z_0)(-a_{UV})  {\bf C}_{IR}^{AC} \biggr\} \CR
  & &   \hskip 0.6in \cdot ({\bf C}^{-1}_{IR})^{CB}
  G_{+,-b_{UV}}(z',z_0)\biggr] \ . 
\eeqa{secondIR}

The term in braces is 
\beqa
 & & U^{AC} \biggl\{ G_{+, -a_{UV}}(z,z_0) (-a_{UV})
 G_{-c_{IR},+a_{UV}}(z_R,z_0)\CR 
& & \hskip 0.5in -\  G_{+, +a_{UV} } (z,z_0) (-a_{UV}) 
G_{-c_{IR},-a_{UV}} (z_R,z_0) \biggr\}
\CR
& & = U^{AC} \biggl\{  G_{+, +}(z,z_0) G_{-c_{IR} ,-}(z_R,z_0)
- G_{+, - } (z,z_0)  G_{-c_{IR},+} (z_R,z_0) \biggr\} \CR
& & = U^{AC}  {1\over p z_0} G_{+, -c_{IR}}(z,z_R) \ ,
\eeqa{thirdIR}
and again,  in the last line, we have used \leqn{genWronsk}. 

The IR gauge Green's function then reassembles into 
\beq
   \VEV{A^A_m(z) A^B_n(z')}_{IR} = - U^{AC} \eta_{mn} k z z' 
        \big[   G_{+,-c_{IR} }(z,z_R)  {\bf C}^{-1}_{IR}{}^{CB}
           G_{+,-b_{UV}}(z',z_0) \big] \ .
\eeq{finalIR}

To compare \leqn{finalUV} and \leqn{finalIR}, note that
\leqn{Greverse} implies, using the explicit formulae above,
\beq 
               C_{IR} = - C_{UV} \ . 
\eeqn
Then \leqn{finalUV} and \leqn{finalIR} have the same form, except
that, in the latter, 
the matrix $U$ is moved to the right.   This proves \leqn{UVIRagain}.

Notice that, in this calculation, the first  index  $+$  on the $G$
functions for the $A$ fields, the IR boundary condition of $A^A_m$,
and the UV boundary condition of $A^B_n$ play no role in the 
cancellation. The parallel calculation  for $z < z'$ depends on the 
IR boundary condition of $A^B_n$  and the UV boundary condition
of  $A_m^A$ and also goes through for any values of these. 
The $G$ functions in the cancellation are linked
by $U$ matrices and therefore have the same value of $c$.   Thus, the
same argument goes through for any Green's function of RS fields.

Using the same method, one can prove the identity 
\beq
      {\bf D }{\bf C}^\dagger -  {\bf C} {\bf D}^\dagger = 0 
\eeqn
 for both the UV and IR forms of these matrices.   After the use of
 the identity \leqn{genWronsk}, one finds that the $G$ functions
 combine into 
\beq 
         G_{a,a}(z_R,z_R) \quad \mbox{or}\quad   G_{a,a}(z_0,z_0)  \ . 
\eeqn
These expressions are zero by \leqn{Greverse}.   This identity implies
the Hermitian property for the ${\bf A}$ matrices discussed at the end
of Appendix A.

\section{Small $s$ expansion of the
 Coleman-Weinberg potentials}

In this appendix, we discuss the expansion of the Coleman-Weinberg potentials \leqn{VW}, \leqn{VZ}, \leqn{Vt}, and \leqn{VT} for small values of $s$. Here we generalize the discussion on the Coleman-Weinberg potentials in \cite{YPone} and include the effect of the boundary kinetic terms. 

Analogously to the definition of $G_{\alpha \beta}$ in \leqn{Gdef}, we define the Green's functions $G_{E\alpha \beta}$ in Euclidean momentum:  
\beq
    G_{E\alpha\beta}(z_1,z_2) =  K_{\alpha}(p_E z_1)  I_{\beta} (p_E z_2)
   - (-1)^\delta I_\alpha(p_E z_1)  K_\beta (p_E z_2) \ , 
\eeq{euclideanG}
where $(-1)^\delta = 1$ for $\alpha = \beta$ and $-1$ for $\alpha \neq \beta$. The Green's functions are positive definite. For large $p_E$, we have 
\beq
      G_{E\alpha\beta}(z_0,z_R) \sim  e^{ p_E (z_R - z_0) } \ .
\eeq{Gasympt}

First we consider the potential $V_T$ in \leqn{VT}. Note that 
\beq
s_2^2(2-s_2^2) = {1\over 2}s^2 + {1\over 16}s^4 + \O (s^6) \ .
\eeqn
Then the integrand can be expanded about $s^2 = 0$ under the integral sign. After the expansion, we get
\beq
    V_{T}(h) =   {N_T k_R^4 \over 4\pi ^2} \left[  A_T(c_T) \left( {1 \over 2} s^2 + { 1 \over 16 } s^4 \right)  + {1 \over 8} B_T(c_T) s^4  + \O (s^6) \right] \ ,
\eeq{approxVT}
where $N_T$ is the number of QCD colors of $\Psi_T$. Whether $N_T = 3$ or 1 is a model-building choice. The coefficients $A_T$ and $B_T$ are given by
\beqa
A_T(c) & = & \int_0^\infty dp_E\, p_E^3  \ { z_R^4 \over  p_E^2 z_0 z_R G_{E-+}  G_{E+-}}  \CR
B_T(c)  & = & \int_0^\infty  dp_E\,  p_E^3\   { z_R^4 \over ( p_E^2 z_0 z_R G_{E-+} G_{E+-})^2}  \ , 
\eeqa{repulsive}
and both are positive definite for all values of $c$. For $p_E \to 0$,
\beq 
    G_{E-+} G_{E+-}  =    {1\over p_E^2 z_0 z_R} (1  + \O (p_E^2)) \ , 
\eeqn
and therefore together with \leqn{Gasympt}, the integrals are
convergent. Rescaling $p \rightarrow p z_R$ in \leqn{repulsive} shows
that $A_T$ and $B_T$ depend only on the ratio $z_R/z_0$, not on $z_0$
or $z_R$ individually. For the representative case  $z_R/z_0 = 100$,
the values of these coefficients at $c = 0 $
 are
\beq
  A_T(0) = 1.4078,     \qquad     B_T(0) = 0.21694 \ ,
\eeqn
and they decrease as $c$ increases. Note that $B_T$ is much smaller than $A_T$.

We can similarly proceed for the top quark contribution $V_t$ in
\leqn{Vt}, but for this case more care is necessary due to IR
divergence of the integrand.
 Following the prescription given in \cite{YPone}, we get 

\beq
    V_{t}(h) =  {3k_R^4 \over 4\pi ^2} \left[ - {1 \over 2} A_t(c_t)
      s^2 
+ {1 \over 8} B_t(c_t) s^4 + {1 \over 8} C_t(c_t) s^4 \log {1 \over
  s^2/2 } 
+ \O (s^6) \right] 
\eeq{approxVt}

where we define
\beqa
A_t(c) & = & \int_0^\infty dp_E\,  p_E^3 \ { z_R^4 \over  p_E^2 {\bf G}_t(p_E)}  \CR
B_t(c)  & = & z_R^4 \biggl[ {1\over {\bf G}_t(0)^2} [ {1\over 4} -
  {\gamma\over 2} ] + \int_0^\infty {dp_E\over p_E} \ \bigl\{  {1\over
   {\bf G}_t(p_E)^2} - {1\over {\bf G}_t(0)^2} e^{- {\bf G}_t(0) p_E^2} \bigr\}
   \biggr] \CR
C_t(c) &=&  {z_R^4 \over 2 {\bf G}_t(0)^2 }
\eeqa{Vtcoeff}
and 
\beq
   {\bf G}_t(p_E) =    z_0 z_R G_{E++} ( G_{E--} + a_t p_E z_0 G_{E+-}) \ .
\eeqn
For $z_R/z_0 = 100$ and $a_t = 0$, the values of these coefficients at $c = 0 $ are 
\beq
  A_t(0) = 1.8771,      \qquad     B_t(0) = 0.19585, \qquad C_t(0) =  0.52051 \ ,
\eeqn
and they decrease as $c$ increases. Here we can also see that a large boundary kinetic term $a_t$ will suppress the potential. Note that $B_t$ and $C_t$ are much smaller than $A_t$.

Finally, for $V_W$ \leqn{VW} and $V_Z$ \leqn{VZ}, we have 
\beqa
    V_W(h) &=&  {3k_R^4 \over 8\pi ^2} \left[ {1 \over 2} A_W s^2 - {1 \over 8} B_W s^4 - {1 \over 8} C_W s^4 \log {1 \over s^2/2 } + \O (s^6) \right]   \CR
    V_Z(h) &=&  {3k_R^4 \over 16\pi ^2} \left[ {1 \over 2} A_Z s^2 - {1 \over 8} B_Z s^4 - {1 \over 8} C_Z s^4 \log {1 \over s^2/2 } + \O (s^6) \right] \ ,
\eeqa{approxVWZ}
where the coefficients can be obtained from \leqn{Vtcoeff} by replacing ${\bf G}_t$ with
\beqa
   {\bf G}_W(p_E) &=& z_0 z_R  G_{E++}( G_{E--}+a_W p_E z_0 G_{E+-} )  \CR
   {\bf G}_Z(p_E) &=&  { z_0 z_R  G_{E++} (G_{E--} + a_B p_E z_0 G_{E+-}) \over (G_{E--} + a_B p_E z_0 G_{E+-}) +  s_\beta^2 
        (G_{E--} + a_W p_E z_0 G_{E+-}) }
\eeqan
where $c=1/2$. It is instructive to note that because of the factor of 3 from $SU(3)_C$ color, the fermion contribution to the Higgs potential is usually larger than that of gauge bosons. In realisitic models, the gauge boson boundary kinetic term further suppresses $V_W$ and $V_Z$ and therefore makes them almost negligible compared to the potential by fermions, especially when $c_t$ and $c_T$ are small.  

Summing up, we get the expansion of the full Higgs potential \leqn{approxV}
\beq
V(h) = {k_R^4 \over 8\pi ^2} \left[ - A s^2 + {1 \over 2} B s^4 + {1 \over 2} C s^4 \log {1 \over s^2 } + \O (s^6) \right] \ ,
\eeqn
where
\beqa
A &=& 3 A_t(c_t) - N_T A_T(c_T) - {3 \over 2} A_W - {3 \over 4} A_Z \CR
B &=& {3 \over 2} (B_t(c_t) + C_t(c_t) \log 2 ) + { N_T \over 4} A_T(c_T) + {N_T \over 2} B_T(c_T) \CR
	& & \hskip 0.3in	 - {3 \over 4} (B_W + C_W \log 2)  - {3 \over 8} (B_Z + C_Z \log 2) \CR
C &=& {3 \over 2} C_t(c_t) - {3 \over 4} C_W - {3 \over 8} C_Z \ .
\eeqa{coeffs}
We can make further approximations on the potential, using that $B_{T,t}$ and $C_{T,t}$ are much smaller than $A_{T,t}$. Furthermore, the gauge boson terms are suppressed if it includes large UV boundary kinetic terms, which indeed is the case for our model. Then, we have
\beqa
A &\sim & 3 A_t(c_t) - N_T A_T(c_T) \CR
B &\sim &  { N_T \over 4} A_T(c_T) \CR
C &\sim & 0 \ .
\eeqan
If we tune $c_t$ and $c_T$ so that $A \sim 0$, we can realize $v \ll f$. In this case, we have $B \sim { 3 \over 4} A_t(c_t)$. With this crude approximation, we can get a simple relationship between the Higgs mass to the top quark mass, which is independent on $c_t$ and $a_t$. 

It should be noted that $\frac{3}{4} A_T$ in $B$ gives a large contribution to the Higgs quartic potential. This term appears as we embed $\Psi_T$ in {\bf 5} of $SO(5)$, as in \leqn{t5}. With the {\bf 4} representaion in \leqn{t4}, we do not have such term and it makes the parameter space of the {\bf 4} model more constrained than that of the {\bf 5}.

\section{Coefficients in the fermion loop correction to $T$ parameter}

In this Appendix, we calculate the coefficients $A$, $B$, and $C$
needed in the calculation of the RS correction to the $T$ parameter
from \leqn{Texpress}. 
For this, we need to compute the $LL$ components of the $t_L$ and
$b_L$ propagators in Euclidean space, as indicated in
\leqn{tbEforms}. 
These coefficients depend on the arguments of the Green's functions
$z$ and $z'$ as well as on the Euclidean momentum $p$ and $m_t$.   

In \leqn{Greensconstructpsi}, we showed that the fermion Green's
functions in the UV gauge  are a sum of  two terms, the first of which contains the
matrix    ${\bf A} = {\bf C}^{-1} {\bf D}$ and the second of which
contains the unit matrix and is independent of boundary mixing.   This
latter term is identical for $t_L$ and $b_L$, since both have $+$
boundary conditions on the IR brane.  So we will ignore this second
term, since it does not contribute to 
the difference of the propagators.

We now need to compute the ${\bf A}$ coefficients for $t_L$ and
$b_L$.  For $b_L$, an unmixed fermion with $++$ boundary conditions,
we did this calculation already in \leqn{findAforpsippwa} and found
\beq 
             {\bf A} =  -   {G_{t-+}\over G_{t--}}  \ ,
\eeq{findAforbE}
where $G_{t-\pm}$ are defined in \leqn{Gtdef}. The same result carries
over to Euclidean space, with $G$ replaced by $G_E$.  In the analysis
below, we will abbreviate $G_{E\alpha\beta}(z_0,z_R)$ by $G_{E\alpha\beta}$.

 It will be useful to
adopt a compact notation for the expansions of the $G$ functions.  We
will write
\beq
G_{E++}(z,z_R; p ) =   G_{E++} (z, z_R;p=0) \bigl[ 1 +  (p z_R)^2  {\cal
  Z}_{++}(z) + \cdots \bigr] \ , 
\eeq{calZdef}
and similarly for the other $G_E$ functions, putting the appropriate
subscript on the ${\cal Z}$ coefficient.  Using this notation, it
follows from \leqn{findAforbE} and the Euclidean version of 
\leqn{Greensconstructpsi} that
\beq
     C =  \Z_{t-+}(z_0) - \Z_{t--}(z_0) + \Z_{+-}(z) + 
\Z_{+-}(z') \    . 
\eeq{Cvalue}

To evaluate the $t_L$ propagator, we  need to compute the $3\times 3$
matrix ${\bf A}$ for this case.  We find 
\beq
  {\bf A}^{11} =  -  G_{E++} \left( {1\over p^2 z_0 z_R} + G_{E++} G_{Et--}+
  {\cal O}(s^4) \right)/
  \det {\bf C}  \ . 
\eeq{firstAoneone}
Using the Euclidean space form of the Wronskian identity
\beq
    G_{E+ + }(z_1,z_2) G_{E--}(z_1,z_2) -  G_{E+ -
    }(z_1,z_2) G_{E-+}(z_1,z_2) = -   {1\over p^2 z_1 z_2}\ ,
\eeq{myEWronsk}
this becomes
\beq
     {\bf A}^{11} =  -  G_{E++}  G_{E+-} G_{Et-+} /   \det {\bf C}  \ . 
\eeqn
Further,
\beq
   {1 \over \det {\bf C}} ={1\over G_{E++}  G_{E+-} G_{Et--}}
  {p^2\over p^2 + m_t^2} \left( 1 + (m_t z_R)^2 (\Z_{++} + \Z_{t--} ) \right)\ ,
\eeqn
so 
\beq
  {\bf A}_{11} =  - { G_{Et-+}\over G_{Et--}} {p^2\over p^2 + m_t^2} 
 \left( 1 + (m_t z_R)^2 (\Z_{++} + \Z_{t--} ) \right)\ ,
\eeq{secondAoneone}
in parallel with \leqn{findAforbE}.   Similarly,
\beq
{\bf A}_{13} = {\bf A}_{31} = {s/\sqrt{2} \over p^2 z_0 z_R G_{E++} G_{Et--}}
{p^2\over p^2 + m_t^2} \ , 
\eeqn
up to ${\cal O}(s^2)$. 

We now transform to the IR gauge using \leqn{UVIR}.  Up to ${\cal
  O}(s^2)$, the relevant terms are
\beqa
\VEV{(t_L)_L (t_L)_L^\dagger}_{IR,11} &=&  (1 - {s^2\over 2}) 
\VEV{(t_L)_L (t_L)_L^\dagger}_{UV,11} \CR & & \hskip -0.1in
+ ( -{s\over \sqrt{2}}) \VEV{(t_L)_L
   ( t_L)_L^\dagger}_{UV,13}  +  ( -{s\over \sqrt{2}}) \VEV{(t_L)_L
    (t_L)_L^\dagger}_{UV,31} 
\eeqan
We must now expand this expression and set the result into the form
\leqn{Stbdecomp}.   The terms explicitly proportional to $s^2$ are
contributions to the $A$ coefficient, since, from \leqn{mtfirstval}
\beq
      m_t^2 z_R^2 =   s^2  {2c_t+1\over 2 L_t }
\biggl({z_0 \over z_R} \biggr)^{c_t-1/2}  \ .
\eeq{smtrelation}
We then find for the  $A$ coefficient
\beqa
 A& =& {L_t\over (2c_t+1) }\biggl({z_R \over z_0}\biggr)^{c_t-1/2} \left\{
 -1 + (2c_t + 1) \left[ ({z\over z_R})^{c_t+1/2} R(z) +  ({z'\over
   z_R})^{c_t+1/2} R(z') \right] \right\}  \CR
    & & \hskip  0.3in  + \Z_{++}(z_0) + \Z_{t--}(z_0)  \   ,
\eeqa{Avalue}
where $R(z) = G_{E++}(z, z_R; p = 0)$.   The $B$ coefficient is 
\beq
B = C =  \Z_{t-+}(z_0) - \Z_{t--}(z_0) + \Z_{+-}(z) + 
\Z_{+-}(z') \    . 
\eeq{Bvalue}

To evaluate the expressions for $A$, $B$, and $C$, we need the
expansions
\beqa
   R(z) &=& {1\over 2c+1} \biggl[ \bigl(  {z_R\over z} \bigr)^{c+1/2} -( {z\over
     z_R}\bigr)^{c+1/2} \biggr] \CR 
   \Z_{++}(z_0) &=& {1\over 2 (2c+3)} \CR
\Z_{t--}(z_0) &=& {1\over 2(2c+1) L_t }\biggl({1\over (2c-3)}\biggl\{ \bigl[
 ( {z_R\over z_0})^{c-1/2} +( {z_0\over z_R})^{c-1/2} \bigr]  \CR
 & & \hskip 0.2in - {2\over 2c-1} \bigl[
 ( {z_R\over z_0})^{c-1/2} -( {z_0\over z_R})^{c-1/2} \bigr]\biggr\} + 
 a_t  ( {z_R\over z_0})^{c-1/2} \biggr) \CR
\Z_{t-+} (z_0) &=&  - {1\over 2 (2c+1) } \biggl[ 1 + {2\over 2c-1}
\bigl( 1 - ( {z_R\over z_0})^{2c-1} \bigr) \biggr] + {a_t\over 2c+1} ( {z_R\over z_0})^{2c-1}
\CR
\Z_{+-}(z) &=& {1\over 2(2c+1)} \biggl( 1 - {z^2\over z_R^2} \bigl[ 1
+ {2 \over 2c-1} (1 -  ( {z\over z_R})^{2c-1} ) \bigr] \biggr) \ . 
\eeqan
We have made the above formulae somewhat simpler 
by ignoring factors of
$(z_0/z_R)$ and $(z_0/z_R)^{c + 1/2}$ (but not $(z_0/z_R)^{c - 1/2}$)
for the relevant values  $c > 0.3$. 
Also note that there is an identity between the expressions above, 
\beq 
  {z_0 \over z_R} R(z_0)  L_t =  \Z_{+-}(z_0) + \Z_{t-+} (z_0)
\eeq{RLidentity}
which follows from the Wronskian identity \leqn{myEWronsk}. 

Using these formulae, our estimate of the correction to $T$ can be
written as
\beqa
T &\approx& {3 m_t^2  \over 16 \pi s_w^2 c_w^2 m_Z^2}
 \Biggl\{ 2 (m_t z_R)^2 \biggl[\VEV{\Z_{+-}(z) +
     \Z_{+-}(z')} - 2 \Z_{t--}(z_0) - \Z_{++}(z_0) -\Z_{+-}(z_0)   \biggr] \CR
   & & \hskip 1.0in +  s^2 \VEV{ ( {z\over z_R})^{1+2c_t} + ( {z'\over
       z_R })^{1+2c_t} } 
\Biggr\} \cdot \log \left(\Lambda^2/m_t^2\right)\    .
\eeqa{fullTvalue}

In our discussion of parameters, we saw that $a_t$ has a large value,
of order 10.   Then it makes sense to extract the terms in
\leqn{fullTvalue} that are enhanced by a power of $a_t$.   These 
come from the term with $s^2$, which is proportional to $L_t$ through
\leqn{smtrelation}.  Keeping only this term, we find a much simpler
expression,
which is quoted in \leqn{finalTresult}. 
However, the small values of the expectation values of $z$ and $z'$
counterbalance the large value of $a_t$, so this parametrically large 
contributin is not actually dominant.
   The values of $T$ in 
Fig.~\ref{fig:ST} are evaluated with the full expression
\leqn{fullTvalue}.

\Acknowledgements

We are grateful to Christophe Grojean, Howard Haber, Hitoshi Murayama,
Yael Shadmi, and
the members of the SLAC Theory Group for informative discussions of the
topics presented in this paper. 
This work was supported by the U.S. Department 
of Energy under contract DE--AC02--76SF00515. JY is supported by a Kwanjeong Graduate
Fellowship.

\end{document}